\documentclass[%
 preprint,
 aps,
 prd,
 superscriptaddress
]{revtex4-2}
\usepackage{tocloft,graphicx,bm,dcolumn,url,textcase,natbib}
\usepackage{ifpdf,amsmath, amssymb,textcomp,mathrsfs,mathtools,xcolor,hyperref}
\usepackage{listofitems}
\usepackage{tikz}
\usetikzlibrary{calc,decorations.markings,shapes.geometric}

\definecolor{mhvblue2}{rgb}{0.3,0.3,0.575}

\definecolor{mhvblue}{rgb}{0.6,0.6,0.7765}
\definecolor{nmhvred}{rgb}{0.6765,0.15,0.3}
\definecolor{ampgrey}{rgb}{0.9,0.9,0.9}
\definecolor{unord}{rgb}{0,0,0}
\definecolor{ord}{rgb}{0,0,0.575}
\definecolor{anchorLeg}{rgb}{0.575,0.0,0.225}
\definecolor{labelcolor}{rgb}{0,0,0}
\definecolor{extColor}{rgb}{0,0,0}

\newcounter{legSteps}
\newcounter{offset}

\def\figScale{0.9}
\def\legSpread{4}
\def\extLegLen{0.9*0.32*\figScale}
\def\edgeLen{1*\figScale}

\pgfmathsetmacro{\pLen}{\edgeLen/(2*sin(72/2))}
\def\legLen{\edgeLen*0.45}

\def\labelDist{\legLen*1.5}
\def\lineThickness{(1pt)}
\def\dotSize{(\figScale*12pt)}
\def\ampSize{(1*\figScale*12pt)}
\def\eph{0.4}

\tikzset{fullamp/.style={coordinate,minimum size=0.7*\ampSize,ball color=black!20,circle,text=white,inner sep=0}}
\tikzset{fullmhv/.style={coordinate,minimum size=0.8*\ampSize,ball color=mhvblue,circle,text=white,inner sep=0}}
\tikzset{fullmhvBig/.style={coordinate,minimum size=1*\ampSize,ball color=mhvblue,circle,text=white,inner sep=0}}
\tikzset{fullnmhv/.style={coordinate,minimum size=0.8*\ampSize,ball color=nmhvred,circle,text=white,inner sep=0}}
\tikzset{fullmhvBar/.style={coordinate,minimum size=0.8*\ampSize,ball color=white,circle,text=white,inner sep=0}}
\tikzset{ordAmp/.style={fill=ampgrey,circle,draw=black,line width=\lineThickness,minimum size=0.6*\ampSize,text=white,inner sep=0}}
\tikzset{mhv/.style={fill=mhvblue,circle,draw=black,line width=\lineThickness,minimum size=0.7*\ampSize,text=white,inner sep=0}}
\tikzset{nmhv/.style={fill=nmhvred,circle,draw=black,line width=\lineThickness,minimum size=0.7*\ampSize,text=white,inner sep=0}}
\tikzset{mhvBar/.style={fill=white,circle,draw=black,line width=\lineThickness,minimum size=0.7*\ampSize,text=white,inner sep=0}}
\tikzset{rEdge/.style={anchorLeg,line width=\lineThickness,line cap=round}}
\tikzset{fgraphEdge/.style={anchorLeg,line width=0.8*\lineThickness,line cap=round}}
\tikzset{fgraphExt/.style={ord,line width=\lineThickness,line cap=round}}
\tikzset{fgraphOpt/.style={ord,dotted,line width=\lineThickness,line cap=round}}
\tikzset{fdot/.style={fill=anchorLeg,circle,minimum size=0.35*\ampSize,inner sep=0}}
\tikzset{bdot/.style={fill=black,circle,minimum size=0.35*\ampSize,inner sep=0}}
\tikzset{ephdot/.style={fill=black,circle,minimum size=0.125*\ampSize,inner sep=0}}
\tikzset{ext/.style={black,line width=\lineThickness,line cap=round,rounded corners=0.1pt}}
\tikzset{under/.style={white,line width=4*\lineThickness,line cap=round}}
\tikzset{optExt/.style={black,dotted,line width=\lineThickness,line cap=round,rounded corners=10pt}}
\tikzset{optExtSc/.style={black,line width=\lineThickness,line cap=round,rounded corners=10pt}}
\tikzset{dashed/.style={black!70,dotted,line width=\lineThickness,line cap=round,rounded corners=10pt}}
\tikzset{ddot/.style={fill=black,circle,minimum size=0.275*\dotSize,inner sep=0}}
\tikzset{rdot/.style={fill=hred,circle,minimum size=0.275*\dotSize,inner sep=0}}
\tikzset{bldot/.style={fill=hblue,circle,minimum size=0.275*\dotSize,inner sep=0}}
\tikzset{int/.style={black,line width=\lineThickness,line cap=round,rounded corners=1.5pt}}

\tikzset{intInfR/.style={nmhvred,line width=\lineThickness,line cap=round,rounded corners=1.5pt}}
\tikzset{blueDot/.style={fill=mhvblue,circle,draw=black,line width=\lineThickness,minimum size=0.5*\ampSize,text=white,inner sep=0}}
\tikzset{whiteDot/.style={fill=white,circle,draw=black,line width=\lineThickness,minimum size=0.5*\ampSize,text=white,inner sep=0}}
\tikzset{blackDot/.style={fill=black,circle,minimum size=0.5*\ampSize,inner sep=0}}
\tikzset{compositeDot/.style={fill=none,draw=black,line width=\lineThickness,circle,minimum size=0.75*\ampSize,inner sep=0}}

\tikzset{directedEdge/.style={draw=none,decoration={markings,mark connection node=connode,mark=at position 0.5 with {\node[transform shape, scale=0.25*\figScale,shape=dart,aspect=0.5,fill=black,draw] (connode) {};}},postaction={decorate}}}

\tikzset{directedEdgeBend/.style={rounded corners=10pt,draw=none,decoration={markings,mark connection node=connode,mark=at position 0.5 with {\node[transform shape, scale=0.205,shape=dart,aspect=0.5,fill=black,draw] (connode) {};}},postaction={decorate}}}

\newcommand{\dimLines}{\tikzset{int/.style={black!25,line width=\lineThickness,line cap=round,rounded corners=1.5pt}}
\tikzset{ext/.style={black!25,line width=\lineThickness,line cap=round,rounded corners=0.1pt}}\tikzset{directedEdge/.style={draw=none,decoration={markings,mark connection node=connode,mark=at position 0.5 with {\node[transform shape, scale=0.205*\figScale,shape=dart,black!25,aspect=0.5,fill=black!25,draw] (connode) {};}},postaction={decorate}}}
\definecolor{extColor}{rgb}{0.75,0.75,0.75}}

\newcommand{\restoreDark}{
\tikzset{int/.style={black,line width=\lineThickness,line cap=round,rounded corners=1.5pt}}
\tikzset{ext/.style={black,line width=\lineThickness,line cap=round,rounded corners=0.1pt}}\tikzset{directedEdge/.style={draw=none,decoration={markings,mark connection node=connode,mark=at position 0.5 with {\node[transform shape, scale=0.205*\figScale,shape=dart,aspect=0.5,fill=black,draw] (connode) {};}},postaction={decorate}}}
\definecolor{extColor}{rgb}{0,0,0}}

\newcommand{\leg}[3]{\draw[ext] #1--($#1+(#2:\legLen)$);\node at ($#1+(#2:\labelDist)$)[]{{\footnotesize #3}};}

\newcommand{\legMassive}[3]{\draw[ext,fill=extColor] #1--($#1+(#2+\legSpread*2:\legLen)$)--($#1+(#2-\legSpread*2:\legLen)$)--#1;\node at ($#1+(#2:\labelDist)$)[]{{\footnotesize #3}};}

\def\boundingDraw{red}
\def\boundingDraw{none}

\newcommand{\contourVerts}[7]{
\ifthenelse{#1=1}{\node at (v1) [whiteDot] {};}{\ifthenelse{#1=2}{\node at (v1) [blueDot] {};}{\ifthenelse{#1=4}{\node at (v1) [blackDot] {};}{\ifthenelse{#1=3}{\node at (v1) [bdot] {};\node at (v1) [compositeDot] {};
}{\ifthenelse{#1=0}{\node at (v1) [bdot] {};}{}}}}}

\ifthenelse{#2=1}{\node at (v2) [whiteDot] {};}{\ifthenelse{#2=2}{\node at (v2) [blueDot] {};}{\ifthenelse{#2=4}{\node at (v2) [blackDot] {};}{\ifthenelse{#2=3}{\node at (v2) [bdot] {};\node at (v2) [compositeDot] {};
}{\ifthenelse{#2=0}{\node at (v2) [bdot] {};}{}}}}}

\ifthenelse{#3=1}{\node at (v3) [whiteDot] {};}{\ifthenelse{#3=2}{\node at (v3) [blueDot] {};}{\ifthenelse{#3=4}{\node at (v3) [blackDot] {};}{\ifthenelse{#3=3}{\node at (v3) [bdot] {};\node at (v3) [compositeDot] {};
}{\ifthenelse{#3=0}{\node at (v3) [bdot] {};}{}}}}}

\ifthenelse{#4=1}{\node at (v4) [whiteDot] {};}{\ifthenelse{#4=2}{\node at (v4) [blueDot] {};}{\ifthenelse{#4=4}{\node at (v4) [blackDot] {};}{\ifthenelse{#4=3}{\node at (v4) [bdot] {};\node at (v4) [compositeDot] {};
}{\ifthenelse{#4=0}{\node at (v4) [bdot] {};}{}}}}}

\ifthenelse{#5=1}{\node at (v5) [whiteDot] {};}{\ifthenelse{#5=2}{\node at (v5) [blueDot] {};}{\ifthenelse{#5=4}{\node at (v5) [blackDot] {};}{\ifthenelse{#5=3}{\node at (v5) [bdot] {};\node at (v5) [compositeDot] {};
}{\ifthenelse{#5=0}{\node at (v5) [bdot] {};}{}}}}}
\ifthenelse{#6=1}{\node at (v6) [whiteDot] {};}{\ifthenelse{#6=2}{\node at (v6) [blueDot] {};}{\ifthenelse{#6=4}{\node at (v6) [blackDot] {};}{\ifthenelse{#6=3}{\node at (v6) [bdot] {};\node at (v6) [compositeDot] {};
}{\ifthenelse{#6=0}{\node at (v6) [bdot] {};}{}}}}}
\ifthenelse{#7=1}{\node at (v7) [whiteDot] {};}{\ifthenelse{#7=2}{\node at (v7) [blueDot] {};}{\ifthenelse{#7=4}{\node at (v7) [blackDot] {};}{\ifthenelse{#7=3}{\node at (v7) [bdot] {};\node at (v7) [compositeDot] {};
}{\ifthenelse{#7=0}{\node at (v7) [bdot] {};}{}}}}}
}

\definecolor{legColour}{rgb}{0.35,0.35,0.35}
\definecolor{ndotColor}{rgb}{0.65,0.25,0.25}
\def\markStroke{0.65}

\tikzset{ndot/.style={transform shape,scale=0.65*\figScale,aspect=0.75,draw=ndotColor,line width=1.4*\markStroke*\figScale,shape=circle,fill=none}}

\tikzset{markedEdgeR/.style={draw=none,decoration={markings,mark connection node=connode,mark=at position 0.5 with {\node[ndotR] (connode) {};}},postaction={decorate}}}

\tikzset{ndotR/.style={transform shape,scale=0.65*\figScale,aspect=0.65,draw=ndotColor,line width=\markStroke*\figScale,shape=circle,fill=ndotColor}}

\tikzset{markedEdge/.style={draw=none,decoration={markings,mark connection node=connode,mark=at position 0.5 with {\node[ndot] (connode) {};}},postaction={decorate}}}

\newcommand{\legAlt}[2]{
\fill[legColour] #1--($#1+(#2+\legSpread*3:\extLegLen)$)--($#1+(#2-\legSpread*3:\extLegLen)$);
\node at #1 [ddot]{};
}

\newcommand{\oneLoopGraphElement}[2][0]{\def\rotn{-360/#2}\def\edgeLen{0.75*\figScale}
\ifthenelse{#2=3}{\coordinate (v0) at (0,-\edgeLen/9)}{\coordinate (v0) at (0,0)};
\ifthenelse{#2=1}{\draw[int]($(v0)+(\edgeLen/2*0.3,0)$) arc (0:360:\edgeLen/2*0.3 and \edgeLen/1.125*0.3)coordinate[pos=0.25](e1);\legAlt{($(v0)-(0,\edgeLen/1.125*0.3)$)}{-90}}{\ifthenelse{#2=2}{
\draw[int]($(v0)+(\edgeLen/1.125*0.45,0)$) arc (0:360:\edgeLen/1.125*0.45 and \edgeLen/2*0.45)coordinate[pos=0.5](a1)coordinate[pos=0](a0)coordinate[pos=0.25](e1)coordinate[pos=0.75](e2);\legAlt{(a1)}{180};\legAlt{(a0)}{0};
}{
\foreach\a in {0,...,#2}{\coordinate (a\a) at ($(v0)+(\a*\rotn-90-\rotn/2:\edgeLen/2)$);};
\foreach\a[remember=\a as \la] in {0,...,#2}{\ifthenelse{\a=0}{}{\draw[int](a\la)--(a\a)coordinate (e\a) at ($(a\la)!0.5!(a\a)$);}};
\foreach\a in {1,...,#2}{\legAlt{(a\a)}{\a*\rotn-90-\rotn/2};};
}};
}

\newcommand{\oneLoopGraphElementLS}[2][0]{\def\extLegLen{1.1*0.32*\figScale}
\def\rotn{-360/#2}\def\edgeLen{0.75*\figScale}
\ifthenelse{#2=3}{\coordinate (v0) at (0,-\edgeLen/9)}{\coordinate (v0) at (0,0)};
\ifthenelse{#2=1}{\draw[int]($(v0)+(\edgeLen/2*0.3,0)$) arc (0:360:\edgeLen/2*0.3 and \edgeLen/1.125*0.3)coordinate[pos=0.25](e1);\legAlt{($(v0)-(0,\edgeLen/1.125*0.3)$)}{-90}}{\ifthenelse{#2=2}{
\draw[int]($(v0)+(\edgeLen/1.125*0.45,0)$) arc (0:360:\edgeLen/1.125*0.45 and \edgeLen/2*0.45)coordinate[pos=0.5](a1)coordinate[pos=0](a0)coordinate[pos=0.25](e1)coordinate[pos=0.75](e2);\legAlt{(a1)}{180};\legAlt{(a0)}{0};
}{
\foreach\a in {0,...,#2}{\coordinate (a\a) at ($(v0)+(\a*\rotn-90-\rotn/2:\edgeLen/2)$);};
\foreach\a[remember=\a as \la] in {0,...,#2}{\ifthenelse{\a=0}{}{\draw[int](a\la)--(a\a)coordinate (e\a) at ($(a\la)!0.5!(a\a)$);}};
\foreach\a in {1,...,#2}{\legAlt{(a\a)}{\a*\rotn-90-\rotn/2};};
}};\def\extLegLen{0.9*0.32*\figScale}
}

\newcommand{\lsVerts}[7]{
\ifthenelse{#1=1}{\node at (v1) [fullmhvBar] {};}{\ifthenelse{#1=2}{\node at (v1) [fullmhv] {};}{\ifthenelse{#1=3}{\node at (v1) [fullnmhv] {};}{\ifthenelse{#1=0}{\node at (v1) [bdot] {};}{}}}}
\ifthenelse{#2=1}{\node at (v2) [fullmhvBar] {};}{\ifthenelse{#2=2}{\node at (v2) [fullmhv] {};}{\ifthenelse{#2=3}{\node at (v2) [fullnmhv] {};}{\ifthenelse{#2=0}{\node at (v2) [bdot] {};}{}}}}
\ifthenelse{#3=1}{\node at (v3) [fullmhvBar] {};}{\ifthenelse{#3=2}{\node at (v3) [fullmhv] {};}{\ifthenelse{#3=3}{\node at (v3) [fullnmhv] {};}{\ifthenelse{#3=0}{\node at (v3) [bdot] {};}{}}}}
\ifthenelse{#4=1}{\node at (v4) [fullmhvBar] {};}{\ifthenelse{#4=2}{\node at (v4) [fullmhv] {};}{\ifthenelse{#4=3}{\node at (v4) [fullnmhv] {};}{\ifthenelse{#4=0}{\node at (v4) [bdot] {};}{}}}}
\ifthenelse{#5=1}{\node at (v5) [fullmhvBar] {};}{\ifthenelse{#5=2}{\node at (v5) [fullmhv] {};}{\ifthenelse{#5=3}{\node at (v5) [fullnmhv] {};}{\ifthenelse{#5=0}{\node at (v5) [bdot] {};}{}}}}
\ifthenelse{#6=1}{\node at (v6) [fullmhvBar] {};}{\ifthenelse{#6=2}{\node at (v6) [fullmhv] {};}{\ifthenelse{#6=3}{\node at (v6) [fullnmhv] {};}{\ifthenelse{#6=0}{\node at (v6) [bdot] {};}{}}}}
\ifthenelse{#7=1}{\node at (v7) [fullmhvBar] {};}{\ifthenelse{#7=2}{\node at (v7) [fullmhv] {};}{\ifthenelse{#7=3}{\node at (v7) [fullnmhv] {};}{\ifthenelse{#7=0}{\node at (v7) [bdot] {};}{}}}}}

\newcommand{\lsVertsOrd}[7]{
\ifthenelse{#1=1}{\node at (v1) [mhvBar] {};}{\ifthenelse{#1=2}{\node at (v1) [mhv] {};}{\ifthenelse{#1=3}{\node at (v1) [nmhv] {};}{\ifthenelse{#1=0}{\node at (v1) [bdot] {};}{}}}}
\ifthenelse{#2=1}{\node at (v2) [mhvBar] {};}{\ifthenelse{#2=2}{\node at (v2) [mhv] {};}{\ifthenelse{#2=3}{\node at (v2) [nmhv] {};}{\ifthenelse{#2=0}{\node at (v2) [bdot] {};}{}}}}
\ifthenelse{#3=1}{\node at (v3) [mhvBar] {};}{\ifthenelse{#3=2}{\node at (v3) [mhv] {};}{\ifthenelse{#3=3}{\node at (v3) [nmhv] {};}{\ifthenelse{#3=0}{\node at (v3) [bdot] {};}{}}}}
\ifthenelse{#4=1}{\node at (v4) [mhvBar] {};}{\ifthenelse{#4=2}{\node at (v4) [mhv] {};}{\ifthenelse{#4=3}{\node at (v4) [nmhv] {};}{\ifthenelse{#4=0}{\node at (v4) [bdot] {};}{}}}}
\ifthenelse{#5=1}{\node at (v5) [mhvBar] {};}{\ifthenelse{#5=2}{\node at (v5) [mhv] {};}{\ifthenelse{#5=3}{\node at (v5) [nmhv] {};}{\ifthenelse{#5=0}{\node at (v5) [bdot] {};}{}}}}
\ifthenelse{#6=1}{\node at (v6) [mhvBar] {};}{\ifthenelse{#6=2}{\node at (v6) [mhv] {};}{\ifthenelse{#6=3}{\node at (v6) [nmhv] {};}{\ifthenelse{#6=0}{\node at (v6) [bdot] {};}{}}}}
\ifthenelse{#7=1}{\node at (v7) [mhvBar] {};}{\ifthenelse{#7=2}{\node at (v7) [mhv] {};}{\ifthenelse{#7=3}{\node at (v7) [fullnmhv] {};}{\ifthenelse{#7=0}{\node at (v7) [bdot] {};}{}}}}}

\newcommand{\kBoxLegs}[7]{
\setcounter{legSteps}{0}
\def\zeroAngle{-90}\def\spread{45}\setcounter{offset}{-1}\addtocounter{offset}{#1}
\ifthenelse{#1=0}{}{\ifthenelse{#1=1}{\stepcounter{legSteps}\leg{(v1)}{\zeroAngle}{\arabic{legSteps}};}{
\foreach\n in {1,...,#1}{\def\eph{\arabic{offset}}\def\angle{\zeroAngle-2*\n*\spread/\eph+#1*\spread/\eph+\spread/\eph}\stepcounter{legSteps}\leg{(v1)}{\angle}{\arabic{legSteps}}}}}

\def\zeroAngle{180}\def\spread{45}\setcounter{offset}{-1}\addtocounter{offset}{#2}
\ifthenelse{#2=0}{}{\ifthenelse{#2=1}{\stepcounter{legSteps}\leg{(v2)}{\zeroAngle}{\arabic{legSteps}};}{
\foreach\n in {1,...,#2}{\def\eph{\arabic{offset}}\def\angle{\zeroAngle-2*\n*\spread/\eph+#2*\spread/\eph+\spread/\eph}\stepcounter{legSteps}\leg{(v2)}{\angle}{\arabic{legSteps}}}}}

\def\zeroAngle{90}\def\spread{45}\setcounter{offset}{-1}\addtocounter{offset}{#3}
\ifthenelse{#3=0}{}{\ifthenelse{#3=1}{\stepcounter{legSteps}\leg{(v3)}{\zeroAngle}{\arabic{legSteps}};}{
\foreach\n in {1,...,#3}{\def\eph{\arabic{offset}}\def\angle{\zeroAngle-2*\n*\spread/\eph+#3*\spread/\eph+\spread/\eph}\stepcounter{legSteps}\leg{(v3)}{\angle}{\arabic{legSteps}}}}}

\def\zeroAngle{90}\def\spread{15}\setcounter{offset}{-1}\addtocounter{offset}{#4}
\ifthenelse{#4=0}{}{\ifthenelse{#4=1}{\stepcounter{legSteps}\leg{(v4)}{\zeroAngle}{\arabic{legSteps}};}{
\foreach\n in {1,...,#4}{\def\eph{\arabic{offset}}\def\angle{\zeroAngle-2*\n*\spread/\eph+#4*\spread/\eph+\spread/\eph}\stepcounter{legSteps}\leg{(v4)}{\angle}{\arabic{legSteps}}}}}

\def\zeroAngle{90}\def\spread{45}\setcounter{offset}{-1}\addtocounter{offset}{#5}
\ifthenelse{#5=0}{}{\ifthenelse{#5=1}{\stepcounter{legSteps}\leg{(v5)}{\zeroAngle}{\arabic{legSteps}};}{
\foreach\n in {1,...,#5}{\def\eph{\arabic{offset}}\def\angle{\zeroAngle-2*\n*\spread/\eph+#5*\spread/\eph+\spread/\eph}\stepcounter{legSteps}\leg{(v5)}{\angle}{\arabic{legSteps}}}}}

\def\zeroAngle{0}\def\spread{45}\setcounter{offset}{-1}\addtocounter{offset}{#6}
\ifthenelse{#6=0}{}{\ifthenelse{#6=1}{\stepcounter{legSteps}\leg{(v6)}{\zeroAngle}{\arabic{legSteps}};}{
\foreach\n in {1,...,#6}{\def\eph{\arabic{offset}}\def\angle{\zeroAngle-2*\n*\spread/\eph+#6*\spread/\eph+\spread/\eph}\stepcounter{legSteps}\leg{(v6)}{\angle}{\arabic{legSteps}}}}}

\def\zeroAngle{-90}\def\spread{45}\setcounter{offset}{-1}\addtocounter{offset}{#7}
\ifthenelse{#7=0}{}{\ifthenelse{#7=1}{\stepcounter{legSteps}\leg{(v7)}{\zeroAngle}{\arabic{legSteps}};}{
\foreach\n in {1,...,#7}{\def\eph{\arabic{offset}}\def\angle{\zeroAngle-2*\n*\spread/\eph+#7*\spread/\eph+\spread/\eph}\stepcounter{legSteps}\leg{(v7)}{\angle}{\arabic{legSteps}}}}}
}

\newcommand{\kTboxLegs}[6]{
\setcounter{legSteps}{0}
\def\zeroAngle{-135}\def\spread{45}\setcounter{offset}{-1}\addtocounter{offset}{#1}
\ifthenelse{#1=0}{}{\ifthenelse{#1=1}{\stepcounter{legSteps}\leg{(v1)}{\zeroAngle}{\arabic{legSteps}};}{
\foreach\n in {1,...,#1}{\def\eph{\arabic{offset}}\def\angle{\zeroAngle-2*\n*\spread/\eph+#1*\spread/\eph+\spread/\eph}\stepcounter{legSteps}\leg{(v1)}{\angle}{\arabic{legSteps}}}}}

\def\zeroAngle{135}\def\spread{45}\setcounter{offset}{-1}\addtocounter{offset}{#2}
\ifthenelse{#2=0}{}{\ifthenelse{#2=1}{\stepcounter{legSteps}\leg{(v2)}{\zeroAngle}{\arabic{legSteps}};}{
\foreach\n in {1,...,#2}{\def\eph{\arabic{offset}}\def\angle{\zeroAngle-2*\n*\spread/\eph+#2*\spread/\eph+\spread/\eph}\stepcounter{legSteps}\leg{(v2)}{\angle}{\arabic{legSteps}}}}}

\def\zeroAngle{90}\def\spread{15}\setcounter{offset}{-1}\addtocounter{offset}{#3}
\ifthenelse{#3=0}{}{\ifthenelse{#3=1}{\stepcounter{legSteps}\leg{(v3)}{\zeroAngle}{\arabic{legSteps}};}{
\foreach\n in {1,...,#3}{\def\eph{\arabic{offset}}\def\angle{\zeroAngle-2*\n*\spread/\eph+#3*\spread/\eph+\spread/\eph}\stepcounter{legSteps}\leg{(v3)}{\angle}{\arabic{legSteps}}}}}

\def\zeroAngle{90}\def\spread{45}\setcounter{offset}{-1}\addtocounter{offset}{#4}
\ifthenelse{#4=0}{}{\ifthenelse{#4=1}{\stepcounter{legSteps}\leg{(v4)}{\zeroAngle}{\arabic{legSteps}};}{
\foreach\n in {1,...,#4}{\def\eph{\arabic{offset}}\def\angle{\zeroAngle-2*\n*\spread/\eph+#4*\spread/\eph+\spread/\eph}\stepcounter{legSteps}\leg{(v4)}{\angle}{\arabic{legSteps}}}}}

\def\zeroAngle{0}\def\spread{45}\setcounter{offset}{-1}\addtocounter{offset}{#5}
\ifthenelse{#5=0}{}{\ifthenelse{#5=1}{\stepcounter{legSteps}\leg{(v5)}{\zeroAngle}{\arabic{legSteps}};}{
\foreach\n in {1,...,#5}{\def\eph{\arabic{offset}}\def\angle{\zeroAngle-2*\n*\spread/\eph+#5*\spread/\eph+\spread/\eph}\stepcounter{legSteps}\leg{(v5)}{\angle}{\arabic{legSteps}}}}}

\def\zeroAngle{-90}\def\spread{45}\setcounter{offset}{-1}\addtocounter{offset}{#6}
\ifthenelse{#6=0}{}{\ifthenelse{#6=1}{\stepcounter{legSteps}\leg{(v6)}{\zeroAngle}{\arabic{legSteps}};}{
\foreach\n in {1,...,#6}{\def\eph{\arabic{offset}}\def\angle{\zeroAngle-2*\n*\spread/\eph+#6*\spread/\eph+\spread/\eph}\stepcounter{legSteps}\leg{(v6)}{\angle}{\arabic{legSteps}}}}}
}

\newcommand{\kTLegs}[5]{
\setcounter{legSteps}{0}
\def\zeroAngle{-135}\def\spread{45}\setcounter{offset}{-1}\addtocounter{offset}{#1}
\ifthenelse{#1=0}{}{\ifthenelse{#1=1}{\stepcounter{legSteps}\leg{(v1)}{\zeroAngle}{\arabic{legSteps}};}{
\foreach\n in {1,...,#1}{\def\eph{\arabic{offset}}\def\angle{\zeroAngle-2*\n*\spread/\eph+#1*\spread/\eph+\spread/\eph}\stepcounter{legSteps}\leg{(v1)}{\angle}{\arabic{legSteps}}}}}

\def\zeroAngle{135}\def\spread{45}\setcounter{offset}{-1}\addtocounter{offset}{#2}
\ifthenelse{#2=0}{}{\ifthenelse{#2=1}{\stepcounter{legSteps}\leg{(v2)}{\zeroAngle}{\arabic{legSteps}};}{
\foreach\n in {1,...,#2}{\def\eph{\arabic{offset}}\def\angle{\zeroAngle-2*\n*\spread/\eph+#2*\spread/\eph+\spread/\eph}\stepcounter{legSteps}\leg{(v2)}{\angle}{\arabic{legSteps}}}}}

\def\zeroAngle{90}\def\spread{15}\setcounter{offset}{-1}\addtocounter{offset}{#3}
\ifthenelse{#3=0}{}{\ifthenelse{#3=1}{\stepcounter{legSteps}\leg{(v3)}{\zeroAngle}{\arabic{legSteps}};}{
\foreach\n in {1,...,#3}{\def\eph{\arabic{offset}}\def\angle{\zeroAngle-2*\n*\spread/\eph+#3*\spread/\eph+\spread/\eph}\stepcounter{legSteps}\leg{(v3)}{\angle}{\arabic{legSteps}}}}}

\def\zeroAngle{45}\def\spread{45}\setcounter{offset}{-1}\addtocounter{offset}{#4}
\ifthenelse{#4=0}{}{\ifthenelse{#4=1}{\stepcounter{legSteps}\leg{(v4)}{\zeroAngle}{\arabic{legSteps}};}{
\foreach\n in {1,...,#4}{\def\eph{\arabic{offset}}\def\angle{\zeroAngle-2*\n*\spread/\eph+#4*\spread/\eph+\spread/\eph}\stepcounter{legSteps}\leg{(v4)}{\angle}{\arabic{legSteps}}}}}

\def\zeroAngle{-45}\def\spread{45}\setcounter{offset}{-1}\addtocounter{offset}{#5}
\ifthenelse{#5=0}{}{\ifthenelse{#5=1}{\stepcounter{legSteps}\leg{(v5)}{\zeroAngle}{\arabic{legSteps}};}{
\foreach\n in {1,...,#5}{\def\eph{\arabic{offset}}\def\angle{\zeroAngle-2*\n*\spread/\eph+#5*\spread/\eph+\spread/\eph}\stepcounter{legSteps}\leg{(v5)}{\angle}{\arabic{legSteps}}}}}
}

\newcommand{\pBoxLegs}[7]{
\setcounter{legSteps}{0}
\def\zeroAngle{-108}\def\spread{45}\setcounter{offset}{-1}\addtocounter{offset}{#1}
\ifthenelse{#1=0}{}{\ifthenelse{#1=1}{\stepcounter{legSteps}\leg{(v1)}{\zeroAngle}{\arabic{legSteps}};}{
\foreach\n in {1,...,#1}{\def\eph{\arabic{offset}}\def\angle{\zeroAngle-2*\n*\spread/\eph+#1*\spread/\eph+\spread/\eph}\stepcounter{legSteps}\leg{(v1)}{\angle}{\arabic{legSteps}}}}}

\def\zeroAngle{180}\def\spread{45}\setcounter{offset}{-1}\addtocounter{offset}{#2}
\ifthenelse{#2=0}{}{\ifthenelse{#2=1}{\stepcounter{legSteps}\leg{(v2)}{\zeroAngle}{\arabic{legSteps}};}{
\foreach\n in {1,...,#2}{\def\eph{\arabic{offset}}\def\angle{\zeroAngle-2*\n*\spread/\eph+#2*\spread/\eph+\spread/\eph}\stepcounter{legSteps}\leg{(v2)}{\angle}{\arabic{legSteps}}}}}

\def\zeroAngle{108}\def\spread{45}\setcounter{offset}{-1}\addtocounter{offset}{#3}
\ifthenelse{#3=0}{}{\ifthenelse{#3=1}{\stepcounter{legSteps}\leg{(v3)}{\zeroAngle}{\arabic{legSteps}};}{
\foreach\n in {1,...,#3}{\def\eph{\arabic{offset}}\def\angle{\zeroAngle-2*\n*\spread/\eph+#3*\spread/\eph+\spread/\eph}\stepcounter{legSteps}\leg{(v3)}{\angle}{\arabic{legSteps}}}}}

\def\zeroAngle{81}\def\spread{20}\setcounter{offset}{-1}\addtocounter{offset}{#4}
\ifthenelse{#4=0}{}{\ifthenelse{#4=1}{\stepcounter{legSteps}\leg{(v4)}{\zeroAngle}{\arabic{legSteps}};}{
\foreach\n in {1,...,#4}{\def\eph{\arabic{offset}}\def\angle{\zeroAngle-2*\n*\spread/\eph+#4*\spread/\eph+\spread/\eph}\stepcounter{legSteps}\leg{(v4)}{\angle}{\arabic{legSteps}}}}}

\def\zeroAngle{45}\def\spread{45}\setcounter{offset}{-1}\addtocounter{offset}{#5}
\ifthenelse{#5=0}{}{\ifthenelse{#5=1}{\stepcounter{legSteps}\leg{(v5)}{\zeroAngle}{\arabic{legSteps}};}{
\foreach\n in {1,...,#5}{\def\eph{\arabic{offset}}\def\angle{\zeroAngle-2*\n*\spread/\eph+#5*\spread/\eph+\spread/\eph}\stepcounter{legSteps}\leg{(v5)}{\angle}{\arabic{legSteps}}}}}

\def\zeroAngle{-45}\def\spread{45}\setcounter{offset}{-1}\addtocounter{offset}{#6}
\ifthenelse{#6=0}{}{\ifthenelse{#6=1}{\stepcounter{legSteps}\leg{(v6)}{\zeroAngle}{\arabic{legSteps}};}{
\foreach\n in {1,...,#6}{\def\eph{\arabic{offset}}\def\angle{\zeroAngle-2*\n*\spread/\eph+#6*\spread/\eph+\spread/\eph}\stepcounter{legSteps}\leg{(v6)}{\angle}{\arabic{legSteps}}}}}

\def\zeroAngle{-81}\def\spread{20}\setcounter{offset}{-1}\addtocounter{offset}{#7}
\ifthenelse{#7=0}{}{\ifthenelse{#7=1}{\stepcounter{legSteps}\leg{(v7)}{\zeroAngle}{\arabic{legSteps}};}{
\foreach\n in {1,...,#7}{\def\eph{\arabic{offset}}\def\angle{\zeroAngle-2*\n*\spread/\eph+#7*\spread/\eph+\spread/\eph}\stepcounter{legSteps}\leg{(v7)}{\angle}{\arabic{legSteps}}}}}
}

\newcommand{\hBoxLegs}[7]{
\setcounter{legSteps}{0}
\def\zeroAngle{-115}\def\spread{45}\setcounter{offset}{-1}\addtocounter{offset}{#1}
\ifthenelse{#1=0}{}{\ifthenelse{#1=1}{\stepcounter{legSteps}\leg{(v1)}{\zeroAngle}{\arabic{legSteps}};}{
\foreach\n in {1,...,#1}{\def\eph{\arabic{offset}}\def\angle{\zeroAngle-2*\n*\spread/\eph+#1*\spread/\eph+\spread/\eph}\stepcounter{legSteps}\leg{(v1)}{\angle}{\arabic{legSteps}}}}}

\def\zeroAngle{180}\def\spread{45}\setcounter{offset}{-1}\addtocounter{offset}{#2}
\ifthenelse{#2=0}{}{\ifthenelse{#2=1}{\stepcounter{legSteps}\leg{(v2)}{\zeroAngle}{\arabic{legSteps}};}{
\foreach\n in {1,...,#2}{\def\eph{\arabic{offset}}\def\angle{\zeroAngle-2*\n*\spread/\eph+#2*\spread/\eph+\spread/\eph}\stepcounter{legSteps}\leg{(v2)}{\angle}{\arabic{legSteps}}}}}

\def\zeroAngle{115}\def\spread{45}\setcounter{offset}{-1}\addtocounter{offset}{#3}
\ifthenelse{#3=0}{}{\ifthenelse{#3=1}{\stepcounter{legSteps}\leg{(v3)}{\zeroAngle}{\arabic{legSteps}};}{
\foreach\n in {1,...,#3}{\def\eph{\arabic{offset}}\def\angle{\zeroAngle-2*\n*\spread/\eph+#3*\spread/\eph+\spread/\eph}\stepcounter{legSteps}\leg{(v3)}{\angle}{\arabic{legSteps}}}}}

\def\zeroAngle{61.5}\def\spread{30}\setcounter{offset}{-1}\addtocounter{offset}{#4}
\ifthenelse{#4=0}{}{\ifthenelse{#4=1}{\stepcounter{legSteps}\leg{(v4)}{\zeroAngle}{\arabic{legSteps}};}{
\foreach\n in {1,...,#4}{\def\eph{\arabic{offset}}\def\angle{\zeroAngle-2*\n*\spread/\eph+#4*\spread/\eph+\spread/\eph}\stepcounter{legSteps}\leg{(v4)}{\angle}{\arabic{legSteps}}}}}

\def\zeroAngle{0}\def\spread{45}\setcounter{offset}{-1}\addtocounter{offset}{#5}
\ifthenelse{#5=0}{}{\ifthenelse{#5=1}{\stepcounter{legSteps}\leg{(v5)}{\zeroAngle}{\arabic{legSteps}};}{
\foreach\n in {1,...,#5}{\def\eph{\arabic{offset}}\def\angle{\zeroAngle-2*\n*\spread/\eph+#5*\spread/\eph+\spread/\eph}\stepcounter{legSteps}\leg{(v5)}{\angle}{\arabic{legSteps}}}}}

\def\zeroAngle{-61.5}\def\spread{30}\setcounter{offset}{-1}\addtocounter{offset}{#6}
\ifthenelse{#6=0}{}{\ifthenelse{#6=1}{\stepcounter{legSteps}\leg{(v6)}{\zeroAngle}{\arabic{legSteps}};}{
\foreach\n in {1,...,#6}{\def\eph{\arabic{offset}}\def\angle{\zeroAngle-2*\n*\spread/\eph+#6*\spread/\eph+\spread/\eph}\stepcounter{legSteps}\leg{(v6)}{\angle}{\arabic{legSteps}}}}}

\def\zeroAngle{180}\def\spread{25}\setcounter{offset}{-1}\addtocounter{offset}{#7}
\ifthenelse{#7=0}{}{\ifthenelse{#7=1}{\stepcounter{legSteps}\leg{(v7)}{\zeroAngle}{\arabic{legSteps}};}{
\foreach\n in {1,...,#7}{\def\eph{\arabic{offset}}\def\angle{\zeroAngle-2*\n*\spread/\eph+#7*\spread/\eph+\spread/\eph}\stepcounter{legSteps}\leg{(v7)}{\angle}{\arabic{legSteps}}}}}
}

\newcommand{\npPboxLegs}[6]{
\setcounter{legSteps}{0}
\def\zeroAngle{-135}\def\spread{45}\setcounter{offset}{-1}\addtocounter{offset}{#1}
\ifthenelse{#1=0}{}{\ifthenelse{#1=1}{\stepcounter{legSteps}\leg{(v1)}{\zeroAngle}{{\color{labelcolor}\arabic{legSteps}}};}{
\foreach\n in {1,...,#1}{\def\eph{\arabic{offset}}\def\angle{\zeroAngle-2*\n*\spread/\eph+#1*\spread/\eph+\spread/\eph}\stepcounter{legSteps}\leg{(v1)}{\angle}{{\color{labelcolor}\arabic{legSteps}}}}}}

\def\zeroAngle{135}\def\spread{45}\setcounter{offset}{-1}\addtocounter{offset}{#2}
\ifthenelse{#2=0}{}{\ifthenelse{#2=1}{\stepcounter{legSteps}\leg{(v2)}{\zeroAngle}{{\color{labelcolor}\arabic{legSteps}}};}{
\foreach\n in {1,...,#2}{\def\eph{\arabic{offset}}\def\angle{\zeroAngle-2*\n*\spread/\eph+#2*\spread/\eph+\spread/\eph}\stepcounter{legSteps}\leg{(v2)}{\angle}{{\color{labelcolor}\arabic{legSteps}}}}}}

\def\zeroAngle{61.5}\def\spread{30}\setcounter{offset}{-1}\addtocounter{offset}{#3}
\ifthenelse{#3=0}{}{\ifthenelse{#3=1}{\stepcounter{legSteps}\leg{(v3)}{\zeroAngle}{{\color{labelcolor}\arabic{legSteps}}};}{
\foreach\n in {1,...,#3}{\def\eph{\arabic{offset}}\def\angle{\zeroAngle-2*\n*\spread/\eph+#3*\spread/\eph+\spread/\eph}\stepcounter{legSteps}\leg{(v3)}{\angle}{{\color{labelcolor}\arabic{legSteps}}}}}}

\def\zeroAngle{0}\def\spread{45}\setcounter{offset}{-1}\addtocounter{offset}{#4}
\ifthenelse{#4=0}{}{\ifthenelse{#4=1}{\stepcounter{legSteps}\leg{(v4)}{\zeroAngle}{{\color{labelcolor}\arabic{legSteps}}};}{
\foreach\n in {1,...,#4}{\def\eph{\arabic{offset}}\def\angle{\zeroAngle-2*\n*\spread/\eph+#4*\spread/\eph+\spread/\eph}\stepcounter{legSteps}\leg{(v4)}{\angle}{{\color{labelcolor}\arabic{legSteps}}}}}}

\def\zeroAngle{-61.5}\def\spread{30}\setcounter{offset}{-1}\addtocounter{offset}{#5}
\ifthenelse{#5=0}{}{\ifthenelse{#5=1}{\stepcounter{legSteps}\leg{(v5)}{\zeroAngle}{{\color{labelcolor}\arabic{legSteps}}};}{
\foreach\n in {1,...,#5}{\def\eph{\arabic{offset}}\def\angle{\zeroAngle-2*\n*\spread/\eph+#5*\spread/\eph+\spread/\eph}\stepcounter{legSteps}\leg{(v5)}{\angle}{{\color{labelcolor}\arabic{legSteps}}}}}}

\def\zeroAngle{180}\def\spread{25}\setcounter{offset}{-1}\addtocounter{offset}{#6}
\ifthenelse{#6=0}{}{\ifthenelse{#6=1}{\stepcounter{legSteps}\leg{(v6)}{\zeroAngle}{{\color{labelcolor}\arabic{legSteps}}};}{
\foreach\n in {1,...,#6}{\def\eph{\arabic{offset}}\def\angle{\zeroAngle-2*\n*\spread/\eph+#6*\spread/\eph+\spread/\eph}\stepcounter{legSteps}\leg{(v6)}{\angle}{{\color{labelcolor}\arabic{legSteps}}}}}}
}

\newcommand{\pTLegs}[6]{
\setcounter{legSteps}{0}
\def\zeroAngle{-115}\def\spread{45}\setcounter{offset}{-1}\addtocounter{offset}{#1}
\ifthenelse{#1=0}{}{\ifthenelse{#1=1}{\stepcounter{legSteps}\leg{(v1)}{\zeroAngle}{\arabic{legSteps}};}{
\foreach\n in {1,...,#1}{\def\eph{\arabic{offset}}\def\angle{\zeroAngle-2*\n*\spread/\eph+#1*\spread/\eph+\spread/\eph}\stepcounter{legSteps}\leg{(v1)}{\angle}{\arabic{legSteps}}}}}

\def\zeroAngle{180}\def\spread{45}\setcounter{offset}{-1}\addtocounter{offset}{#2}
\ifthenelse{#2=0}{}{\ifthenelse{#2=1}{\stepcounter{legSteps}\leg{(v2)}{\zeroAngle}{\arabic{legSteps}};}{
\foreach\n in {1,...,#2}{\def\eph{\arabic{offset}}\def\angle{\zeroAngle-2*\n*\spread/\eph+#2*\spread/\eph+\spread/\eph}\stepcounter{legSteps}\leg{(v2)}{\angle}{\arabic{legSteps}}}}}

\def\zeroAngle{115}\def\spread{45}\setcounter{offset}{-1}\addtocounter{offset}{#3}
\ifthenelse{#3=0}{}{\ifthenelse{#3=1}{\stepcounter{legSteps}\leg{(v3)}{\zeroAngle}{\arabic{legSteps}};}{
\foreach\n in {1,...,#3}{\def\eph{\arabic{offset}}\def\angle{\zeroAngle-2*\n*\spread/\eph+#3*\spread/\eph+\spread/\eph}\stepcounter{legSteps}\leg{(v3)}{\angle}{\arabic{legSteps}}}}}

\def\zeroAngle{61.5}\def\spread{30}\setcounter{offset}{-1}\addtocounter{offset}{#4}
\ifthenelse{#4=0}{}{\ifthenelse{#4=1}{\stepcounter{legSteps}\leg{(v4)}{\zeroAngle}{\arabic{legSteps}};}{
\foreach\n in {1,...,#4}{\def\eph{\arabic{offset}}\def\angle{\zeroAngle-2*\n*\spread/\eph+#4*\spread/\eph+\spread/\eph}\stepcounter{legSteps}\leg{(v4)}{\angle}{\arabic{legSteps}}}}}

\def\zeroAngle{0}\def\spread{45}\setcounter{offset}{-1}\addtocounter{offset}{#5}
\ifthenelse{#5=0}{}{\ifthenelse{#5=1}{\stepcounter{legSteps}\leg{(v5)}{\zeroAngle}{\arabic{legSteps}};}{
\foreach\n in {1,...,#5}{\def\eph{\arabic{offset}}\def\angle{\zeroAngle-2*\n*\spread/\eph+#5*\spread/\eph+\spread/\eph}\stepcounter{legSteps}\leg{(v5)}{\angle}{\arabic{legSteps}}}}}

\def\zeroAngle{-61.5}\def\spread{30}\setcounter{offset}{-1}\addtocounter{offset}{#6}
\ifthenelse{#6=0}{}{\ifthenelse{#6=1}{\stepcounter{legSteps}\leg{(v6)}{\zeroAngle}{\arabic{legSteps}};}{
\foreach\n in {1,...,#6}{\def\eph{\arabic{offset}}\def\angle{\zeroAngle-2*\n*\spread/\eph+#6*\spread/\eph+\spread/\eph}\stepcounter{legSteps}\leg{(v6)}{\angle}{\arabic{legSteps}}}}}
}

\newcommand{\dPentLegs}[7]{
\setcounter{legSteps}{0}
\def\zeroAngle{-90-45}\def\spread{45}\setcounter{offset}{-1}\addtocounter{offset}{#1}
\ifthenelse{#1=0}{}{\ifthenelse{#1=1}{\stepcounter{legSteps}\leg{(v1)}{\zeroAngle}{\arabic{legSteps}};}{
\foreach\n in {1,...,#1}{\def\eph{\arabic{offset}}\def\angle{\zeroAngle-2*\n*\spread/\eph+#1*\spread/\eph+\spread/\eph}\stepcounter{legSteps}\leg{(v1)}{\angle}{\arabic{legSteps}}}}}

\def\zeroAngle{90+45}\def\spread{45}\setcounter{offset}{-1}\addtocounter{offset}{#2}
\ifthenelse{#2=0}{}{\ifthenelse{#2=1}{\stepcounter{legSteps}\leg{(v2)}{\zeroAngle}{\arabic{legSteps}};}{
\foreach\n in {1,...,#2}{\def\eph{\arabic{offset}}\def\angle{\zeroAngle-2*\n*\spread/\eph+#2*\spread/\eph+\spread/\eph}\stepcounter{legSteps}\leg{(v2)}{\angle}{\arabic{legSteps}}}}}

\def\zeroAngle{90}\def\spread{25}\setcounter{offset}{-1}\addtocounter{offset}{#3}
\ifthenelse{#3=0}{}{\ifthenelse{#3=1}{\stepcounter{legSteps}\leg{(v3)}{\zeroAngle}{\arabic{legSteps}};}{
\foreach\n in {1,...,#3}{\def\eph{\arabic{offset}}\def\angle{\zeroAngle-2*\n*\spread/\eph+#3*\spread/\eph+\spread/\eph}\stepcounter{legSteps}\leg{(v3)}{\angle}{\arabic{legSteps}}}}}

\def\zeroAngle{45}\def\spread{45}\setcounter{offset}{-1}\addtocounter{offset}{#4}
\ifthenelse{#4=0}{}{\ifthenelse{#4=1}{\stepcounter{legSteps}\leg{(v4)}{\zeroAngle}{\arabic{legSteps}};}{
\foreach\n in {1,...,#4}{\def\eph{\arabic{offset}}\def\angle{\zeroAngle-2*\n*\spread/\eph+#4*\spread/\eph+\spread/\eph}\stepcounter{legSteps}\leg{(v4)}{\angle}{\arabic{legSteps}}}}}

\def\zeroAngle{-45}\def\spread{45}\setcounter{offset}{-1}\addtocounter{offset}{#5}
\ifthenelse{#5=0}{}{\ifthenelse{#5=1}{\stepcounter{legSteps}\leg{(v5)}{\zeroAngle}{\arabic{legSteps}};}{
\foreach\n in {1,...,#5}{\def\eph{\arabic{offset}}\def\angle{\zeroAngle-2*\n*\spread/\eph+#5*\spread/\eph+\spread/\eph}\stepcounter{legSteps}\leg{(v5)}{\angle}{\arabic{legSteps}}}}}

\def\zeroAngle{-90}\def\spread{25}\setcounter{offset}{-1}\addtocounter{offset}{#6}
\ifthenelse{#6=0}{}{\ifthenelse{#6=1}{\stepcounter{legSteps}\leg{(v6)}{\zeroAngle}{\arabic{legSteps}};}{
\foreach\n in {1,...,#6}{\def\eph{\arabic{offset}}\def\angle{\zeroAngle-2*\n*\spread/\eph+#6*\spread/\eph+\spread/\eph}\stepcounter{legSteps}\leg{(v6)}{\angle}{\arabic{legSteps}}}}}

\def\zeroAngle{180}\def\spread{20}\setcounter{offset}{-1}\addtocounter{offset}{#7}
\ifthenelse{#7=0}{}{\ifthenelse{#7=1}{\stepcounter{legSteps}\leg{(v7)}{\zeroAngle}{\arabic{legSteps}};}{
\foreach\n in {1,...,#7}{\def\eph{\arabic{offset}}\def\angle{\zeroAngle-2*\n*\spread/\eph+#7*\spread/\eph+\spread/\eph}\stepcounter{legSteps}\leg{(v7)}{\angle}{\arabic{legSteps}}}}}
}

\newcommand{\dBoxLegs}[6]{
\setcounter{legSteps}{0}
\def\zeroAngle{-90-45}\def\spread{45}\setcounter{offset}{-1}\addtocounter{offset}{#1}
\ifthenelse{#1=0}{}{\ifthenelse{#1=1}{\stepcounter{legSteps}\leg{(v1)}{\zeroAngle}{\arabic{legSteps}};}{
\foreach\n in {1,...,#1}{\def\eph{\arabic{offset}}\def\angle{\zeroAngle-2*\n*\spread/\eph+#1*\spread/\eph+\spread/\eph}\stepcounter{legSteps}\leg{(v1)}{\angle}{\arabic{legSteps}}}}}

\def\zeroAngle{90+45}\def\spread{45}\setcounter{offset}{-1}\addtocounter{offset}{#2}
\ifthenelse{#2=0}{}{\ifthenelse{#2=1}{\stepcounter{legSteps}\leg{(v2)}{\zeroAngle}{\arabic{legSteps}};}{
\foreach\n in {1,...,#2}{\def\eph{\arabic{offset}}\def\angle{\zeroAngle-2*\n*\spread/\eph+#2*\spread/\eph+\spread/\eph}\stepcounter{legSteps}\leg{(v2)}{\angle}{\arabic{legSteps}}}}}

\def\zeroAngle{90}\def\spread{25}\setcounter{offset}{-1}\addtocounter{offset}{#3}
\ifthenelse{#3=0}{}{\ifthenelse{#3=1}{\stepcounter{legSteps}\leg{(v3)}{\zeroAngle}{\arabic{legSteps}};}{
\foreach\n in {1,...,#3}{\def\eph{\arabic{offset}}\def\angle{\zeroAngle-2*\n*\spread/\eph+#3*\spread/\eph+\spread/\eph}\stepcounter{legSteps}\leg{(v3)}{\angle}{\arabic{legSteps}}}}}

\def\zeroAngle{45}\def\spread{45}\setcounter{offset}{-1}\addtocounter{offset}{#4}
\ifthenelse{#4=0}{}{\ifthenelse{#4=1}{\stepcounter{legSteps}\leg{(v4)}{\zeroAngle}{\arabic{legSteps}};}{
\foreach\n in {1,...,#4}{\def\eph{\arabic{offset}}\def\angle{\zeroAngle-2*\n*\spread/\eph+#4*\spread/\eph+\spread/\eph}\stepcounter{legSteps}\leg{(v4)}{\angle}{\arabic{legSteps}}}}}

\def\zeroAngle{-45}\def\spread{45}\setcounter{offset}{-1}\addtocounter{offset}{#5}
\ifthenelse{#5=0}{}{\ifthenelse{#5=1}{\stepcounter{legSteps}\leg{(v5)}{\zeroAngle}{\arabic{legSteps}};}{
\foreach\n in {1,...,#5}{\def\eph{\arabic{offset}}\def\angle{\zeroAngle-2*\n*\spread/\eph+#5*\spread/\eph+\spread/\eph}\stepcounter{legSteps}\leg{(v5)}{\angle}{\arabic{legSteps}}}}}

\def\zeroAngle{-90}\def\spread{25}\setcounter{offset}{-1}\addtocounter{offset}{#6}
\ifthenelse{#6=0}{}{\ifthenelse{#6=1}{\stepcounter{legSteps}\leg{(v6)}{\zeroAngle}{\arabic{legSteps}};}{
\foreach\n in {1,...,#6}{\def\eph{\arabic{offset}}\def\angle{\zeroAngle-2*\n*\spread/\eph+#6*\spread/\eph+\spread/\eph}\stepcounter{legSteps}\leg{(v6)}{\angle}{\arabic{legSteps}}}}}
}

\newcommand{\bTLegs}[5]{
\setcounter{legSteps}{0}
\def\zeroAngle{-90-45}\def\spread{45}\setcounter{offset}{-1}\addtocounter{offset}{#1}
\ifthenelse{#1=0}{}{\ifthenelse{#1=1}{\stepcounter{legSteps}\leg{(v1)}{\zeroAngle}{{\color{labelcolor}\arabic{legSteps}}};}{
\foreach\n in {1,...,#1}{\def\eph{\arabic{offset}}\def\angle{\zeroAngle-2*\n*\spread/\eph+#1*\spread/\eph+\spread/\eph}\stepcounter{legSteps}\leg{(v1)}{\angle}{{\color{labelcolor}\arabic{legSteps}}}}}}

\def\zeroAngle{90+45}\def\spread{45}\setcounter{offset}{-1}\addtocounter{offset}{#2}
\ifthenelse{#2=0}{}{\ifthenelse{#2=1}{\stepcounter{legSteps}\leg{(v2)}{\zeroAngle}{{\color{labelcolor}\arabic{legSteps}}};}{
\foreach\n in {1,...,#2}{\def\eph{\arabic{offset}}\def\angle{\zeroAngle-2*\n*\spread/\eph+#2*\spread/\eph+\spread/\eph}\stepcounter{legSteps}\leg{(v2)}{\angle}{{\color{labelcolor}\arabic{legSteps}}}}}}

\def\zeroAngle{70}\def\spread{25}\setcounter{offset}{-1}\addtocounter{offset}{#3}
\ifthenelse{#3=0}{}{\ifthenelse{#3=1}{\stepcounter{legSteps}\leg{(v3)}{\zeroAngle}{{\color{labelcolor}\arabic{legSteps}}};}{
\foreach\n in {1,...,#3}{\def\eph{\arabic{offset}}\def\angle{\zeroAngle-2*\n*\spread/\eph+#3*\spread/\eph+\spread/\eph}\stepcounter{legSteps}\leg{(v3)}{\angle}{{\color{labelcolor}\arabic{legSteps}}}}}}

\def\zeroAngle{0}\def\spread{45}\setcounter{offset}{-1}\addtocounter{offset}{#4}
\ifthenelse{#4=0}{}{\ifthenelse{#4=1}{\stepcounter{legSteps}\leg{(v4)}{\zeroAngle}{{\color{labelcolor}\arabic{legSteps}}};}{
\foreach\n in {1,...,#4}{\def\eph{\arabic{offset}}\def\angle{\zeroAngle-2*\n*\spread/\eph+#4*\spread/\eph+\spread/\eph}\stepcounter{legSteps}\leg{(v4)}{\angle}{{\color{labelcolor}\arabic{legSteps}}}}}}

\def\zeroAngle{-70}\def\spread{25}\setcounter{offset}{-1}\addtocounter{offset}{#5}
\ifthenelse{#5=0}{}{\ifthenelse{#5=1}{\stepcounter{legSteps}\leg{(v5)}{\zeroAngle}{{\color{labelcolor}\arabic{legSteps}}};}{
\foreach\n in {1,...,#5}{\def\eph{\arabic{offset}}\def\angle{\zeroAngle-2*\n*\spread/\eph+#5*\spread/\eph+\spread/\eph}\stepcounter{legSteps}\leg{(v5)}{\angle}{{\color{labelcolor}\arabic{legSteps}}}}}}
}

\newcommand{\dTLegs}[4]{
\setcounter{legSteps}{0}
\def\zeroAngle{180}\def\spread{45}\setcounter{offset}{-1}\addtocounter{offset}{#1}
\ifthenelse{#1=0}{}{\ifthenelse{#1=1}{\stepcounter{legSteps}\leg{(v1)}{\zeroAngle}{{\color{labelcolor}\arabic{legSteps}}};}{
\foreach\n in {1,...,#1}{\def\eph{\arabic{offset}}\def\angle{\zeroAngle-2*\n*\spread/\eph+#1*\spread/\eph+\spread/\eph}\stepcounter{legSteps}\leg{(v1)}{\angle}{{\color{labelcolor}\arabic{legSteps}}}}}}

\def\zeroAngle{90}\def\spread{55}\setcounter{offset}{-1}\addtocounter{offset}{#2}
\ifthenelse{#2=0}{}{\ifthenelse{#2=1}{\stepcounter{legSteps}\leg{(v2)}{\zeroAngle}{{\color{labelcolor}\arabic{legSteps}}};}{
\foreach\n in {1,...,#2}{\def\eph{\arabic{offset}}\def\angle{\zeroAngle-2*\n*\spread/\eph+#2*\spread/\eph+\spread/\eph}\stepcounter{legSteps}\leg{(v2)}{\angle}{{\color{labelcolor}\arabic{legSteps}}}}}}

\def\zeroAngle{0}\def\spread{60}\setcounter{offset}{-1}\addtocounter{offset}{#3}
\ifthenelse{#3=0}{}{\ifthenelse{#3=1}{\stepcounter{legSteps}\leg{(v3)}{\zeroAngle}{{\color{labelcolor}\arabic{legSteps}}};}{
\foreach\n in {1,...,#3}{\def\eph{\arabic{offset}}\def\angle{\zeroAngle-2*\n*\spread/\eph+#3*\spread/\eph+\spread/\eph}\stepcounter{legSteps}\leg{(v3)}{\angle}{{\color{labelcolor}\arabic{legSteps}}}}}}

\def\zeroAngle{-90}\def\spread{55}\setcounter{offset}{-1}\addtocounter{offset}{#4}
\ifthenelse{#4=0}{}{\ifthenelse{#4=1}{\stepcounter{legSteps}\leg{(v4)}{\zeroAngle}{{\color{labelcolor}\arabic{legSteps}}};}{
\foreach\n in {1,...,#4}{\def\eph{\arabic{offset}}\def\angle{\zeroAngle-2*\n*\spread/\eph+#4*\spread/\eph+\spread/\eph}\stepcounter{legSteps}\leg{(v4)}{\angle}{{\color{labelcolor}\arabic{legSteps}}}}}}
}

\newcommand{\tardiLegs}[5]{
\setcounter{legSteps}{0}
\def\zeroAngle{180}\def\spread{45}\setcounter{offset}{-1}\addtocounter{offset}{#1}
\ifthenelse{#1=0}{}{\ifthenelse{#1=1}{\stepcounter{legSteps}\leg{(v1)}{\zeroAngle}{{\color{labelcolor}\arabic{legSteps}}};}{
\foreach\n in {1,...,#1}{\def\eph{\arabic{offset}}\def\angle{\zeroAngle-2*\n*\spread/\eph+#1*\spread/\eph+\spread/\eph}\stepcounter{legSteps}\leg{(v1)}{\angle}{{\color{labelcolor}\arabic{legSteps}}}}}}

\def\zeroAngle{90}\def\spread{55}\setcounter{offset}{-1}\addtocounter{offset}{#2}
\ifthenelse{#2=0}{}{\ifthenelse{#2=1}{\stepcounter{legSteps}\leg{(v2)}{\zeroAngle}{{\color{labelcolor}\arabic{legSteps}}};}{
\foreach\n in {1,...,#2}{\def\eph{\arabic{offset}}\def\angle{\zeroAngle-2*\n*\spread/\eph+#2*\spread/\eph+\spread/\eph}\stepcounter{legSteps}\leg{(v2)}{\angle}{{\color{labelcolor}\arabic{legSteps}}}}}}

\def\zeroAngle{0}\def\spread{45}\setcounter{offset}{-1}\addtocounter{offset}{#3}
\ifthenelse{#3=0}{}{\ifthenelse{#3=1}{\stepcounter{legSteps}\leg{(v3)}{\zeroAngle}{{\color{labelcolor}\arabic{legSteps}}};}{
\foreach\n in {1,...,#3}{\def\eph{\arabic{offset}}\def\angle{\zeroAngle-2*\n*\spread/\eph+#3*\spread/\eph+\spread/\eph}\stepcounter{legSteps}\leg{(v3)}{\angle}{{\color{labelcolor}\arabic{legSteps}}}}}}

\def\zeroAngle{-90}\def\spread{55}\setcounter{offset}{-1}\addtocounter{offset}{#4}
\ifthenelse{#4=0}{}{\ifthenelse{#4=1}{\stepcounter{legSteps}\leg{(v4)}{\zeroAngle}{{\color{labelcolor}\arabic{legSteps}}};}{
\foreach\n in {1,...,#4}{\def\eph{\arabic{offset}}\def\angle{\zeroAngle-2*\n*\spread/\eph+#4*\spread/\eph+\spread/\eph}\stepcounter{legSteps}\leg{(v4)}{\angle}{{\color{labelcolor}\arabic{legSteps}}}}}}

\def\zeroAngle{180}\def\spread{35}\setcounter{offset}{-1}\addtocounter{offset}{#5}
\ifthenelse{#5=0}{}{\ifthenelse{#5=1}{\stepcounter{legSteps}\leg{(v5)}{\zeroAngle}{{\color{labelcolor}\arabic{legSteps}}};}{\ifthenelse{#5=2}{\stepcounter{legSteps}\leg{(v5)}{180}{{\color{labelcolor}\arabic{legSteps}}};\stepcounter{legSteps}\leg{(v5)}{0}{{\color{labelcolor}\arabic{legSteps}}};}{
\foreach\n in {1,...,#5}{\def\eph{\arabic{offset}}\def\angle{\zeroAngle-2*\n*\spread/\eph+#5*\spread/\eph+\spread/\eph}\stepcounter{legSteps}\leg{(v5)}{\angle}{{\color{labelcolor}\arabic{legSteps}}}}}}}

}

\let\olditemize\itemize\renewcommand{\itemize}{\vspace{-2pt}\olditemize\setlength{\itemsep}{1pt}\setlength{\parskip}{0pt}\setlength{\parsep}{-0pt}}
\let\oldenumerate\enumerate\renewcommand{\enumerate}{\vspace{-4pt}\oldenumerate\setlength{\itemsep}{1pt}\setlength{\parskip}{0pt}\setlength{\parsep}{0pt}}
\makeatletter\renewcommand\section{\addtocontents{toc}{\protect\addvspace{-2.25\p@}}\@startsection {section}{1}{\z@}{-0.0ex \@plus .2ex \@minus 0.2ex}{1ex \@plus.1ex\@minus .5ex}{\normalfont\large\bfseries}}
\renewcommand\subsection{\addtocontents{toc}{\protect\addvspace{-2.5\p@}}\@startsection {subsection}{1}{\z@}{0.5ex \@plus .2ex \@minus 0.2ex}{0.75ex \@plus.1ex\@minus .5ex}{\normalfont\bfseries}}
\renewcommand\subsubsection{\addtocontents{toc}{\protect\addvspace{-2.5\p@}}\@startsection {subsubsection}{1}{\z@}{0.5ex \@plus .2ex \@minus 0.2ex}{0.75ex \@plus.1ex\@minus .5ex}{\normalfont\bfseries}}
\definecolor{rindou1}{rgb}{0.4431,0.2862,0.7960}
\definecolor{rindou2}{rgb}{0.0078,0.1215,0.4392}
\definecolor{lapis}{rgb}{0.0.0470,0.2941,0.5568}
\definecolor{emerald}{rgb}{0.31, 0.78, 0.47}
\definecolor{pinegreen}{rgb}{0.0, 0.47, 0.44}
\definecolor{jade}{rgb}{0.0, 0.66, 0.42}
\definecolor{teal}{rgb}{0.0, 0.5, 0.5}
\definecolor{dim}{rgb}{0.55,0.55,0.55}
\definecolor{deemph}{rgb}{0.25,0.25,0.25}
\definecolor{hblue}{rgb}{0,0,0.575}
\definecolor{hred}{rgb}{0.575,0.0,0.225}
\definecolor{hgreen}{rgb}{0.0,0.4,0.2}
\definecolor{hteal}{rgb}{0.0,0.445,0.6451}
\renewcommand{\r}[1]{{\color{hred}#1}}
\renewcommand{\b}[1]{{\color{hblue}#1}}
\newcommand{\g}[1]{{\color{hteal}#1}}
\newcommand{\eq}[1]{\vspace{-0.5pt}\begin{equation}#1\vspace{-0.5pt}\end{equation}}
\newcommand{\fwbox}[2]{\text{\makebox[#1][c]{$\hspace{-150pt}\displaystyle#2\hspace{-150pt}$}}}
\newcommand{\fwboxL}[2]{\text{\makebox[#1][l]{$#2$}}}
\newcommand{\fwboxR}[2]{\text{\makebox[#1][r]{$#2$}}}
\newcommand{\equivR}{\fwbox{14.5pt}{\hspace{-0pt}\fwboxR{0pt}{\raisebox{0.47pt}{\hspace{1.25pt}:\hspace{-4pt}}}=\fwboxL{0pt}{}}}
\newcommand{\equivL}{\fwbox{14.5pt}{\fwboxR{0pt}{}=\fwboxL{0pt}{\raisebox{0.47pt}{\hspace{-4pt}:\hspace{1.25pt}}}}}
\renewcommand{\hat}{\widehat}
\newcommand{\newcap}{\mathrm{\raisebox{0.75pt}{{$\,\bigcap\,$}}}}
\newcommand{\tcap}{\scalebox{0.9}{$\!\newcap\!$}}

\newcommand{\x}[2]{{\color{black}(}\hspace{-0.85pt}{\color{black}#1}\hspace{-0.25pt}{\color{black}|}\hspace{-0.25pt}{\color{black}#2}\hspace{-0.85pt}{\color{black})}}

\newcommand{\bigger}[1]{\raisebox{-0.95pt}{\scalebox{1.25}{$#1$}}}

\newcommand{\ab}[1]{\langle #1\rangle}
\renewcommand{\tilde}{\widetilde}

\newcommand{\myTwistors}[1]{%
\setsepchar{,}
\readlist\arg{#1}

\begin{align}
    Z_1&=\Big(\arg[1],\,\arg[2],\,\arg[3],\,\arg[4]\Big)\\
    Z_2&=\Big(\arg[5],\,\arg[6],\,\arg[7],\,\arg[8]\Big)\\
    Z_3&=\Big(\arg[9],\,\arg[10],\,\arg[11],\,\arg[12]\Big)\\
    Z_4&=\Big(\arg[13],\,\arg[14],\,\arg[15],\,\arg[16]\Big)\\
    Z_5&=\Big(\arg[17],\,\arg[18],\,\arg[19],\,\arg[20]\Big)\\
    Z_6&=\Big(\arg[21],\,\arg[22],\,\arg[23],\,\arg[24]\Big)\\
    Z_7&=\Big(\arg[25],\,\arg[26],\,\arg[27],\,\arg[28]\Big)\\
    Z_8&=\Big(\arg[29],\,\arg[30],\,\arg[31],\,\arg[32]\Big)\\
    Z_9&=\Big(\arg[33],\,\arg[34],\,\arg[35],\,\arg[36]\Big)\\
    Z_{10}&=\Big(\arg[37],\,\arg[38],\,\arg[39],\,\arg[40]\Big)\\
    Z_{11}&=\Big(\arg[41],\,\arg[42],\,\arg[43],\,\arg[44]\Big)
\end{align}

}

\begin{document}

\title{Landau Singularities and Higher-Order Roots}

\author{Jacob L. Bourjaily}
\email{bourjaily@psu.edu}
\affiliation{Institute for Gravitation and the Cosmos, Department of Physics,\\Pennsylvania State University, University Park, PA 16802, USA}
\affiliation{Niels Bohr International Academy and Discovery Center, Niels Bohr Institute,\\University of Copenhagen, Blegdamsvej 17, DK-2100, Copenhagen \O, Denmark}
\author{Cristian Vergu}
\email{c.vergu@nbi.ku.dk}
\affiliation{Institute for Gravitation and the Cosmos, Department of Physics,\\Pennsylvania State University, University Park, PA 16802, USA}
\affiliation{Niels Bohr International Academy and Discovery Center, Niels Bohr Institute,\\University of Copenhagen, Blegdamsvej 17, DK-2100, Copenhagen \O, Denmark}
\author{Matt von Hippel}
\email{matthew.vonhippel@ipht.fr}
\affiliation{Niels Bohr International Academy and Discovery Center, Niels Bohr Institute,\\University of Copenhagen, Blegdamsvej 17, DK-2100, Copenhagen \O, Denmark}
\affiliation{Institut de Physique Th\'{e}orique, CEA Paris-Saclay\\ F–91191 Gif-sur-Yvette cedex, France}

\date{\today}

\begin{abstract}
Landau's work on the singularities of Feynman diagrams suggests that they can only be of three types: either poles, logarithmic divergences, or the roots of \emph{quadratic} polynomials. On the other hand, many Feynman integrals exist whose singularities involve arbitrarily higher-order polynomial roots. We investigate this apparent paradox using concrete examples involving cube-roots and roots of a degree-eight polynomial in four dimensions and roots of a degree-six polynomial in two dimensions, and suggest that these higher-order singularities can only be approached via kinematic limits of higher co-dimension than one, thus evading Landau's argument.
\end{abstract}

\maketitle

\vspace{-5pt}\section{Introduction}\vspace{-2pt}
\label{sec:introduction}

This paper tries to reconcile two seemingly contradictory facts.  First, there is the folk wisdom that Feynman diagrams have only three types of singularities: poles, square roots, and logarithms. This folk wisdom can be traced back to Landau, who demonstrated it in Ref.~\cite{landau1959}, though his argument included one oft-forgotten caveat: the restriction only holds for singularities approachable at kinematic configurations of \emph{co-dimension one}. Second, we now have something Landau did not: an extensive body of work on the singularities of perturbative scattering amplitudes---especially in the case of planar, maximally supersymmetric ($\mathcal{N}\!=\!4$) super Yang-Mills (`sYM') \mbox{\cite{Cachazo:2008vp,ArkaniHamed:2009dn,Kaplan:2009mh,Mason:2009sa,ArkaniHamed:2009sx,ArkaniHamed:2009dg,Ashok:2010ie,ArkaniHamed:book,Bourjaily:2012gy}}---where examples are known to involve roots of arbitrarily high-order polynomials in the kinematics. Consider for example the following on-shell diagram involving 40 external particles at 37 loops:
\vspace{0pt}\eq{\begin{split}~\\[-20pt]\fwbox{0pt}{\hspace{-0pt}\raisebox{-90pt}{\includegraphics[scale=0.575]{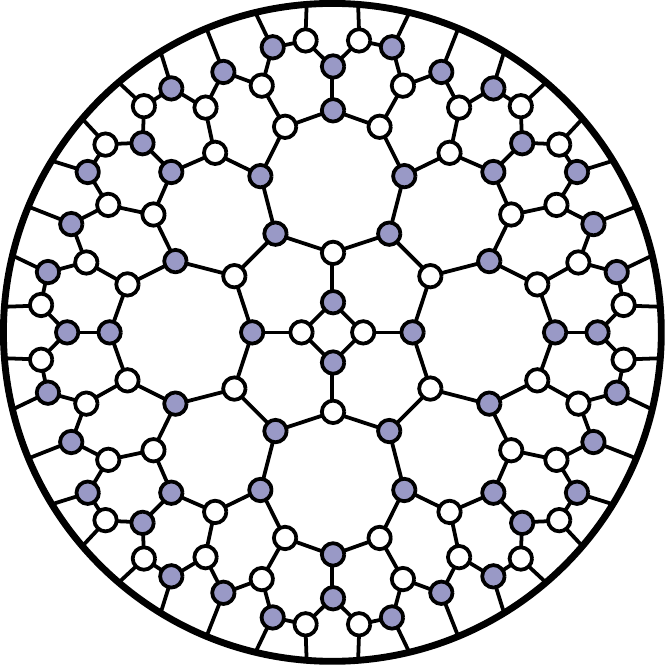}}\hspace{-0pt}}\\[-20pt]~\end{split}\vspace{-5pt}}
This diagram encodes a function where all internal lines are taken to be on-shell, with the particular parities of each three-point vertex indicated in the diagram. But it turns out that there are $2^{10}$(!) \emph{distinct} solutions to the cut conditions with this particular set of three-point parities---with each encoding a different, specific on-shell function involving the various roots of some 1024th-degree polynomial \cite{ArkaniHamed:book,Bourjaily:2012gy}.

Simpler examples of higher-than-quadratic roots arise at much lower loop-orders. In massless theories, nested quadratic roots make their first appearance at two loops, but the first example of an irreducible cubic root appears at three loops involving eleven massless particles. For massive Feynman integrals, integrals involving sextic roots also appear at three loops. This raises a natural question: do these examples contradict Landau's analysis? Or if not, why not?\\

In this work, we suggest a simple resolution to the apparent paradox: cubic- (or higher-) root singularities can and do occur in Feynman integrals, but correspond to singularities of amplitudes at co-dimension two or higher. Specifically, this means that such roots cannot be accessed by any \emph{co-dimension one} kinematic limit. To access such higher-order singularities, one may first consider starting from a set of restricted kinematics in which some discriminant locus vanishes. This may be possible in some cases, but is not in the cases we investigate here: the discriminant locus is not merely vanishing, but singular in these restricted kinematics, which makes further monodromies ill-defined.\\

Our work is organized as follows. We begin in section~\ref{sec:prelims} with some useful background material and review; specifically, we review how one finds singularities via the Landau equations in subsection~\ref{subsec:where}, and revisit Landau's argument regarding the types of singularities that arise in subsection~\ref{subsec:what}, before finishing the section by discussing some useful facts about the roots of higher-degree polynomials. We then study the implications of these ideas for three concrete examples in considerable detail. In section~\ref{sec:cubic} we describe the first example of a cube root arising in the case of theories of massless particles in four dimensions. We show how this root is required when one tries to represent the on-shell space of the diagram in terms of a rational parametrization of the external kinematics, and how any codimension-one kinematic limit will nonetheless isolate a pole, not a cubic root. We also discuss how one could set up the Landau equations in this kinematic parametrization. In section~\ref{sec:three-quadrics} we discuss a similar example with more generic kinematics, where the on-shell space is described by two degree-eight polynomials. In this example the Landau equations can be fully solved, and we describe a parametrization of the kinematics which rationalizes the roots present, analogously to the way one can rationalize the square root in the one-loop box. Finally, in section~\ref{sec:sextic} we describe an example involving sextic roots that arises for massive theories in two dimensions, where we can once again solve the Landau equations. 

\newpage
\section{Preliminaries}
\label{sec:prelims}

\subsection{Where do Amplitudes have Singularities?}
\label{subsec:where}

It has been known since the work of Landau (see Ref.~\cite{landau1959} but also Ref.~\cite{BSMF_1959__87__81_0}) that the singularities of scattering amplitudes or Feynman integrals are given by the solutions to Landau equations. The scalar Feynman integral 
\begin{equation}
    \label{eq:integral}
    I(p) = \int \frac{d^n k}{s_1(p, k) \cdots s_m(p, k)},
\end{equation}
has singularities when a subset of the denominators vanish \emph{and} there exist a not all-vanishing set $\alpha_1, \dotsc, \alpha_m$ such that 
\begin{equation}
    \label{eq:landau-differential}
    \sum_{e = 1}^m \alpha_e\,d s_e(p,k) = 0,
\end{equation}
where the differential is taken with respect to the integration variables $k$ only.

The perhaps more familiar form of the Landau equations is obtained by using $s_e(p, k) = q_e^2 - m_e^2$ where $q_e$ is a linear combination of independent external momenta \mbox{$p\equivR\{p_a\}$} and independent loop momenta $k$ which depends on the graph we are studying.  Then eq.~\eqref{eq:landau-differential} yields a vector equation for each independent loop momentum, which are normally called the Landau equations. 

Mathematically, the condition in eq.~\eqref{eq:landau-differential} arises as a critical value condition for a map between the on-shell space defined by the vanishing of a subset of denominators and the space of external momenta (see Refs.~\cite{pham2011singularities, pham1968singularities, pham}).  Ref.~\cite{Coleman:1965xm} introduced a physical way of understanding Landau diagrams as real scattering processes of on-shell particles---where the Landau equations arise as the condition for the closure of the loops.

\subsection{What Kind of Singularities Can Amplitudes Have?}
\label{subsec:what}

In the original paper Ref.~\cite{landau1959}, Landau studied not only the location of the singularities, but also their nature: how scattering amplitudes behave when approaching these singularities.  Similar results were obtained by Leray in Ref.~\cite{BSMF_1959__87__81_0} (see also a review of Leray's results in Ref.~\cite[p.~109, chap.~VI]{pham2011singularities}).  A more detailed derivation following Landau's original idea was presented in Ref.~\cite{PolkinghorneScreaton}.  The asymptotic expansion around these singularities has been used recently in Ref.~\cite{Hannesdottir:2021kpd} to impose constraints on the symbol of polylogarithmic integrals.

The derivation of the behavior of a Feynman integral in the neighborhood of a Landau singularity is slightly too technical for our current needs, so we will not present it in detail. Instead, we refer the reader to the result of Ref.~\cite{pham2011singularities} (which uses slightly different notation from ours).

Consider an integral of the form,\footnote{In Ref.~\cite{pham2011singularities} one can find a more general case where the integration cycle is any element of the relative homology.}
\begin{equation}
    I(p) = \int \frac{\omega}{\prod_{e = 1}^m s_e(p, k)^{\delta_e}},
\end{equation}
where $\delta_1, \dotsc, \delta_m > 0$ are integers and $\omega$ is some holomorphic differential $n$-form on the $k$'s independent of the external momenta $p$. We will find it useful to define $\delta\equivR \sum_{e = 1}^m \delta_e$.\footnote{In the context of analytic regularization, one would like the exponents $\delta_i$ to be rational or even complex-valued; but this goes outside the hypothesis of the theorem proved by Leray---although for fractional exponents it should be possible to extend the original argument by going to a covering space as discussed in Ref.~\cite{pham1965formules, pham}.}

The integral is regular unless the contour of integration is trapped (or pinched) by the singularity hypersurfaces $s_1(p, k) = \cdots = s_m(p, k) = 0$, so that it cannot be deformed away from them.  A detailed discussion of the geometry of the contour pinching can be found in Ref.~\cite{pham2011singularities}.  The integral is potentially singular for $p$ such that we have a set of not-all-vanishing variables $\alpha_1, \dotsc, \alpha_m$ such that there exists a function $\ell(p)$ with
\begin{equation}
    d \ell(p) = \sum_{e = 1}^m \alpha_e d s_e(p, k),
\end{equation}
where now on the right-hand side the differential is taken with respect to both $p$ and $k$.  The left-hand side is independent of $k$ by virtue of eq.~\eqref{eq:landau-differential}.  We can then choose a constant factor so that the equation of the Landau singularities is  $\ell(p) = 0$ (see Ref.~\cite{pham} for more details).

Some technical assumptions are necessary here.  First, we assume that the Landau locus $\ell(p) = 0$ arises from a \emph{single} Landau diagram.

Second, we take $p \to p^*$ where $p^*$ is smooth point of the Landau locus ($\ell(p^*) = 0$).  This is necessary to be able to define homotopy paths going around the complexified Landau locus.  Roughly, one needs to be able to define a complex transversal space where the Landau locus sits at the origin.  Then in this transversal space one can construct a homotopy path as a loop around the origin.  However, at a singular point the tangent space is not well-defined and therefore one can not define a transversal space either.  The Landau loci are themselves generically singular so this restriction is necessary (we present an example at the end of sec.~\ref{subsec:roots}).  Third, the pinching has to happen for a unique value of internal momenta.  Related to this, a certain Hessian (see Ref.~\cite[Appendix~E]{Hannesdottir:2021kpd} for a more detailed discussion) which appears in the denominator of the coefficient $A$ below must not vanish.

Finally, we will approach the Landau locus only at generic points, which are not also singular points for \emph{other} Landau singularities.  Indeed, Landau loci generically intersect other Landau loci.  These intersections have been the subject of study, for example in connection with Steinmann relations (see Ref.~\cite{Hannesdottir:2021kpd}).

In this notation and under the conditions described above, we have the following asymptotic behavior for $I(p)$ when $\ell(p) \to 0$:
\begin{enumerate}
    \item if $n + m - 1$ is odd, then
    \begin{equation}
        \label{eq:asy1}
        I(p) = -\frac N 2 \frac{(2 \pi i)^m A(p) \prod_{i = 1}^m (-\alpha_i)^{\delta_i}}{\prod_{i = 1}^m (\alpha_i - 1)!} \frac{\ell(p)^{\frac{n + m - 1}{2} - \delta}}{\Gamma(1 + \frac{n + m - 1}{2} - \delta)} (1 + o(\ell(p))) + \text{hf};
    \end{equation}
    \item if $n + m - 1$ is even and $n + m - 1 \geq 2 \delta$, then
    \begin{equation}
        \label{eq:asy2}
        I(p) = N \frac{(2 \pi i)^{m - 1} A(p) \prod_{i = 1}^m (-\alpha_i)^{\delta_i}}{\prod_{i = 1}^m (\alpha_i - 1)!} \frac{\ell(p)^{\frac{n + m - 1}{2} - \delta}}{(\frac{n + m - 1}{2} - \delta)!} \log (\ell(p)) (1 + o(\ell(p))) + \text{hf};
    \end{equation}
    \item if $n + m - 1$ is even and $n + m - 1 < 2 \delta$, then
    \begin{align}
        \label{eq:asy3}
        I(p) =& -N \frac{(2 \pi i)^{m - 1} A(p) \prod_{i = 1}^m (-\alpha_i)^{\delta_i}}{\prod_{i = 1}^m (\alpha_i - 1)!} \frac{(-\frac{n + m - 1}{2} + \delta - 1)!}{(-\ell(p))^{-\frac{n + m - 1}{2} + \delta}} (1 + o(\ell(p))) \nonumber\\&+ N \log (\ell(p)) \text{hf} + \text{hf'};
    \end{align}
    \item if $m = n + 1$, we have the more specific result that
    \begin{equation}
        \label{eq:asy4}
        I(p) = (-1)^{n + 1} N \frac{(2 \pi i)^n A(p) \prod_{i = 1}^m \alpha_i^{\delta_i}}{\prod_{i = 1}^m (\alpha_i - 1)!} \frac{(\delta - n - 1)!}{\ell(p)^{\delta - n}} (1 + o(\ell(p))) + \text{hf}.
    \end{equation}
\end{enumerate}
Here, $n$ is the number of integrations (see eq.~\eqref{eq:integral}), ``$\text{hf}$'' (and $\text{hf'}$) denote any holomorphic function and $o(\ell(p))$ is any holomorphic function which vanishes at the Landau locus.  The quantity $A(p)$ is essentially the inverse of the square root of a Hessian, while $N$ is an intersection index which is purely numerical. Explicit expressions can be given for $N$ and $A(p)$, but we will not need their detailed form in what follows.

From this discussion we may conclude that there are three basic types of singularities that may arise in the neighborhood of a Landau locus: polar, square root, and logarithmic.  The ramification type of each in the neighborhood of a Landau locus can also be determined from a homological analysis, as was done in Ref.~\cite[p.~96, sec.~2.6]{pham2011singularities}.

\subsection{Non-examples}
\label{sec:nonexamples}

Let us list a number of cases which might look like they violate the result in the previous subsection, but do not satisfy the conditions we require.

First, one might think about removing the square root singularities by a change of coordinates, which amounts to going to a double cover.  Such transformations are sometimes useful but we want to study the singularities in the original coordinates, which can be either Mandelstam invariants or momentum components in a special frame.

We do not study second type singularities, for which the pinch happens at infinite values of loop momenta.  In principle this case can also be studied after an appropriate compactification of the internal kinematic space has been made.  Then one can can change coordinates to parametrize the points at infinity and do the same analysis.  The numerator enters in an essential way in this analysis; for pinches at finite values of momenta the numerators can only cancel singularities, not create new ones.

Another seeming counterexample is that of the massless box in six dimensions.  This is a finite integral, which up to a global factor can be computed to be
\begin{equation}
    \frac{\log^2 \frac{s}{t} + \pi^2}{s + t}.
\end{equation}
It might look like terms of type $\log^2$ contradict the general results in eqs.~\eqref{eq:asy1}, \eqref{eq:asy2}, \eqref{eq:asy3}, \eqref{eq:asy4}.

However, this example violates several of the requirements we have imposed above.  First, the kinematics is such that the momentum of each external massless particle is at the threshold.  Therefore, the integral has a permanent pinch.  For integrals which have such permanent pinches even the \emph{existence} of an analytic continuation in kinematic variables is not guaranteed.  Another problem is that $s = 0$ or $t = 0$ correspond to multiple Landau loci.  They arise in bubble Landau singularities as well as in triangle and box singularities.  The bubble and triangle singularities are particularly troublesome since the pinching in internal momenta does not happen for some fixed values of the momenta, but for a one-parameter family of values.  This precludes the usual construction of vanishing homology classes which arise in the application of Picard--Lefschetz theory (see Ref.~\cite{pham2011singularities} and also Ref.~\cite{Hannesdottir:2021kpd}).

Despite these problems, at least the prefactor of the integral (which is a second type singularity in six dimensions), can be analyzed in dimensional regularization and computed to all orders in $\epsilon$.  It is also interesting to note that, since this second type singularity can not appear in the physical region, when $s = -t$ the numerator vanishes so as to cancel the pole.  This uniquely fixes the $\pi^2$ term once we know there is a $\log^2 \frac{s}{t}$ term.  We believe this kind of constraint should provide a handle on ``beyond the symbol terms'' (or ``initial conditions'' in a differential equations language).

In general, we expect that cases such as these should give rise to higher powers of the singularities described in the previous section (higher poles, powers of square roots, and powers of logarithms), as they correspond to factorizable singularities. We do not expect them to give rise to new types of singularities, such as higher roots.

\subsection{Roots of Polynomial Equations}
\label{subsec:roots}

Consider what happens to the roots of an algebraic equation when taking its coefficients along some closed path. As the coefficients change, the various roots follow always continuous paths themselves, and return to the same locations as at the beginning---but possibly up to permutation.

If we have a square root singularity at co-dimension one, this means that going \emph{twice} around the singularity must always return the roots to where they began: if they had been exchanged in going once around, then going twice would return them to their initial locations. 

When the discriminant of their defining polynomial vanishes, the two roots coincide, and the singularity is therefore at co-dimension one.

A cubic root singularity would correspond to a situation where \emph{three} roots simultaneously coincide. This however can only happen at higher than co-dimension one for an intuitive reason: the equality of any pair of roots imposes one constraint on the coefficients of the polynomial, and collapsing the pair with the remaining one requires one further constraint. Thus, all three roots collide only at co-dimension two.

Consider for example the depressed cubic equation $x^3 + p x + q = 0$. Its discriminant is (up to a sign) $\Delta = 4 p^3 + 27 q^2$. $\Delta = 0$ implies only that two of the roots have collided. One could easily make all three collide by requiring $p=q=0$, but this demand clearly constitutes two conditions. Incidentally, the vanishing of the discriminant locus is itself a cuspidal cubic in terms of $p, q$, and it has a cuspidal singularity when $p = q = 0$---precisely at the locus where the three roots coincide.

The case of a general cubic hypersurface $0=x^3{+}a_2 x^2{+}a_1 x{+}a_0$ can be understood similarly. If two roots coincide and equal $u$ while the third equals $v$ we have $2 u + v = -a_2$, $u^2 + 2 u v = a_1$ and $u^2 v = -a_0$.  After eliminating $u$ and $v$ we find
\begin{equation}
    \label{eq:cubic-coincident-root-locus-21}
    a_1^2 a_2^2 - 4 a_0 a_2^3 - 4 a_1^3 + 18 a_0 a_1 a_2 - 27 a_0^3 = 0.
\end{equation}
It turns out that this surface has a singularity when three roots coincide.  To see this, we set $v = u$ and after eliminating $u$ we find $a_0 = \frac{a_2^3}{27}$ and $a_1 = \frac{a_2^2}{3}$, which is the equation of a rational normal curve.  It is now easy to see that if we take the differential of eq.~\eqref{eq:cubic-coincident-root-locus-21} we find zero, upon replacing the solutions for $a_0$ and $a_1$ in terms of $a_2$.\footnote{This means that the cotangent space along the locus where three roots coincide is ill-defined.  In turn this implies that the tangent space is ill-defined and therefore the surface has a singularity.}  This means that the surface defined by eq.~\eqref{eq:cubic-coincident-root-locus-21} is singular along the curve where three roots coincide.

To illustrate the complexities that can arise for higher degrees, let us discuss the quartic equation.  To find when the depressed quartic $x^4 + a_2 x^2 + a_1 x + a_0 = 0$ has multiple solutions we take the roots $r_1, r_2, r_3, r_4$, which are such that $r_1 + r_2 + r_3 + r_4 = 0$. If we set $r_3 = r_4$ we can solve $r_3 = r_4 = -\frac{1}{2} (r_1 + r_2)$.  Plugging back in the expressions for $a_0, a_1, a_2$ and eliminating $r_1$ and $r_2$ we find
\begin{equation}
    \label{eq:quartic-coincident-root-locus-211}
    4 a_1^2 a_2^3 - 16 a_0 a_2^4 + 27 a_1^4 - 144 a_0 a_1^2 a_2 + 128 a_0^2 a_2^2 - 256 a_0^3 = 0.
\end{equation}
For the quartic we can also have three roots coinciding or two pairs of roots coinciding.  In general, the type of coincident root locus is specified by a partition of the degree of the equation. In the case $(3, 1)$, that is when three roots coincide, we have
\begin{gather}
    a_2^2 + 12 a_0 = 0, \qquad
    9 a_1^2 - 32 a_0 a_2 = 0.
\end{gather}
In the case $(2, 2)$, that is when two pairs of roots coincide, we have
\begin{gather}
    a_1 = 0, \qquad
    a_2^2 - 4 a_0 = 0.
\end{gather}
Again, it can explicitly checked that the differential of eq.~\eqref{eq:quartic-coincident-root-locus-211} vanishes along the $(3, 1)$ locus and along the $(2, 2)$ locus, which means that the surface defined by eq.~\eqref{eq:quartic-coincident-root-locus-211} is singular there.

The $(3, 1)$ locus intersects the $(2, 2)$ locus in the $(4)$ locus, where all the roots coincide.  In the case of the depressed quartic, this means that all the roots vanish (since their sum vanishes).  In the neighborhood where all four roots coincide the vanishing discriminant locus has a swallow-tail singularity.

\begin{figure}
  \centering
\includegraphics{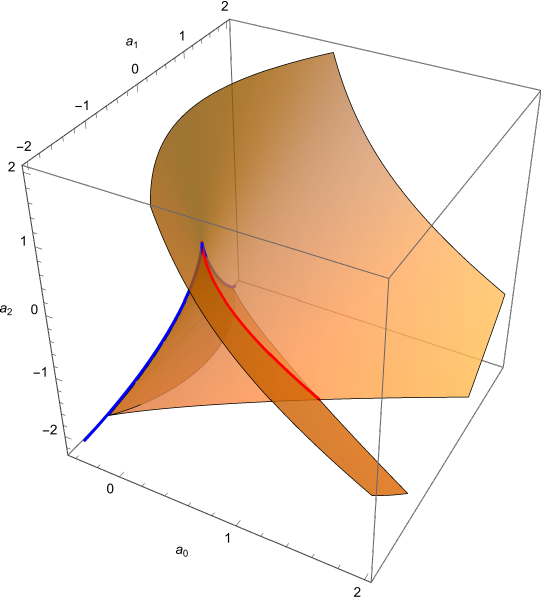}
\label{fig:swallowtail}
\caption{The swallow-tail singularity of eq.~\eqref{eq:quartic-coincident-root-locus-211}. The blue curves depict the locus where three roots coincide, while the red curve depicts the locus where two pairs of roots coincide. At the point where these curves intersect, all roots coincide.}
\end{figure}

\newpage
\section{\texorpdfstring{\emph{Exempli Gratia}}{Exempli Gratia}: A Cube-Root at Three Loops}
\label{sec:cubic}

Consider the scalar three-loop scalar Feynman integral
\eq{\hspace{-10pt}\begin{tikzpicture}[scale=0.9*\figScale,baseline=-3.05]\useasboundingbox ($(-2.,-1.5)$) rectangle ($(2.,1.5)$);\draw[int,line width=0.1,red,draw=\boundingDraw] ($(-2.,-1.5)$) rectangle ($(2.,1.5)$);
\coordinate (v0) at (0,0);
\coordinate (v1) at ($(v0)+(150:0.75)$);
\coordinate (v2) at ($(v0)+(30:0.75)$);
\coordinate (v3) at ($(v0)+(-90:0.75)$);
\coordinate (a1) at ($(v0)+(110:1.2)$);
\coordinate (a2) at ($(v0)+(70:1.2)$);
\coordinate (b1) at ($(v0)+(-10:1.2)$);
\coordinate (b2) at ($(v0)+(-50:1.2)$);
\coordinate (c1) at ($(v0)+(-130:1.2)$);
\coordinate (c2) at ($(v0)+(-170:1.2)$);
\draw[int] (v0)--(v1);\draw[int] (v0)--(v2);\draw[int] (v0)--(v3);
\draw[int] (v2)--(a2);\draw[int] (v1)--(a1);\draw[int] (a1)--(a2);\draw[int] (v2)--(b1);\draw[int] (v3)--(b2);\draw[int] (b1)--(b2);\draw[int] (v3)--(c1);\draw[int] (v1)--(c2);\draw[int] (c1)--(c2);
\leg{(a1)}{135}{$2$}\leg{(a2)}{53}{$3$}\leg{(v2)}{40}{$4$}\leg{(b1)}{5+45}{$5$}\leg{(b1)}{5-45}{$6$}\leg{(b2)}{-67+45}{$7$}\leg{(b2)}{-67-45}{$8$}\leg{(c1)}{-113+45}{$9$}\leg{(c1)}{-113-45}{$10$}\leg{(c2)}{175+45}{$11$}\leg{(c2)}{175-45}{$1$}
\node at (v0) [bdot] {};\node at (v1) [bdot] {};\node at (v2) [bdot] {};\node at (v3) [bdot] {};\node at (a1) [bdot] {};\node at (a2) [bdot] {};\node at (b1) [bdot] {};\node at (b2) [bdot] {};\node at (c1) [bdot] {};\node at (c2) [bdot] {};\node at ($(v0)+(90:0.6)$) [] {$\ell_1$};\node at ($(v0)+(-30:0.6)$) [] {$\ell_2$};\node at ($(v0)+(-150:0.6)$) [] {$\ell_3$};
\end{tikzpicture}\!\!\!\bigger{\Leftrightarrow}\,\frac{d^4\ell_1 d^4\ell_2 d^4\ell_3}{\x{\ell_1}{2}\x{\ell_1}{3}\x{\ell_1}{4}\x{\ell_1}{\ell_2}\x{\ell_2}{5}\x{\ell_2}{7}\x{\ell_2}{9}\x{\ell_2}{\ell_3}\x{\ell_3}{9}\x{\ell_3}{11}\x{\ell_3}{2}\x{\ell_3}{\ell_1}}\,,\nonumber}
where we have used dual momentum coordinates with $p_a\equivL x_{a+1}{-}x_a$ and where we have defined the shorthand $\x{a}{b}\equivR(x_a{-}x_b)^2$. This integral contributes to eleven-particle amplitudes in both pure and maximally supersymmetric $(\mathcal{N}\!=\!4)$ super Yang-Mills theory. 

Here, we will establish that, in order to represent this integral in a reasonable parametrization of the kinematic space, we require cubic roots. We will primarily establish this by examining the on-shell space, and in particular the value of the integral on this space (i.e. at the leading singularity). We will confirm this by parametrizing the integrals in a non-redundant, rational manner, to show that this is not just an artifact of a poor choice of parametrization. However, as suggested by the discussion in section~\ref{sec:prelims}, we will find that in this non-redundant parametrization it is impossible to approach these roots in a codimension-one limit\footnote{While this case also evades Landau's classification of singularities by being massless, as discussed at the end of section~\ref{sec:nonexamples} this is not enough by itself to allow for a codimension-one approach to a cubic root singularity.}. Having established what we set out to, we briefly discuss how one might set up the problem of determining the Landau singularities of the diagram, leaving a full derivation for future work.

\subsection{On-Shell Space}

To begin, let us discuss the on-shell space. This integral has 16 locations in the (complexified) space of loop momenta which put all its 12 propagators on-shell---corresponding to the solutions to its `maximal-cut' equations. These solutions may be organized according to which \emph{particular} solution to the cut equations is involved at each three-particle vertex, which can take one of two forms: with all spinor-helicity variables $\lambda$ of the participating on-shell momenta being proportional (white) or all their conjugate $\widetilde{\lambda}$'s proportional (blue). When represented in momentum-twistor space \cite{Hodges:2009hk}, these correspond to the cases where the lines (representing the loop momenta $\ell_i$ and the external dual-momentum points $x_a$) pass through each other at a single point (white) or when they are co-planar (blue). 

As there is a three-particle vertex in the middle of the loop integrand, we can divide the 16 solutions into two groups of 8 according to whether the lines representing $(\ell_i)$ in momentum-twistor space intersect at a point or not (white and blue, respectively). Of these eight, there are 3 rational solutions, 2 (a pair) that involve quadratic roots, and 3 which involve the roots of an irreducible cubic. We'd like to analyze and describe these cube-root solutions to the maximal cut equations explicitly.

Let us start with a concrete parameterization of a subset of solutions to the \emph{next-to-maximal} cut (that for which $\x{\ell_1}{3}\neq0$)---those associated with the parity/coloring of three-particle vertices as indicated in the following contour diagram:
\vspace{-6pt}\eq{\begin{tikzpicture}[scale=1.4*\figScale,baseline=-3.05]\useasboundingbox ($(-2.,-1.5)$) rectangle ($(2.,1.5)$);\draw[int,line width=0.1,red,draw=\boundingDraw] ($(-2.,-1.5)$) rectangle ($(2.,1.5)$);
\coordinate (v0) at (0,0);
\coordinate (v1) at ($(v0)+(150:0.75)$);
\coordinate (v2) at ($(v0)+(30:0.75)$);
\coordinate (v3) at ($(v0)+(-90:0.75)$);
\coordinate (a1) at ($(v0)+(90:0.9)$);
\coordinate (b1) at ($(v0)+(-10:1.2)$);
\coordinate (b2) at ($(v0)+(-50:1.2)$);
\coordinate (c1) at ($(v0)+(-130:1.2)$);
\coordinate (c2) at ($(v0)+(-170:1.2)$);
\draw[int] (v0)--(v1);\draw[int] (v0)--(v2);\draw[int] (v0)--(v3);
\draw[int] (v1)--(a1);\draw[int] (v2)--(a1);\draw[int] (v2)--(b1);\draw[int] (v3)--(b2);\draw[int] (b1)--(b2);\draw[int] (v3)--(c1);\draw[int] (v1)--(c2);\draw[int] (c1)--(c2);
\leg{(a1)}{135}{$2$}\leg{(a1)}{45}{$3$}\leg{(v2)}{40}{$4$}\leg{(b1)}{5+45}{$5$}\leg{(b1)}{5-45}{$6$}\leg{(b2)}{-67+45}{$7$}\leg{(b2)}{-67-45}{$8$}\leg{(c1)}{-113+45}{$9$}\leg{(c1)}{-113-45}{$10$}\leg{(c2)}{175+45}{$11$}\leg{(c2)}{175-45}{$1$}
\node at (v0) [blueDot] {};\node at (v1) [whiteDot] {};\node at (v2) [bdot] {};\node at (v3) [whiteDot] {};\node at (a1) [bdot] {};\node at (b1) [bdot] {};\node at (b2) [bdot] {};\node at (c1) [bdot] {};\node at (c2) [bdot] {};\node at ($(v0)+(90:0.45)$) [] {$\ell_1$};\node at ($(v0)+(-30:0.6)$) [] {$\ell_2$};\node at ($(v0)+(-150:0.6)$) [] {$\ell_3$};
\end{tikzpicture}\label{eleven_cut_diagram}\vspace{0pt}}
We may parameterize this one-dimensional family of solutions to the cut equations in momentum-twistor space by writing $(\ell_1)\equivR(\hat{2}\,\hat{4})$, $(\ell_2)\equivR(\hat{5}\,\hat{9})$, $(\ell_3)\equivR(\hat{9}\,\hat{2})$, where 
\eq{\begin{array}{ll@{$\hspace{20pt}$}lllll}\hat{2}(\r{\alpha})&\equivR z_{2}+\r{\alpha}\,z_1&\hat{5}(\r{\alpha})&\equivR\big(5\,4\big)\tcap\big(6\,7\,\hat{9}\big)\\
\hat{4}(\r{\alpha})&\equivR \big(4\,3\big)\tcap\big(\hat{5}\,\hat{9}\,\hat{2}\big)&\hat{9}(\r{\alpha})&\equivR\big(9\,8\big)\tcap\big(10\,11\,\hat{2}\big)\end{array}\,.\label{eleven_cut_parameterization}}
It is not hard to verify that this one-parameter family of $\ell_i(\r{\alpha})$ satisfy the eleven cut equations, and that they correspond to the particular branch indicated by the colouring of vertices in (\ref{eleven_cut_diagram}). Readers interested in the details on how such a parameterization can be constructed may consult appendix~\ref{app:cussextdis}.
To access the \emph{leading} singularities (that is the singularities arising from the pinch of all the internal lines) of the initial integral from this eleven-cut, we must cut the final propagator---that is, take residues about the solutions to $\x{\ell_1}{3}\!\propto\!\ab{\ell_1\,2\,3}\!=\!\ab{\hat{2}\,\hat{4}\,2\,3}\!=\!\r{\alpha}\ab{1\,\hat{4}\,2\,3}\!=\!0$. There are 4 solutions to this final cut equation: one rational ($\r{\alpha}\!=\!0$), and three roots of an irreducible cubic. The $\r{\alpha}\!=\!0$ solution corresponds to a rational leading singularity with the following colouring of vertices:
\eq{\begin{tikzpicture}[scale=1.4*\figScale,baseline=-3.05]\useasboundingbox ($(-2.,-1.5)$) rectangle ($(2.,1.5)$);\draw[int,line width=0.1,red,draw=\boundingDraw] ($(-2.,-1.5)$) rectangle ($(2.,1.5)$);
\coordinate (v0) at (0,0);
\coordinate (v1) at ($(v0)+(150:0.75)$);
\coordinate (v2) at ($(v0)+(30:0.75)$);
\coordinate (v3) at ($(v0)+(-90:0.75)$);
\coordinate (a1) at ($(v0)+(90:0.9)$);
\coordinate (b1) at ($(v0)+(-10:1.2)$);
\coordinate (b2) at ($(v0)+(-50:1.2)$);
\coordinate (c1) at ($(v0)+(-130:1.2)$);
\coordinate (c2) at ($(v0)+(-170:1.2)$);
\draw[int] (v0)--(v1);\draw[int] (v0)--(v2);\draw[int] (v0)--(v3);
\draw[int] (v1)--(a1);\draw[int] (v2)--(a1);\draw[int] (v2)--(b1);\draw[int] (v3)--(b2);\draw[int] (b1)--(b2);\draw[int] (v3)--(c1);\draw[int] (v1)--(c2);\draw[int] (c1)--(c2);
\leg{(a1)}{135}{$2$}\leg{(a1)}{45}{$3$}\leg{(v2)}{40}{$4$}\leg{(b1)}{5+45}{$5$}\leg{(b1)}{5-45}{$6$}\leg{(b2)}{-67+45}{$7$}\leg{(b2)}{-67-45}{$8$}\leg{(c1)}{-113+45}{$9$}\leg{(c1)}{-113-45}{$10$}\leg{(c2)}{175+45}{$11$}\leg{(c2)}{175-45}{$1$}
\node at (v0) [blueDot] {};\node at (v1) [whiteDot] {};\node at (v2) [bdot] {};\node at (v3) [whiteDot] {};\node at (a1) [bdot] {};\node at (b1) [bdot] {};\node at (b2) [bdot] {};\node at (c1) [bdot] {};\node at (c2) [bdot] {};\node at ($(v0)+(90:0.45)$) [] {$\ell_1$};\node at ($(v0)+(-30:0.6)$) [] {$\ell_2$};\node at ($(v0)+(-150:0.6)$) [] {$\ell_3$};
\end{tikzpicture}\bigger{\underset{\r{\alpha}=0}{\Rightarrow}}\begin{tikzpicture}[scale=1.4*\figScale,baseline=-3.05]\useasboundingbox ($(-2.,-1.5)$) rectangle ($(2.,1.5)$);\draw[int,line width=0.1,red,draw=\boundingDraw] ($(-2.,-1.5)$) rectangle ($(2.,1.5)$);
\coordinate (v0) at (0,0);
\coordinate (v1) at ($(v0)+(150:0.75)$);
\coordinate (v2) at ($(v0)+(30:0.75)$);
\coordinate (v3) at ($(v0)+(-90:0.75)$);
\coordinate (a1) at ($(v0)+(110:1.2)$);
\coordinate (a2) at ($(v0)+(70:1.2)$);
\coordinate (b1) at ($(v0)+(-10:1.2)$);
\coordinate (b2) at ($(v0)+(-50:1.2)$);
\coordinate (c1) at ($(v0)+(-130:1.2)$);
\coordinate (c2) at ($(v0)+(-170:1.2)$);
\draw[int] (v0)--(v1);\draw[int] (v0)--(v2);\draw[int] (v0)--(v3);
\draw[int] (v2)--(a2);\draw[int] (v1)--(a1);\draw[int] (a1)--(a2);\draw[int] (v2)--(b1);\draw[int] (v3)--(b2);\draw[int] (b1)--(b2);\draw[int] (v3)--(c1);\draw[int] (v1)--(c2);\draw[int] (c1)--(c2);
\leg{(a1)}{135}{$2$}\leg{(a2)}{53}{$3$}\leg{(v2)}{40}{$4$}\leg{(b1)}{5+45}{$5$}\leg{(b1)}{5-45}{$6$}\leg{(b2)}{-67+45}{$7$}\leg{(b2)}{-67-45}{$8$}\leg{(c1)}{-113+45}{$9$}\leg{(c1)}{-113-45}{$10$}\leg{(c2)}{175+45}{$11$}\leg{(c2)}{175-45}{$1$}
\node at (v0) [blueDot] {};\node at (v1) [whiteDot] {};\node at (v2) [bdot] {};\node at (v3) [whiteDot] {};\node at (a1) [whiteDot] {};\node at (a2) [blueDot] {};\node at (b1) [bdot] {};\node at (b2) [bdot] {};\node at (c1) [bdot] {};\node at (c2) [bdot] {};\node at ($(v0)+(90:0.6)$) [] {$\ell_1$};\node at ($(v0)+(-30:0.6)$) [] {$\ell_2$};\node at ($(v0)+(-150:0.6)$) [] {$\ell_3$};
\end{tikzpicture}\fwboxL{0pt}{.}}%
To see this, notice that when $\r{\alpha}\!=\!0$, the line in momentum-twistor space corresponding to $(\ell_1)\equivR(\hat{2}\,\hat{4})$ (as parameterized in (\ref{eleven_cut_parameterization})) passes directly through the point $z_2$ (as $\hat{2}\!\to\!z_2$ when $\r{\alpha}\!\to\!0$).

We are interested in the \emph{other} leading singularities---those involving the three solutions to the final-cut equation, where $\ab{1\,\hat{4}\,2\,3}\!=\!\ab{1\,2\,3\,\big(4\,3\big)\tcap\big(\hat{5}\,\hat{9}\,\hat{2}\big)}\!\propto\!\ab{\hat{2}\,\hat{5}\,\hat{9}\,3}\!=\!0$, which would correspond to leading singularities associated with the following contour diagram:
\eq{\begin{tikzpicture}[scale=1.4*\figScale,baseline=-3.05]\useasboundingbox ($(-2.,-1.5)$) rectangle ($(2.,1.5)$);\draw[int,line width=0.1,red,draw=\boundingDraw] ($(-2.,-1.5)$) rectangle ($(2.,1.5)$);
\coordinate (v0) at (0,0);
\coordinate (v1) at ($(v0)+(150:0.75)$);
\coordinate (v2) at ($(v0)+(30:0.75)$);
\coordinate (v3) at ($(v0)+(-90:0.75)$);
\coordinate (a1) at ($(v0)+(110:1.2)$);
\coordinate (a2) at ($(v0)+(70:1.2)$);
\coordinate (b1) at ($(v0)+(-10:1.2)$);
\coordinate (b2) at ($(v0)+(-50:1.2)$);
\coordinate (c1) at ($(v0)+(-130:1.2)$);
\coordinate (c2) at ($(v0)+(-170:1.2)$);
\draw[int] (v0)--(v1);\draw[int] (v0)--(v2);\draw[int] (v0)--(v3);
\draw[int] (v2)--(a2);\draw[int] (v1)--(a1);\draw[int] (a1)--(a2);\draw[int] (v2)--(b1);\draw[int] (v3)--(b2);\draw[int] (b1)--(b2);\draw[int] (v3)--(c1);\draw[int] (v1)--(c2);\draw[int] (c1)--(c2);
\leg{(a1)}{135}{$2$}\leg{(a2)}{53}{$3$}\leg{(v2)}{40}{$4$}\leg{(b1)}{5+45}{$5$}\leg{(b1)}{5-45}{$6$}\leg{(b2)}{-67+45}{$7$}\leg{(b2)}{-67-45}{$8$}\leg{(c1)}{-113+45}{$9$}\leg{(c1)}{-113-45}{$10$}\leg{(c2)}{175+45}{$11$}\leg{(c2)}{175-45}{$1$}
\node at (v0) [blueDot] {};\node at (v1) [whiteDot] {};\node at (v2) [bdot] {};\node at (v3) [whiteDot] {};\node at (a1) [blueDot] {};\node at (a2) [whiteDot] {};\node at (b1) [bdot] {};\node at (b2) [bdot] {};\node at (c1) [bdot] {};\node at (c2) [bdot] {};\node at ($(v0)+(90:0.6)$) [] {$\ell_1$};\node at ($(v0)+(-30:0.6)$) [] {$\ell_2$};\node at ($(v0)+(-150:0.6)$) [] {$\ell_3$};
\end{tikzpicture}\fwboxL{0pt}{.}}
It is not hard to check that this final cut equation is (irreducibly) cubic in the parameter $\r{\alpha}$. In particular, it is given by:
\eq{q(\r{\alpha})\equivR\ab{\hat{2}\,\hat{5}\,\hat{9}\,3}\equivL c_0{+}c_1\,\r{\alpha}{+}c_2\,\r{\alpha}^2{+}c_3\,\r{\alpha}^3\label{eq:lscubic}}
with 
\eq{\begin{array}{l@{}l@{$\,\,\,\;$}l@{}@{}l}c_0\equivR&\ab{2\,3\,\hat{5}_2\,\hat{9}_2}&c_1\equivR&\ab{1\,3\,\hat{5}_2\,\hat{9}_2}{+}\ab{2\,3\,\hat{5}_1\,\hat{9}_2}{+}\ab{2\,3\,\hat{5}_2\,\hat{9}_1}\\
c_2\equivR&\ab{2\,3\,\hat{5}_1\,\hat{9}_1}{+}\ab{1\,3\,\hat{5}_2\,\hat{9}_1}{+}\ab{1\,3\,\hat{5}_1\,\hat{9}_2}
&c_3\equivR&\ab{1\,3\,\hat{5}_1\,\hat{9}_1}
\end{array}}
where we have introduced $\hat{5}_i\equivR\big(5\,4)\tcap\big(6\,7\,\hat{9}_i\big)$ and $\hat{9}_i\equivR\big(9\,8)\tcap\big(10\,11\,i\big)$.

\subsection{Cube-Root Leading Singularities of an Amplitude in \texorpdfstring{$\mathcal{N}\!=\!4$}{N=4} sYM}
For maximally supersymmetric Yang-Mills theory (`sYM'), we can give an explicit formula for the three on-shell functions  associated with these cube-root-dependent leading singularities of the original Feynman integral. To do this, we start by first computing the on-shell function for the 11-cut of (\ref{eleven_cut_diagram}). Representing MHV and $\overline{\text{MHV}}$ tree amplitudes by blue and white vertices, respectively, the eleven-cut on-shell diagram corresponds to:%
\eq{\begin{tikzpicture}[scale=1.4*\figScale,baseline=-1
]\useasboundingbox ($(-2.,-1.5)$) rectangle ($(2.,1.5)$);\draw[int,line width=0.1,red,draw=\boundingDraw] ($(-2.,-1.5)$) rectangle ($(2.,1.5)$);
\coordinate (v0) at (0,0);
\coordinate (v1) at ($(v0)+(150:0.75)$);
\coordinate (v2) at ($(v0)+(30:0.75)$);
\coordinate (v3) at ($(v0)+(-90:0.75)$);
\coordinate (a1) at ($(v0)+(90:0.9)$);
\coordinate (b1) at ($(v0)+(-10:1.2)$);
\coordinate (b2) at ($(v0)+(-50:1.2)$);
\coordinate (c1) at ($(v0)+(-130:1.2)$);
\coordinate (c2) at ($(v0)+(-170:1.2)$);
\draw[int] (v0)--(v1);\draw[int] (v0)--(v2);\draw[int] (v0)--(v3);
\draw[int] (v1)--(a1);\draw[int] (v2)--(a1);\draw[int] (v2)--(b1);\draw[int] (v3)--(b2);\draw[int] (b1)--(b2);\draw[int] (v3)--(c1);\draw[int] (v1)--(c2);\draw[int] (c1)--(c2);
\leg{(a1)}{135}{$2$}\leg{(a1)}{45}{$3$}\leg{(v2)}{40}{$4$}\leg{(b1)}{5+45}{$5$}\leg{(b1)}{5-45}{$6$}\leg{(b2)}{-67+45}{$7$}\leg{(b2)}{-67-45}{$8$}\leg{(c1)}{-113+45}{$9$}\leg{(c1)}{-113-45}{$10$}\leg{(c2)}{175+45}{$11$}\leg{(c2)}{175-45}{$1$}
\node at (v0) [mhv] {};\node at (v1) [mhvBar] {};\node at (v2) [mhv] {};\node at (v3) [mhvBar] {};\node at (a1) [mhv] {};\node at (b1) [mhv] {};\node at (b2) [mhv] {};\node at (c1) [mhv] {};\node at (c2) [mhv] {};
\end{tikzpicture}\bigger{\Leftrightarrow}\begin{tikzpicture}[scale=1.4*\figScale,baseline=-3.05]\useasboundingbox ($(-2.,-1.5)$) rectangle ($(2.,1.5)$);\draw[int,line width=0.1,red,draw=\boundingDraw] ($(-2.,-1.5)$) rectangle ($(2.,1.5)$);
\coordinate (v0) at (0,0);
\coordinate (v1) at ($(v0)+(150:0.75)$);
\coordinate (v2) at ($(v0)+(30:0.75)$);
\coordinate (v3) at ($(v0)+(-90:0.75)$);
\coordinate (a1) at ($(v0)+(90:1)$);
\coordinate (b1) at ($(v0)+(-10:1.2)$);
\coordinate (b2) at ($(v0)+(-50:1.2)$);
\coordinate (c1) at ($(v0)+(-130:1.2)$);
\coordinate (c2) at ($(v0)+(-170:1.2)$);
\coordinate (a11) at ($(a1)+(-135:0.3)$);
\coordinate (a12) at ($(a1)+(135:0.3)$);
\coordinate (a13) at ($(a1)+(45:0.3)$);
\coordinate (a14) at ($(a1)+(-45:0.3)$);
\draw[int] (v0)--(v1);\draw[int] (v0)--(v2);\draw[int] (v0)--(v3);
\draw[int] (v1)--(a11);\draw[int] (a11)--(a12);\draw[int] (a12)--(a13);\draw[int] (a13)--(a14);\draw[int] (a14)--(v2);\draw[int] (a11)--(a14);\draw[int] (v2)--(b1);\draw[int] (v3)--(b2);\draw[int] (b1)--(b2);\draw[int] (v3)--(c1);\draw[int] (v1)--(c2);\draw[int] (c1)--(c2);
\leg{(a12)}{135}{$2$}\leg{(a13)}{45}{$3$}
\leg{(v2)}{40}{$4$}\leg{(b1)}{5+45}{$5$}\leg{(b1)}{5-45}{$6$}\leg{(b2)}{-67+45}{$7$}\leg{(b2)}{-67-45}{$8$}\leg{(c1)}{-113+45}{$9$}\leg{(c1)}{-113-45}{$10$}\leg{(c2)}{175+45}{$11$}\leg{(c2)}{175-45}{$1$}
\node at (v0) [mhv] {};\node at (v1) [mhvBar] {};\node at (v2) [mhv] {};\node at (v3) [mhvBar] {};\node at (a14) [mhv] {};\node at (a12) [mhv] {};\node at (a11) [mhvBar] {};\node at (a13) [mhvBar] {};\node at (b1) [mhv] {};\node at (b2) [mhv] {};\node at (c1) [mhv] {};\node at (c2) [mhv] {};
\end{tikzpicture}}
where the latter is one realization of the former, using one of the (two) BCFW representations of the four-particle MHV amplitude appearing on the LHS. Thus, we can recognize the on-shell function for the eleven-cut as the one-parameter BCFW shift of the following, purely rational leading singularity in sYM:
\eq{\fwbox{0pt}{\hspace{-20pt}\begin{tikzpicture}[scale=1.4*\figScale,baseline=-3.05]\useasboundingbox ($(-2.,-1.75)$) rectangle ($(2.,1.75)$);\draw[int,line width=0.1,red,draw=\boundingDraw] ($(-2.,-1.5)$) rectangle ($(2.,1.5)$);
\coordinate (v0) at (0,0);
\coordinate (v1) at ($(v0)+(150:0.75)$);
\coordinate (v2) at ($(v0)+(30:0.75)$);
\coordinate (v3) at ($(v0)+(-90:0.75)$);
\coordinate (a1) at ($(v0)+(110:1.2)$);
\coordinate (a2) at ($(v0)+(70:1.2)$);
\coordinate (b1) at ($(v0)+(-10:1.2)$);
\coordinate (b2) at ($(v0)+(-50:1.2)$);
\coordinate (c1) at ($(v0)+(-130:1.2)$);
\coordinate (c2) at ($(v0)+(-170:1.2)$);
\draw[int] (v0)--(v1);\draw[int] (v0)--(v2);\draw[int] (v0)--(v3);
\draw[int] (v2)--(a2);\draw[int] (v1)--(a1);\draw[int] (a1)--(a2);\draw[int] (v2)--(b1);\draw[int] (v3)--(b2);\draw[int] (b1)--(b2);\draw[int] (v3)--(c1);\draw[int] (v1)--(c2);\draw[int] (c1)--(c2);
\leg{(a1)}{135}{$2$}\leg{(a2)}{53}{$3$}\leg{(v2)}{40}{$4$}\leg{(b1)}{5+45}{$5$}\leg{(b1)}{5-45}{$6$}\leg{(b2)}{-67+45}{$7$}\leg{(b2)}{-67-45}{$8$}\leg{(c1)}{-113+45}{$9$}\leg{(c1)}{-113-45}{$10$}\leg{(c2)}{175+45}{$11$}\leg{(c2)}{175-45}{$1$}
\node at (v0) [mhv] {};\node at (v1) [mhvBar] {};\node at (v2) [mhv] {};\node at (v3) [mhvBar] {};\node at (a1) [mhvBar] {};\node at (a2) [mhv] {};\node at (b1) [mhv] {};\node at (b2) [mhv] {};\node at (c1) [mhv] {};\node at (c2) [mhv] {};
\end{tikzpicture}\hspace{-5pt}{=}\;\;\begin{array}{l}~\\R[8\,9\,10\,11\,2]R[4\,5\,6\,7\,\big(9\,8\big)\tcap\big(10\,11\,2\big)]\\R[3\,4\,\big(9\,8\big)\tcap\big(10\,11\,2\big)\,\big(5\,4\big)\tcap\big(6\,7\,\big(9\,8\big)\tcap\big(10\,11\,2\big)\big)]\end{array}}\label{other_pole_ls_formula}}
This on-shell function has a permutation label given by $\{9,6,7,11,8,3,10,5,1,4,2\}$ \cite{ArkaniHamed:book,Bourjaily:2012gy} and has a representation in terms of $R$-invariants given by the expression above---where 
\eq{R[a\,b\,c\,d\,e]\equivR\frac{\delta^{1\times4}\hspace{-1.5pt}\big(\eta_a\ab{b\,c\,d\,e}{+}\eta_b\ab{c\,d\,e\,a}{+}\eta_c\ab{d\,e\,a\,b}{+}\eta_d\ab{e\,a\,b\,c}{+}\eta_e\ab{a\,b\,c\,d}
}{\ab{a\,b\,c\,d}\ab{b\,c\,d\,e}\ab{c\,d\,e\,a}\ab{d\,e\,a\,b}\ab{e\,a\,b\,c}}\,.}
(The formula given in (\ref{other_pole_ls_formula}) was found using \cite{Bourjaily:2012gy} and  recognizing it (via its permutation label) as a series of inverse-soft factors; in general, any leading singularity that could have arisen via BCFW-bridges may be constructed recursively by recognizing factorized graphs among its bridge-boundaries.) 

To implement the BCFW-shift as required to represent the 11-cut on-shell function, we merely need to replace $z_2\mapsto\hat{z_2}\equivR z_2{+}\r{\alpha}\,z_1$ and include the prefactor of $1/\r{\alpha}$ in the bosonic part of the superfunction. Thus, 
\eq{\fwbox{0pt}{\hspace{-20pt}\begin{tikzpicture}[scale=1.4*\figScale,baseline=-3.05]\useasboundingbox ($(-2.,-1.75)$) rectangle ($(2.,1.75)$);\draw[int,line width=0.1,red,draw=\boundingDraw] ($(-2.,-1.5)$) rectangle ($(2.,1.5)$);
\coordinate (v0) at (0,0);
\coordinate (v1) at ($(v0)+(150:0.75)$);
\coordinate (v2) at ($(v0)+(30:0.75)$);
\coordinate (v3) at ($(v0)+(-90:0.75)$);
\coordinate (a1) at ($(v0)+(90:1)$);
\coordinate (b1) at ($(v0)+(-10:1.2)$);
\coordinate (b2) at ($(v0)+(-50:1.2)$);
\coordinate (c1) at ($(v0)+(-130:1.2)$);
\coordinate (c2) at ($(v0)+(-170:1.2)$);
\coordinate (a11) at ($(a1)+(-135:0.3)$);
\coordinate (a12) at ($(a1)+(135:0.3)$);
\coordinate (a13) at ($(a1)+(45:0.3)$);
\coordinate (a14) at ($(a1)+(-45:0.3)$);
\draw[int] (v0)--(v1);\draw[int] (v0)--(v2);\draw[int] (v0)--(v3);
\draw[int] (v1)--(a11);\draw[int] (a11)--(a12);\draw[int] (a12)--(a13);\draw[int] (a13)--(a14);\draw[int] (a14)--(v2);\draw[int] (a11)--(a14);\draw[int] (v2)--(b1);\draw[int] (v3)--(b2);\draw[int] (b1)--(b2);\draw[int] (v3)--(c1);\draw[int] (v1)--(c2);\draw[int] (c1)--(c2);
\leg{(a12)}{135}{$2$}\leg{(a13)}{45}{$3$}
\leg{(v2)}{40}{$4$}\leg{(b1)}{5+45}{$5$}\leg{(b1)}{5-45}{$6$}\leg{(b2)}{-67+45}{$7$}\leg{(b2)}{-67-45}{$8$}\leg{(c1)}{-113+45}{$9$}\leg{(c1)}{-113-45}{$10$}\leg{(c2)}{175+45}{$11$}\leg{(c2)}{175-45}{$1$}
\node at (v0) [mhv] {};\node at (v1) [mhvBar] {};\node at (v2) [mhv] {};\node at (v3) [mhvBar] {};\node at (a14) [mhv] {};\node at (a12) [mhv] {};\node at (a11) [mhvBar] {};\node at (a13) [mhvBar] {};\node at (b1) [mhv] {};\node at (b2) [mhv] {};\node at (c1) [mhv] {};\node at (c2) [mhv] {};
\end{tikzpicture}\hspace{-0pt}{=}\;\;\begin{array}{l}\frac{1}{\r{\alpha}}R[8\,9\,10\,11\,\hat{2}]R[4\,5\,6\,7\,\hat{9}]\,R[3\,4\,\hat{9}\,\hat{5}\,\hat{2}]\end{array}}\label{eleven_cut_form}}
where $\hat{2},\hat{5}$, and $\hat{9}$ were defined above in (\ref{eleven_cut_parameterization}). This on-shell function corresponds to a $19$-dimensional cell in the momentum-space Grassmannian $Gr_+(11,5)$ labeled by the permutation $\{9,6,7,11,8,2,10,5,1,4,3\}$.

From this 11-cut on-shell function, the 12-cut on-shell functions involving the cube roots can be obtained by taking a simple residue about one of the poles arising from $\ab{\hat{2}\,\hat{5}\,\hat{9}\,3}\!=\!0$ (which appears in the denominator of the bosonic part of the third $R$-invariant factor appearing in the expression above). Notice that this is \emph{precisely} the same cubic as encountered in the final cut-condition above. When a residue is taken on any one of the cube roots, the resulting on-shell function would be labeled by the path permutation $\{9,7,6,11,8,2,10,5,1,4,3\}$ for a $18\!=\!(2\!\times\!11{-}4)$-dimensional cell in $Gr_+(11,5)$; using the tools of \cite{ArkaniHamed:book,Bourjaily:2012gy}, it can be easily confirmed that this cell in the Grassmannian has `intersection number' (with kinematics) equal to 3---meaning, that there are \emph{three} solutions to the constraints connecting the auxiliary Grassmannian to kinematic data. These three particular leading singularities correspond to the particular roots of the cubic used to define the cell via (\ref{eleven_cut_form}).

The existence of three leading singularities here leads to an unusual situation. Typically, we expect leading singularities to serve as prefactors of polylogarithmic functions in the integrated expression for an amplitude. Here, we have three distinct possible prefactors, corresponding to distinct residues of eq.~\ref{eleven_cut_form} at the roots $\r{\alpha_i^*}$ in eq.~\ref{eq:lscubic}. 

The presence of these distinct possible prefactors raises the question of whether such an integral has differential equations that can be written in canonical form (cf.~\cite{Henn:2013pwa}). This form generally demands pure functions, and there is no normalization under which this integral is pure.

This issue is still present even for a scalar version of the diagram. There the leading singularities are just residues of $\frac{1}{q(\r{\alpha})}$ on its three poles, of the form $\frac{1}{(\r{\alpha_i^*}-\r{\alpha_j^*})(\r{\alpha_i^*}-\r{\alpha_k^*})}$. As these must sum to zero, we do have one identity:
\begin{equation}
    \frac{1}{(\r{\alpha_1^*} - \r{\alpha_2^*})(\r{\alpha_1^*} - \r{\alpha_3^*})}= -\frac{1}{(\r{\alpha_2^*} - \r{\alpha_1^*})(\r{\alpha_2^*} - \r{\alpha_3^*})} - \frac{1}{(\r{\alpha_3^*} - \r{\alpha_1^*})(\r{\alpha_3^*} - \r{\alpha_2^*})}\,.
\end{equation}
Thus, we can express one residue in terms of the others, but there are still two independent prefactors.

This situation does not arise for quadratic equations.  In that case we have the possible residues $\frac{1}{\r{\alpha_1^*} - \r{\alpha_2^*}}$ and $\frac{1}{\r{\alpha_2^*} - \r{\alpha_1^*}}$; but since these can differ only by a sign, we can pull out a global factor $\frac{1}{\r{\alpha_1^*} - \r{\alpha_2^*}}$, for example.  

The appearance of this situation for cubic roots is perhaps not so surprising, and is a behavior we expect to continue to higher orders. The final amplitude is, in any event, not generally pure: it has non-trivial leading singularities, which can in fact depend upon arbitrarily high order algebraic roots involving the kinematics. This behavior is natural from the point of view of unitarity methods, where amplitude integrands are \emph{rational} differential forms on the space of loop momenta; thus, when representing an amplitude using unitarity in terms of any basis of master loop integrands, whatever algebraic normalizations one uses must always conspire with algebraic coefficients (leading singularities) to yield a rational loop integrand.

\subsection{Cluster Coordinates for the Integral}
\label{subsec:cluster}

Following the above, we arrive at an expression for the loop momenta on the leading singularity in terms of momentum-twistor four-brackets. As four-brackets satisfy identities, it is not immediately obvious that the cube root is not reduced to something simpler upon application of these identities. To rule this out, we find expressions on the leading singularity in terms of an explicit twistor chart.

As we are considering an eleven-point diagram, one might naively expect to describe it with $3(11)-15=18$ variables. However, this diagram is simpler than a generic eleven-point dual-conformal diagram because of the presence of pairs of legs at the same corner, forming ``masses'': (5,6), (7,8), (9,10), and (11,1). As such, the diagram only depends on the seven dual points $x_2$, $x_3$, $x_4$, $x_5$, $x_7$, $x_9$, and $x_{11}$. Massless legs contribute three light-like conditions, bringing the correct number of variables to $4(7)-3-15=10$. We would thus ideally want a ten-parameter twistor chart.

In Ref.~\cite{Bourjaily:2018aeq}, two of the present authors found twistor charts for a variety of integrals with appropriate numbers of parameters by specializing to particular positroid cells. Unfortunately, this strategy does not suffice here, as there are no ten-dimensional boundaries of the eleven-point top cell that preserve dependence on the required dual points. Instead, we will follow a strategy outlined in Ref.~\cite{He:2021eec} based on cluster algebras, finding a sub-quiver of $G(4,11)$ with ten $\mathcal{X}$-coordinates that spans the correct space of dual points.

To carry out this strategy, we begin with a quiver for $G(4,11)$, then mutate on all possible nodes. For each mutation, we check to see if there is a subset of its $\mathcal{X}$-coordinates that is independent of the dual points of our integral. We keep only the mutations with the largest sets of $\mathcal{X}$-coordinates that satisfy this condition. We stop when we find at least one quiver which contains eight $\mathcal{X}$-coordinates that are independent from the seven dual points of our diagram, which in this case happens after fourteen mutations. Setting these eight $\mathcal{X}$-coordinates to one in a twistor parameterization of the quiver, we find a twistor parameterization with the minimal ten parameters for our diagram.

In terms of momentum-twistor four-brackets, these ten parameters can be written as follows,
{\allowdisplaybreaks
\eq{\begin{split}
    e_1&=\frac{\ab{\big(3\,5\big)\tcap\big(4\,6\,7\big)4\,8\,9}}{\ab{3\,4\,8\,9}\ab{4\,5\,6\,7}}\,,\\
    e_2&=\frac{\ab{\big(4\,5\big)\tcap\big(3\,8\,9\big)3\,10\,11}\ab{\big(8\,9\big)\tcap\big(2\,10\,11\big)4\,6\,7}}{\ab{2\,3\,10\,11}\ab{8\,9\,10\,11}\ab{\big(3\,5\big)\tcap\big(4\,6\,7\big)4\,8\,9}}\,,\\
    e_3&=\frac{\ab{3\,4\,6\,7}\ab{\big(4\,5\big)\tcap\big(3\,8\,9\big)3\,10\,11}}{\ab{3\,4\,10\,11}\ab{\big(4\,5\big)\tcap\big(3\,6\,7\big)3\,8\,9}}\,, \\
    e_4&=\frac{\ab{3\,4\,10\,11}\ab{8\,9\,10\,11}\ab{\big(3\,5\big)\tcap\big(4\,6\,7\big)4\,8\,9}\ab{\big(4\,5\big)\tcap\big(3\,6\,7\big)3\,8\,9}}{\ab{6\,7\big(3\,4\,5\big)\tcap\big(8\,9\big(10\,11\big)\tcap\big(3\,4\,5\big)\big)}\ab{3\,4\,8\,9}\ab{\big(8\,9\big)\tcap\big(3\,10\,11\big)4\,6\,7}}\,,\\
    e_5&=\frac{\ab{3\,4\,8\,9}\ab{\big(1\,2\big)\tcap\big(3\,4\,5\big)3\,10\,11}}{\ab{1\,2\,3\,4}\ab{\big(4\,5\big)\tcap\big(3\,8\,9\big)3\,10\,11}}\,, \\
    e_6&=-\frac{\ab{1\,2\,10\,11}\ab{2\,3\,4\,5}\ab{\big(8\,9\big)\tcap\big(3\,10\,11\big)4\,6\,7}}{\ab{\big(1\,2\big)\tcap\big(3\,4\,5\big)3\,10\,11}\ab{\big(8\,9\big)\tcap\big(2\,10\,11\big)4\,6\,7}}\,,\\
    e_7&=\frac{\ab{6\,7\big(3\,4\,5\big)\tcap\big(8\,9\big(10\,11\big)\tcap\big(3\,4\,5\big)\big)}\ab{1\,2\,3\,4}\ab{2\,3\,10\,11}}{\ab{6\,7\big(3\,4\,5\big)\tcap\big(8\,9\big(10\,11\big)\tcap\big(1\,2\,3\big)\big)}\ab{2\,3\,4\,5}\ab{3\,4\,10\,11}}\,,\\
    e_8&=\frac{\ab{8\,9\big(3\,10\,11\big)\tcap\big(4\,6\,7\big)}\ab{\big(4\,5\big)\tcap\big(3\,6\,7\big)3\,8\,9}}{\ab{3\,4\,6\,7}\ab{6\,7\,8\,9}\ab{\big(4\,5\big)\tcap\big(3\,8\,9\big)3\,10\,11}}\,,\\
    e_9&=-\frac{\ab{6\,7\big(3\,4\,5\big)\tcap\big(8\,9\big(10\,11\big)\tcap\big(1\,2\,3\big)\big)}\ab{\big(4\,5\big)\tcap\big(3\,8\,9\big)3\,10\,11}}{\ab{8\,9\,10\,11}\ab{\big(1\,2\big)\tcap\big(3\,4\,5\big)3\,10\,11}\ab{\big(4\,5\big)\tcap\big(3\,6\,7\big)3\,8\,9}}\,,\\
    e_{10}&=-\frac{\ab{6\,7\big(3\,4\,5\big)\tcap\big(8\,9\big(10\,11\big)\tcap\big(3\,4\,5\big)\big)}\ab{3\,4\,8\,9}}{\ab{\big(3\,5\big)\tcap\big(4\,6\,7\big)4\,8\,9}\ab{\big(4\,5\big)\tcap\big(3\,8\,9\big)3\,10\,11}}\,.
\end{split}}
}
We give an explicit parameterization of the momentum twistors in these coordinates in appendix~\Ref{app:explicitparam}, and also include them in an ancillary file \texttt{ClustersForCubic.m}.

\subsubsection{Structure of the Cubic Root}

Out of eighteen cluster $\mathcal{X}$-coordinates, we retain ten coordinates $e_i$ in our ten-parameter chart. In this chart, the cubic equation for $\r{\alpha}$ takes the following form:\\[-10pt]
\eq{
\frac{9}{4} e_2^4 e_3 e_4^4 e_5^2 e_7^2 e_8 e_9^2 e_{10}^2\left(e_8 e_2+e_2+5\right)^4  
   \Big[c_0+c_1 \, \r{\alpha}+ c_2 \, \r{\alpha}^2+c_3 \, \r{\alpha}^3\Big]=0\,.
   \label{eq:clustercubic}
}
The coefficients in this expression are quite long, so we omit them from the main text. They are presented in appendix~\Ref{app:explicitparam}, and in an ancillary file \texttt{ClustersForCubic.m}.

For a general cubic equation $c_0+c_1 x+c_2 x^2+c_3 x^3=0$, one can write the three solutions as,
\eq{x_k=-\frac{1}{3c_3}\left(c_2+\zeta^k C+\frac{\Delta_0}{\zeta^k C}\right)\,, \quad \textrm{where} \quad k\in\{0,1,2\}\,.}
Here we have $\zeta$ a third-root of unity and define,
\eq{
C\equivR\left(\frac{\Delta_1\pm\sqrt{\Delta_1^2-4\Delta_0^3}}{2}\right)^{\frac{1}{3}}\,,
}
with
\eq{\begin{split}
    \Delta_0&\equivR c_2^2-4c_3 c_1\\
    \Delta_1&\equivR 2c_2^3-9c_3 c_2 c_1+27 c_3^2 c_0\,.
\end{split}}

It is clear that, in order for the leading singularity of this diagram to have a singularity that goes as $\rho^{1/3}$ as some kinematic parameter $\rho\rightarrow 0$, we must have $C\rightarrow 0$ in this limit. This in turn demands that $\sqrt{\Delta_1^2-4\Delta_0^3}\rightarrow\Delta_1$, and thus that $\Delta_0\rightarrow 0$. Generically, in a limit where $\Delta_0\rightarrow 0$, $C$ behaves as
\eq{\begin{split}
    C\sim&\left(\frac{\Delta_1\pm\left(\Delta_1-2\frac{\Delta_0^3}{\Delta_1}+\mathcal{O}(\Delta_0^6)\right)}{2}\right)^\frac{1}{3}\hspace{-5pt}\sim\left(\frac{\Delta_0^3}{\Delta_1}+\mathcal{O}(\Delta_0^6)\right)^\frac{1}{3}
    \hspace{-5pt}\sim\frac{\Delta_0}{\Delta_1^{1/3}}+\mathcal{O}(\Delta_0^4)\,.
    \end{split}\label{eq:leadingbehavior}}
Thus, $C$ will generically vanish linearly, not as a third power, in such limits. $C$ only vanishes as a third power if both $\Delta_0$ and $\Delta_1$ simultaneously vanish, as suggested by the form of eq.~\ref{eq:leadingbehavior}. This is a co-dimension-two limit, and thus not forbidden by Landau's analysis.

There are two potential loopholes in this general behavior. If $\Delta_0$ vanishes identically for generic kinematics, then we only need a co-dimension one limit to uncover the cubic root. If $\Delta_0$ and $\Delta_1$ both vanish in the same kinematic limit, then it might also be possible to achieve a cubic root singularity.

In our case, these loopholes can be addressed by the explicit forms of $\Delta_0$ and $\Delta_1$. These are too complicated to print in full here, but their relevant structure is easy to display:
\eq{\begin{split}
 \Delta_0&=e_{6}^2 P(e_i) \\
 \Delta_1&=e_{6}^3 Q(e_i) \,.
\end{split}}
$P(e_i)$ and $Q(e_i)$ are both complicated polynomials, but crucially for our purposes they have no common factors. This means that we can only have a cube root singularity when a common factor of $\Delta_0$ and $\Delta_1$ vanishes, namely when $e_{6}\rightarrow 0$. As $e_{6}\rightarrow 0$, we have $\sqrt{\Delta_1^2-4\Delta_0^3}\sim \sqrt{e_{6}^6}$, and thus $C\sim e_{6}$. Thus, there exists no co-dimension one limit of this integral that behaves like $\rho^{1/3}$, as expected.\footnote{One might also wonder about singularities in which $C$ diverges rather than vanishes. A similar calculation shows that these also cannot contribute a cubic root singularity.}

\subsection{Landau Equations}

The behavior established in the previous section should also be able to be made manifest at the level of the Landau equations.  Here we will describe how this calculation may be initiated, though at this time we have found it too computationally intractable to be worth pursuing.

Examining the leading Landau singularity (with all propagators on-shell), one must in addition impose the Landau loop equations.

For the box integral below
\eq{\label{eq:box}%
\begin{tikzpicture}[scale=\figScale,baseline=-3.05]\useasboundingbox ($(-1.75,-1.75)$) rectangle ($(1.75,1.75)$);\draw[int,line width=0.1,red,draw=\boundingDraw] ($(-1.5,-1.5)$) rectangle ($(1.5,1.5)$);
\def\legLen{\edgeLen*0.55}\def\labelDist{\legLen*1.65}\def\legSpread{3}\def\dotSize{(\figScale*16pt)}
\coordinate (v0) at (0,0);
\coordinate (v4) at ($(v0)+(135:1.65)$);
\coordinate (v1) at ($(v0)+(45:1.65)$);
\coordinate (v2) at ($(v0)+(-45:1.65)$);
\coordinate (v3) at ($(v0)+(-135:1.65)$);
\coordinate (x1) at ($(v0)+(90:1.5)$);
\coordinate (x2) at ($(v0)+(0:1.5)$);
\coordinate (x3) at ($(v0)+(-90:1.5)$);
\coordinate (x4) at ($(v0)+(180:1.5)$);
\draw[int] (v1)--(v2);\draw[int] (v2)--(v3);\draw[int] (v3)--(v4);\draw[int] (v4)--(v1);
\draw[directedEdge](v4)--(v1);\draw[directedEdge](v1)--(v2);\draw[directedEdge](v2)--(v3);\draw[directedEdge](v3)--(v4);
\node[anchor=north,inner sep=3pt] at ($(v4)!.35!(v1)$) {\text{{\footnotesize$\g{\lambda_1\tilde{\lambda}_1}$}}};
\node[anchor=east,inner sep=3pt] at ($(v1)!.35!(v2)$) {\text{{\footnotesize$\g{\lambda_2\tilde{\lambda}_2}$}}};
\node[anchor=south,inner sep=3pt] at ($(v2)!.35!(v3)$) {\text{{\footnotesize$\g{\lambda_3\tilde{\lambda}_3}$}}};
\node[anchor=west,inner sep=3pt] at ($(v3)!.35!(v4)$) {\text{{\footnotesize$\g{\lambda_4\tilde{\lambda}_4}$}}};
\legMassive{(v1)}{45}{\text{{\normalsize$\,p_1$}}}\legMassive{(v2)}{-45}{\text{{\normalsize$\,p_2$}}}\legMassive{(v3)}{-135}{\text{{\normalsize$p_3\,$}}}\legMassive{(v4)}{135}{\text{{\normalsize$p_4\,$}}}
\node[bldot]at(v1){};\node[bldot]at(v2){};\node[bldot]at(v3){};\node[bldot]at(v4){};\node[rdot]at(v0){};
\node[rdot]at(x1){};\node[rdot]at(x2){};\node[rdot]at(x3){};\node[rdot]at(x4){};
\node[anchor=north west,inner sep=1pt]at(v0){$\r{x_0}$};\node[anchor=south,inner sep=2pt]at(x1){$\,\r{x_1}$};\node[anchor=west,inner sep=2pt]at(x2){$\,\r{x_2}$};\node[anchor=north,inner sep=3pt]at(x3){$\,\r{x_3}$};\node[anchor=east,inner sep=2pt]at(x4){$\,\r{x_4}$};
\node[anchor=north west,inner sep=2pt]at(v1){$\b{y_2}$};\node[anchor=north east,inner sep=2pt]at(v2){$\b{y_3}$};\node[anchor=south east,inner sep=2pt]at(v3){$\b{y_4}$};\node[anchor=south west,inner sep=2pt]at(v4){$\b{y_1}$};
\def\legLen{\edgeLen*0.45}\def\labelDist{\legLen*1.5}\def\legSpread{4}\def\dotSize{(\figScale*12pt)}
\end{tikzpicture}\hspace{40pt}\begin{array}{@{}r@{}c@{$\,\,$}l@{}}\r{x_a}{-}\r{x_0}&\equivL&\g{\lambda_a\tilde{\lambda}_a}\\\b{\alpha_a}\g{\lambda_a\tilde{\lambda}_a}&\equivL&\b{y_{a+1}}{-}\b{y_a}\end{array}}
the Landau loop equations in dual coordinate language read
\begin{equation}
    \label{eq:landau-dual}
    \alpha_1 (x_1 - x_0) + \dotso + \alpha_4 (x_4 - x_0) = 0.
\end{equation}

This equation is not manifestly dual-conformal invariant. As such, it is not straightforward to represent it in terms of a non-redundant parametrization of the kinematics. To make the equation dual-conformal invariant, we may upgrade it to an equation in embedding space, where each dual point is represented as a six-dimensional null vector $X_i\equivL(x_i^\mu,X_i^+,X_i^-)$ with metric such that
\begin{equation}
X_i\cdot X_j\equivR X_i^+ X_j^- + X_i^- X_j^+ - x_i\cdot x_j
\end{equation}
where for any non-zero real constant $c$ we take $X_i^+ = c$ for all $i$ and $X_i^- = \frac{1}{2 c} x_i^2$.  Indeed, using the Landau equations in eq.~\eqref{eq:landau-dual} and the on-shell conditions it can be shown that in a gauge where $X_i^+$ is a constant independent on $i$,
\begin{equation}
    \label{eq:landau-dual-embedding}
    \alpha_1 (X_1 - X_0) + \dotso + \alpha_4 (X_4 - X_0) = 0\,.
\end{equation}
For example, suppose we gauge $X_i^+=1$ for all $X_i$. Then the equations for the $x_i^\mu$ components are identical to those in eq.~\ref{eq:landau-dual}, while the $X_i^+$ equation is trivial, so we only need to check $X_i^-$. In this gauge $X_i^-=\frac{x_i^2}{2}$, so the equation for this component becomes,
\begin{equation}
    \alpha_1 (x_1^2-x_0^2) + \dotso \alpha_4 (x_4^2-x_0^2) = 0\,.
\end{equation}
Since we have imposed that the propagators are on-shell, we have $(x_i-x_0)^2=0$ for all $i$. Thus,
\begin{align}
    x_i^2-x_0^2&=x_i^2-x_0^2-(x_i-x_0)^2\nonumber\\
    &=2 x_i\cdot x_0 -2 x_0^2\nonumber\\
    &=2x_0\cdot(x_i-x_0)\,,
\end{align}
which shows that the equation for the $X_i^-$ components is implied by our original equation in $x$ space.

Using this type of manifestly dual-conformal invariant form of the Landau equations for the three-loop integral we have discussed in this section, and writing the dual points of the loop momenta as $X_{A},X_{B},X_{C}$, we find
\begin{align}
0=\,&\alpha_1\left( X_{A}{-} X_3\right){+}\alpha_2\left(X_{A}{-}X_4\right){+}\alpha_3\left(X_{A}{-}X_{B}\right){+}\alpha_4\left(X_{A}{-}X_{C}\right){+}\alpha_5\left(X_{A}{-}X_2\right),\\
0=\,&\alpha_6\left(X_{B}{-}X_5\right){+}\alpha_7\left(X_{B}{-} X_7\right){+}\alpha_8\left(X_{B}{-}X_9\right){+}\alpha_9\left(X_{B}{-}X_{C}\right){+}\alpha_3\left(X_{B}{-}X_A\right),\\
0=\,&\alpha_{10}\left( X_{C}{-}X_9 \right){+}\alpha_{11}\left(X_{C}{-}X_{11}\right){+}\alpha_{12}\left(X_{C}{-}X_{2}\right){+}\alpha_4\left(X_{C}{-}X_{A}\right){+}\alpha_9\left(X_{C}{-}X_B\right)\,.
\end{align}

Separating these component by component gives eighteen equations in the twelve $\alpha_i$, with coefficients that are algebraic functions in the cluster $\mathcal{X}$-coordinates. Some of these equations are redundant, but twelve are independent, giving a twelve-by-twelve matrix of coefficients. The leading Landau singularity occurs when the determinant of this matrix vanishes.

It would be interesting to find and factorize the determinant of this matrix. One would expect that it would have a factor in common with $\Delta_0$, and thus that it vanishing would lead a pair of roots of the cubic to coincide. Unfortunately, we are unable to confirm this due to the complexity of the coefficients present.\footnote{Implementing the problem in \texttt{Magma}~\cite{MR1484478} in terms of a matrix in the appropriate quotient ring ran for six days on a computer with a 2.3 GHz 18-Core processor and used approximately 30GB of memory without finishing.}

Before presenting a less computationally intensive approach in the next section, we make a few brief comments about the geometry of the problem.  Much as the conservation of loop momentum is solved by the introduction of dual coordinates, one can formally solve the Landau loop equations by introducing new auxiliary coordinates associated to the vertices of a diagram that incorporate the variables $\alpha_i$. Then the solution to the Landau equations can be characterized as a set of geometrical constraints on both the momentum twistors and a set of twistors corresponding to these new auxiliary coordinates.

We can briefly illustrate this approach with the example of the one-loop box integral (see eq.~\eqref{eq:box}). Here in analogy with the dual coordinates $\r{x_a}{-}\r{x_0}\equivL\g{\lambda_a\tilde{\lambda}_a}$, we have introduced auxiliary coordinates $\b{y_a}$ such that $\b{\alpha_a}\g{\lambda_a\tilde{\lambda}_a}\equivL\b{y_{a+1}}{-}\b{y_a}$. One can then think of the problem as a set of geometric constraints linking the momentum twistors and dual momentum twistors
\eq{
    \g{Z} = (\g{\lambda}, i \g{\lambda} \r{x})\,,\qquad\g{\tilde{Z}} = (-i \r{x}\g{\tilde{\lambda}}, \g{\tilde{\lambda}})\,,
}
with similar objects introduced for the auxiliary variables, $\g{Y_i} = (\g{\lambda}, i \g{\lambda} \b{y_i})$ and $\g{\tilde{Y}_i} = (-i \b{y_i}\g{\tilde{\lambda}}, \g{\tilde{\lambda}})$. This approach appears promising, and we will pursue more applications of it in future work.

\newpage
\section{\texorpdfstring{\emph{Exempli Gratia}}{Exempli Gratia}: A Pair of Octics}
\label{sec:three-quadrics}

In this section we analyze the singularity arising from the Landau diagram in fig.~\ref{fig:three-pentagons}.  We do not take $\langle A_i B_i A_{i + 1} B_{i + 1}\rangle = 0$ so this is a more general case of the kinematics analyzed before, but it has the advantage of being more symmetric.  In this case, we can describe the Landau locus in a particularly compact manner. We analyze only the case where the three-point amplitude in the center is $\overline{\text{MHV}}$, so the $\lambda$ spinors of the internal lines are proportional.  This means we can set the corresponding twistors equal, as a result they are all labeled as $C$ in the diagram. The parity-conjugate case can be obtained by projective duality.  Projective duality exchanges points and planes and sends lines to lines so the dual of three lines intersecting in one point is a configuration of three lines belonging to the same plane.

\begin{figure}
  \centering
  \tikzset{every picture/.style={line width=0.75pt}}
\begin{tikzpicture}[x=0.75pt,y=0.75pt,yscale=-1,xscale=1]
\draw  (275,114) -- (232.5,187.61) -- (147.5,187.61) -- (105,114) -- (147.5,40.39) -- (232.5,40.39) -- cycle ;
\draw    (232.5,187.61) -- (255,224) ;
\draw    (147.5,187.61) -- (127,223) ;
\draw    (105,114) -- (79,115) ;
\draw    (301,113) -- (275,114) ;
\draw    (125,4) -- (147.5,40.39) ;
\draw    (253,5) -- (232.5,40.39) ;
\draw    (190,114) -- (101,164) ;
\draw    (283,166) -- (190,114) ;
\draw    (191,15) -- (190,114) ;
\draw (36,127.4) node [anchor=north west][inner sep=0.75pt]    {$A_{1} \land B_{1}$};
\draw (57,54.4) node [anchor=north west][inner sep=0.75pt]    {$A_{2} \land B_{2}$};
\draw (132,2.4) node [anchor=north west][inner sep=0.75pt]    {$A_{3} \land B_{3}$};
\draw (191,2.4) node [anchor=north west][inner sep=0.75pt]    {$A_{4} \land B_{4}$};
\draw (256,54.4) node [anchor=north west][inner sep=0.75pt]    {$A_{5} \land B_{5}$};
\draw (270,127.4) node [anchor=north west][inner sep=0.75pt]    {$A_{6} \land B_{6}$};
\draw (243,171.4) node [anchor=north west][inner sep=0.75pt]    {$A_{7} \land B_{7}$};
\draw (163,191.4) node [anchor=north west][inner sep=0.75pt]    {$A_{8} \land B_{8}$};
\draw (80,169.4) node [anchor=north west][inner sep=0.75pt]    {$A_{9} \land B_{9}$};
\draw (131,84.4) node [anchor=north west][inner sep=0.75pt]    {$C\land D_{1}$};
\draw (198,85.4) node [anchor=north west][inner sep=0.75pt]    {$C\land D_{2}$};
\draw (162,141.4) node [anchor=north west][inner sep=0.75pt]    {$C\land D_{3}$};
\end{tikzpicture}
\caption{Three-pentagon cubic root singularity.  We have described the dual points of the diagram in terms of lines in twistor space.}
\label{fig:three-pentagons}
\end{figure}

\subsection{On-Shell Space}

We begin by discussing the on-shell space, using some notions from projective geometry. In order for all propagators of the diagram in fig.~\ref{fig:three-pentagons} to be on-shell, we must have that the lines in twistor space $C\wedge D_1$, $C\wedge D_2$, and $C\wedge D_3$ intersect with all of the lines defining their adjacent dual points.

  To begin, we focus on solving the on-shell conditions for the propagators on the outside of the diagram. The skew lines $A_1 \wedge B_1$, $A_2 \wedge B_2$ and $A_3 \wedge B_3$ are contained in a unique quadric surface $Q_1$.  Similarly for $A_4 \wedge B_4$, $A_5 \wedge B_5$, $A_6 \wedge B_6$ which determine a unique quadric $Q_2$ and for $A_7 \wedge B_7$, $A_8 \wedge B_8$, $A_9 \wedge B_9$ which determine a quadric $Q_3$.  Through any point of $Q_1$ passes a line which intersects $A_1 \wedge B_1$, $A_2 \wedge B_2$ and $A_3 \wedge B_3$.  Conversely, any line which intersects these three lines is completely contained in the quadric $Q_1$. Thus, in order for the propagators surrounding this loop to be on-shell, we must have that the line parametrizing the loop momentum is contained in $Q_1$. In order to parametrize this line, we may write a generic point $P_{A_1 B_1}$ on the line $A_1 \wedge B_1$ with
  \begin{equation}
  P_{A_1 B_1}=A_1+\nu_1 B_1\,.
\end{equation}
  Then we can write the space of lines $I_1$ contained in the quadric $Q_1$ as,
  \begin{equation}
  \begin{split}
  I_1= & (P_{A_1 B_1} A_2 B_2) \cap (P_{A_1 B_1} A_3 B_3)
  \\ = & (P_{A_1 B_1}\wedge A_2)  \langle B_2, P_{A_1 B_1}, A_3, B_3 \rangle -
        (P_{A_1 B_1}\wedge B_2) \langle A_2, P_{A_1 B_1}, A_3, B_3 \rangle
        \end{split}
 \end{equation}
  Following this same procedure for the quadrics $Q_2$ and $Q_3$, one obtains three lines $I_1$, $I_2$, and $I_3$ parametrizing the three loop momenta in terms of three parameters $\nu_1, \nu_2, \nu_3$.

  Finally, we must impose the on-shell conditions of the internal lines. These are enforced by demanding that the lines parametrizing the loop momenta intersect. We need $\langle I_1 I_2\rangle=\langle I_1 I_3\rangle=\langle I_2 I_3\rangle=0$. We have evaluated these equations for generic external kinematics using \texttt{SageMath}~\cite{sagemath}. Two of the variables can be eliminated rationally, resulting in a degree-sixteen polynomial in the final $\nu_3$. This polynomial is reducible: it factors into two irreducible degree-eight polynomials. Thus, the on-shell space for this polynomial involves roots of irreducible octics!

\subsection{Landau Equations}

We now proceed to describe the Landau equations for this integral. We introduce dual embedding space coordinates
\begin{gather}
    X_i = \frac{A_i \wedge B_i}{\langle I A_i B_i\rangle}, \qquad
    W_a = \frac{C \wedge D_a}{\langle I C D_a\rangle}.
\end{gather}
The components $(\alpha \dot{\alpha})$ of these embedding space dual coordinates are the usual four-dimensional dual coordinates (which we denote by lowercase letters).

In terms of these four-dimensional dual coordinates the Landau loop equations read
\begin{gather}
    \alpha_1 x_1 + \alpha_2 x_2 + \alpha_3 x_3 + \alpha_2' w_2 - \alpha_3' w_3 = (\alpha_1 + \alpha_2 + \alpha_3 + \alpha_2' - \alpha_3') w_1, \\
    \alpha_4 x_4 + \alpha_5 x_5 + \alpha_6 x_6 + \alpha_3' w_3 - \alpha_1' w_1 = (\alpha_4 + \alpha_5 + \alpha_6 + \alpha_3' - \alpha_1') w_2, \\
    \alpha_7 x_7 + \alpha_8 x_8 + \alpha_9 x_9 + \alpha_1' w_1 - \alpha_2' w_2 = (\alpha_7 + \alpha_8 + \alpha_9 + \alpha_1' - \alpha_2') w_3.
\end{gather}
When upgraded to twistor language we get
\begin{multline}
    \alpha_1 \frac{A_1 \wedge B_1}{\langle I A_1 B_1\rangle} +
    \alpha_2 \frac{A_2 \wedge B_2}{\langle I A_2 B_2\rangle} +
    \alpha_3 \frac{A_3 \wedge B_3}{\langle I A_3 B_3\rangle} +
    \alpha_2' \frac{C \wedge D_2}{\langle I C D_2\rangle} -
    \alpha_3' \frac{C \wedge D_3}{\langle I C D_3\rangle} = \\
    (\alpha_1 + \alpha_2 + \alpha_3 + \alpha_2' - \alpha_3')
    \frac{C \wedge D_1}{\langle I C D_1\rangle}.
\end{multline}
In the following it will be convenient to redefine
\begin{gather}
    \alpha_i \to \alpha_i \langle I A_i B_i\rangle, \qquad
    \alpha_a' \to \alpha_a' \langle I C D_a\rangle.
\end{gather}

The three loop Landau equations are
\begin{gather}
\begin{split}
  \alpha_1 A_1 \wedge B_1 + \alpha_2 A_2 \wedge B_2 + \alpha_3 A_3 \wedge B_3 + \alpha_2' C \wedge D_2 - \alpha_3' C \wedge D_3 =\\
  \frac {\scriptstyle{\alpha_1 \langle I A_1 B_1\rangle +
   \alpha_2 \langle I A_2 B_2\rangle +
   \alpha_3 \langle I A_3 B_3\rangle +
   \alpha_2' \langle I C D_2\rangle -
   \alpha_3' \langle I C D_3\rangle}}{\langle I C D_1\rangle} C \wedge D_1,
\end{split} \label{eq:hexagon-ll1}\\
\begin{split}
  \alpha_4 A_4 \wedge B_4 + \alpha_5 A_5 \wedge B_5 + \alpha_6 A_6 \wedge B_6 + \alpha_3' C \wedge D_3 - \alpha_1' C \wedge D_1 =\\
  \frac{\scriptstyle{\alpha_4 \langle I A_4 B_4\rangle +
  \alpha_5 \langle I A_5 B_5\rangle +
  \alpha_6 \langle I A_6 B_6\rangle +
  \alpha_3' \langle I C D_3\rangle -
  \alpha_1' \langle I C D_1\rangle}}{\langle I C D_2\rangle} C \wedge D_2,
\end{split} \label{eq:hexagon-ll2}\\
\begin{split}
  \alpha_7 A_7 \wedge B_7 + \alpha_8 A_8 \wedge B_8 + \alpha_9 A_9 \wedge B_9 + \alpha_1' C \wedge D_1 - \alpha_2' C \wedge D_2 =\\
  \frac {\scriptstyle{\alpha_7 \langle I A_7 B_7\rangle +
  \alpha_8 \langle I A_8 B_8\rangle +
  \alpha_9 \langle I A_9 B_9\rangle +
  \alpha_1' \langle I C D_1\rangle -
  \alpha_2' \langle I C D_2\rangle}}{\langle I C D_3\rangle} C \wedge D_3.
\end{split} \label{eq:hexagon-ll3}
\end{gather}
By wedging with $C$ it follows that
\begin{gather}
  \alpha_1 A_1 \wedge B_1 \wedge C + \alpha_2 A_2 \wedge B_2 \wedge C +   \alpha_3 A_3 \wedge B_3 \wedge C = 0, \\
  \alpha_4 A_4 \wedge B_4 \wedge C + \alpha_5 A_5 \wedge B_5 \wedge C +   \alpha_6 A_6 \wedge B_6 \wedge C = 0, \\
  \alpha_7 A_7 \wedge B_7 \wedge C + \alpha_8 A_8 \wedge B_8 \wedge C +   \alpha_9 A_9 \wedge B_9 \wedge C = 0.
\end{gather}
From the first equation we obtain that
\begin{gather}
  \alpha_1 \langle A_1 B_1 C A_3\rangle + \alpha_2 \langle A_2 B_2 C A_3\rangle = 0, \\
  \alpha_1 \langle A_1 B_1 C B_3\rangle + \alpha_2 \langle A_2 B_2 C B_3\rangle = 0.
\end{gather}
If we impose the condition that these two equations are compatible for nonvanishing $\alpha_1$ and $\alpha_2$ we obtain that
\begin{equation}
  \langle A_1 B_1 C A_3\rangle \langle A_2 B_2 C B_3\rangle - \langle A_1 B_1 C B_3\rangle \langle A_2 B_2 C A_3\rangle = 0.
\end{equation}
This is just the condition that $C$ belongs to the quadric containing the lines $A_1 \wedge B_1$, $A_2 \wedge B_2$ and $A_3 \wedge B_3$.

There are three such quadrics.  Recall that we have denoted by $Q_1$ the quadric generated by the lines $A_1 \wedge B_1$, $A_2 \wedge B_2$ and $A_3 \wedge B_3$, etc.  The three quadrics $Q_1$, $Q_2$ and $Q_3$ in $\mathbb{P}^3$ intersect in $2 \times 2 \times 2 = 8$ points.  We will discuss the surprisingly rich geometry of these intersection points in sec.~\ref{sec:intersection_three_quadrics}.

We can use the equations above to solve for $\alpha_2$ and $\alpha_3$ in terms of $\alpha_1$, etc.  We obtain
\begin{gather}
  \alpha_2 = -\alpha_1 \frac {\langle A_1 B_1 C A_3\rangle}{\langle A_2 B_2 C A_3\rangle}, \qquad
  \alpha_3 = -\alpha_1 \frac {\langle A_1 B_1 C A_2\rangle}{\langle A_3 B_3 C A_2\rangle}.
\end{gather}

We can also contract the first Landau loop equation~\eqref{eq:hexagon-ll1} with a line which is transversal to $A_1 B_1$, $A_2 B_2$, $A_3 B_4$ and $C D_2$.  Recall that four skew lines always have at least two (possibly coinciding) transversals.  One of the transversals is $C D_1$.  To obtain the other, we use that the quadric which contains the lines $A_1 B_1$, $A_2 B_2$ and $A_3 B_3$ has equation
\begin{equation}
    \langle A_1 B_1 A_3 X\rangle \langle A_2 B_2 B_3 X\rangle -
    (A_3 \leftrightarrow B_3) = 0.
\end{equation}
Even though this expression looks like it is not symmetric under permutations of the lines $A_i B_i$ for $i = 1, 2, 3$, it can be shown (via Pl\"ucker identities) that it is symmetric.  The line $C D_2$ intersects this quadric in two points, which are given by the solutions of the following quadratic equation in $\mu$:
\begin{equation}
    \langle A_1 B_1 A_3 (C + \mu D_2)\rangle \langle A_2 B_2 B_3 (C + \mu D_2)\rangle -
    (A_3 \leftrightarrow B_3) = 0.
\end{equation}
One of the solutions is $\mu = 0$ since $C$ belongs to the quadric (see above).  Then the other solution can be found straightforwardly
\begin{equation}
    \mu = -\frac{\langle A_1 B_1 A_3 D_2\rangle \langle A_2 B_2 B_3 C\rangle +
    \langle A_1 B_1 A_3 C\rangle \langle A_2 B_2 B_3 D_2\rangle - (A_3 \leftrightarrow B_3)}{\langle A_1 B_1 A_3 D_2\rangle \langle A_2 B_2 B_3 D_2\rangle - (A_3 \leftrightarrow B_3)}.
\end{equation}

If we denote this intersection point by $P_{1 2} = C + \mu D_2$, then the second transversal line can be taken to be $L_{1 2} = (P_{1 2} A_1 B_1) \cap (P_{1 2} A_2 B_2)$.  Then, using the transversality property we have
\begin{equation}
    \langle A_1 B_1 L_{1 2}\rangle =
    \langle A_2 B_2 L_{1 2}\rangle =
    \langle A_3 B_3 L_{1 2}\rangle =
    \langle C D_2 L_{1 2}\rangle = 0,
\end{equation}
so upon contraction with the Landau loop eq.~\eqref{eq:hexagon-ll1} with $L_{1 2}$ we find
\begin{equation}
-\alpha_3' \langle C D_3 L_{1 2}\rangle =
  \frac {\scriptstyle{\alpha_1 \langle I A_1 B_1\rangle +
   \alpha_2 \langle I A_2 B_2\rangle +
   \alpha_3 \langle I A_3 B_3\rangle +
   \alpha_2' \langle I C D_2\rangle -
   \alpha_3' \langle I C D_3\rangle}}{\langle I C D_1\rangle} \langle C D_1 L_{1 2}\rangle.
\end{equation}
We can similarly build a transversal $L_{1 3}$ to $A_1 B_1$, $A_2 B_2$, $A_3 B_3$ and $C D_3$ (which is different from $C D_1$).  Contracting the Landau loop eq.~\eqref{eq:hexagon-ll1} with $L_{1 3}$ we find
\begin{equation}
\alpha_2' \langle C D_2 L_{1 3}\rangle =
  \frac {\scriptstyle{\alpha_1 \langle I A_1 B_1\rangle +
   \alpha_2 \langle I A_2 B_2\rangle +
   \alpha_3 \langle I A_3 B_3\rangle +
   \alpha_2' \langle I C D_2\rangle -
   \alpha_3' \langle I C D_3\rangle}}{\langle I C D_1\rangle} \langle C D_1 L_{1 3}\rangle.
\end{equation}
Hence,
\begin{equation}
    \frac{\alpha_2'}{\alpha_3'} = -\frac{\langle C D_1 L_{1 3}\rangle \langle C D_3 L_{1 2}\rangle}{\langle C D_2 L_{1 3}\rangle \langle C D_1 L_{1 2}\rangle}.    
\end{equation}

Similarly, we find
\begin{equation}
    \frac{\alpha_3'}{\alpha_1'} = -\frac{\langle C D_2 L_{2 1}\rangle \langle C D_1 L_{2 3}\rangle}{\langle C D_3 L_{2 1}\rangle \langle C D_2 L_{2 3}\rangle}, \qquad
    \frac{\alpha_1'}{\alpha_2'} = -\frac{\langle C D_3 L_{3 2}\rangle \langle C D_2 L_{3 1}\rangle}{\langle C D_1 L_{3 2}\rangle \langle C D_3 L_{3 1}\rangle}.
\end{equation}
Taking the product of these ratios we find
\begin{multline}
  \label{eq:landau-cond-hexagon}
    \langle C D_1 L_{1 3}\rangle \langle C D_3 L_{1 2}\rangle
    \langle C D_2 L_{2 1}\rangle \langle C D_1 L_{2 3}\rangle
    \langle C D_3 L_{3 2}\rangle \langle C D_2 L_{3 1}\rangle + \\
    \langle C D_2 L_{1 3}\rangle \langle C D_1 L_{1 2}\rangle
    \langle C D_3 L_{2 1}\rangle \langle C D_2 L_{2 3}\rangle
    \langle C D_1 L_{3 2}\rangle \langle C D_3 L_{3 1}\rangle = 0.
\end{multline}

This equation expresses a codimension-one constraint on the external kinematics, written in terms of lines in twistor space parametrizing the loop momenta in the on-shell space and transversals to those lines. It is the leading Landau singularity for this diagram, defining points at which the integral can develop a branch cut.

\subsection{Intersection of three quadrics}
\label{sec:intersection_three_quadrics}

In practice, even when an algebraic root is needed to describe the on-shell space for an initial choice of kinematic variables, it is often possible to find other variables which rationalize the root. One fruitful way to find such variables is to consider the problem geometrically, and find variables that naturally parametrize the on-shell space. These variables will typically be complicated to express in terms of the external kinematics, as they are in some sense derived ``outside-in'', starting from the solution to the on-shell conditions and deriving a parametrization of the external kinematics from that. As such, one should not think of the existence of these parametrizations as evidence that a diagram lacks a given root: rather, they move the complexity of the root to the complexity of the external kinematics.

A familiar example of this kind is that of the four-mass box integral (with massless internal lines).  It is known that this integral can be computed in terms of two quantities $z$ and $\bar{z}$ which are the roots of a quadratic equation in the Mandelstam invariants of external lines (see ref.~\cite{Hodges:2010kq} for a discussion).  Unlike the case of the two-loop six-point planar MHV amplitude in $\mathcal{N}=4$ super Yang-Mills (see ref.~\cite{Goncharov:2010jf}), these square roots do not disappear even if we use momentum twistors to parametrize the external kinematics.  However, we can choose to parametrize the kinematics in a different way instead.  It turns out (see ref.~\cite{Hodges:2010kq}) that the quantities $z$ and $\bar{z}$ have the following geometric interpretation: in momentum twistor space the external kinematics of a four-mass box is represented by four skew lines in $\mathbb{P}^3$.  Such four lines generically have two transversals (lines which intersect all four of them).  On each of these transversals there are four intersection points.  Taking the cross-ratio of these four points on one of the transversals one obtains $z$ and on the other transversal $\bar{z}$.\footnote{Since the two transversals are not ordered the answer is invariant under $z \leftrightarrow \bar{z}$.  Similarly, since the cross-ratio depends on the ordering of the points the answer is independent on the simultaneous permutation of the points on both transversals.  This symmetry is not manifest on the final answer and happens via dilogarithm identities.}  Using this result, we can parametrize the on-shell space of the massive box integral as follows.  We first pick the two transversals, then we pick four pairs of points, with the first member of the pair on the first transversal and the second on the second transversal.  These four lines are the lines corresponding to the dual external points in twistor space and therefore we have specified the kinematics completely.  However, in this parametrization the variables $z$ and $\bar{z}$ arise naturally and do not involve any square roots.

We will now attempt a similar exercise for the three-loop diagram in figure~\ref{fig:three-pentagons}. Consider three quadrics in $\mathbb{P}^3$ which are in general position.  Then, by Bezout's theorem they intersect in eight points.  However, the converse does not hold.  That is, given eight points in general position (that is no four of them co-planar and therefore no three of them collinear) it is not possible to find three quadrics containing all of them.  Indeed, suppose it was possible to find such three quadrics $Q_1, Q_2, Q_3$.  Then we have a two-parameter (parametrized by points in $\mathbb{P}^2$) quadric $\lambda_1 Q_1 + \lambda_2 Q_2 + \lambda_3 Q_3$ which also passes through all eight points.  However, given seven points in general position (meaning that they impose independent conditions on the coefficients of a quadric containing them) they determine seven of the nine independent coefficients (up to scaling) of a generic quadric.  In other words, two coefficients remain undetermined.  But through eight points in generic position passes only a one-parameter family of quadrics.  The resolution of this paradox is that specifying seven of the points uniquely determines the eighth.

This is an instance of what is sometimes called \emph{Cramer's paradox}.  This paradox first arose in the study of the intersection of cubic curves in the projective plane.  The polynomial defining a cubic curve in $\mathbb{P}^2$ has ten terms.  Considered up to rescaling by a non-vanishing quantity, such a cubic is parametrized by a point in $\mathbb{P}^9$.  Imposing that a point is contained in the curve imposes one constraint.  If we impose that the curve contains nine points (and if these conditions are independent) then the cubic is completely determined.  Now consider the intersection of two such cubic curves.  According to Bezout's theorem, they intersect in $3 \times 3 = 9$ points.  However, we have shown that through nine points in generic position passes a single cubic!  The conclusion is that the nine intersection points are not independent.  Indeed, one can show that if two cubic curves pass through eight points, then they pass through a ninth point as well and this ninth point is uniquely determined by the other eight.

If we denote the eight intersection points of three quadrics by $1, \dotsc, 8$, we should be able to express the constraints on the eighth point in terms of four-brackets.  Indeed, Turnbull (see ref.~\cite{turnbull_1925}) has given such constraints.  A necessary condition is
\begin{equation}
    \label{eq:turnbull}
    \det \begin{pmatrix}
    \langle 1 2 5 6\rangle \langle 3 4 7 6\rangle & \langle 1 2 5 8\rangle \langle 3 4 7 8\rangle \\
    \langle 1 2 7 6\rangle \langle 3 4 5 6\rangle & \langle 1 2 7 6\rangle \langle 3 4 5 8\rangle
    \end{pmatrix} = 0.
\end{equation}
If points $1, \dotsc, 7$ are kept fixed, this is the equation of a quadric in the coordinates of the eighth point.  It can be easily checked that the determinant in eq.~\eqref{eq:turnbull} vanishes when point $8$ coincides with any of $1, \dotsc, 7$.  See also ref.~\cite[p.~153]{MR0178392}.

Eight points which are the intersection of three quadrics are sometimes called a \emph{Cayley octad}.  Their geometry has been studied more recently by Dolgachev \& Ortland (see ref.~\cite{MR1007155}).  Interestingly, their description of the space of Cayley octads is in terms of a concept called Gale duality.  The general form of this duality can be described very concretely as follows (see ref.~\cite{MR1774761} for an introduction).   Given $m$ points in $\mathbb{P}^n$ (with $m > n + 2$), we can represent their configuration by a $(n + 1) \times m$ matrix, with a left action of $\operatorname{PGL}(n + 1)$.  If this configuration is generic, then we can, by this left action, make the left $(n + 1) \times (n + 1)$ minor the identity.  Then the configuration is parameterized by a $(n + 1) \times (m - n - 1)$ matrix $A$.  This matrix $A$ fits in $(m - n - 1) \times m$ matrix $(\mathbf{1}_{m - n - 1}, A^T)$.  The matrix $A^T$, in turn, corresponds to a configuration of $m$ points in $\mathbb{P}^{m - n - 2}$.  It is advantageous to apply this duality when the dimension of the embedding projective space decreases.  This has also been described, in a related geometric context, in ref.~\cite[p.~299]{GONCHAROV1995197} (see also refs.~\cite{Goncharov:2010jf}, \cite{Golden:2013xva} for applications).  The constraints linking these eight points can be described as follows: the configuration of the eight intersection points of three quadrics in $\mathbb{P}^3$ is Gale self-dual.

The eighth point can be parametrized rationally in terms of the other seven, as described in ref.~\cite[prop.~7.1]{PLAUMANN2011712}.  Indeed, we can take the first seven points to have homogeneous coordinates
\begin{gather}
    (1 : 0 : 0 : 0), \qquad
    (0 : 1 : 0 : 0), \qquad
    (0 : 0 : 1 : 0), \qquad
    (0 : 0 : 0 : 1), \\
    (1 : 1 : 1 : 1), \qquad
    (\alpha_6 : \beta_6 : \gamma_6 : \delta_6), \qquad
    (\alpha_7 : \beta_7 : \gamma_7 : \delta_7).
\end{gather}
Then, the eighth point has coordinates $(\alpha_8 : \beta_8 : \gamma_8 : \delta_8)$ given by
\begin{gather}
    \label{eq:eightpt-rational}
    \alpha_8 = \frac{-\gamma_6 \beta_7 + \delta_6 \beta_7 + \beta_6 \gamma_7 - \delta_6 \gamma_7 - \beta_6 \delta_7 + \gamma_6 \delta_7}{-\beta_6 \delta_6 \beta_7 \gamma_7 + \gamma_6 \delta_6 \beta_7 \gamma_7 + \beta_6 \gamma_6 \beta_7 \delta_7 - \gamma_6 \delta_6 \beta_7 \delta_7 - \beta_6 \gamma_6 \gamma_7 \delta_7 + \beta_6 \delta_6 \gamma_7 \delta_7}, \\
    \beta_8 = \frac{-\gamma_6 \alpha_7 + \delta_6 \alpha_7 + \alpha_6 \gamma_7 - \delta_6 \gamma_7 - \alpha_6 \delta_7 + \gamma_6 \delta_7}{-\alpha_6 \delta_6 \alpha_7 \gamma_7 + \gamma_6 \delta_6 \alpha_7 \gamma_7 + \alpha_6 \gamma_6 \alpha_7 \delta_7 - \gamma_6 \delta_6 \alpha_7 \delta_7 - \alpha_6 \gamma_6 \gamma_7 \delta_7 + \alpha_6 \delta_6 \gamma_7 \delta_7}, \\
    \gamma_8 = \frac{-\beta_6 \alpha_7 + \delta_6 \alpha_7 + \alpha_6 \beta_7 - \delta_6 \beta_7 - \alpha_6 \delta_7 + \beta_6 \delta_7}{-\alpha_6 \delta_6 \alpha_7 \beta_7 + \beta_6 \delta_6 \alpha_7 \beta_7 + \alpha_6 \beta_6 \alpha_7 \delta_7 - \beta_6 \delta_6 \alpha_7 \delta_7 - \alpha_6 \beta_6 \beta_7 \delta_7 + \alpha_6 \delta_6 \beta_7 \delta_7}, \\
    \delta_8 = \frac{-\beta_6 \alpha_7 + \gamma_6 \alpha_7 + \alpha_6 \beta_7 - \gamma_6 \beta_7 - \alpha_6 \gamma_7 + \beta_6 \gamma_7}{-\alpha_6 \gamma_6 \alpha_7 \beta_7 + \beta_6 \gamma_6 \alpha_7 \beta_7 + \alpha_6 \beta_6 \alpha_7 \gamma_7 - \beta_6 \gamma_6 \alpha_7 \gamma_7 - \alpha_6 \beta_6 \beta_7 \gamma_7 + \alpha_6 \gamma_6 \beta_7 \gamma_7}.
\end{gather}

To check that these eight points belong to three independent quadrics we proceed as follows.  We start with a quadric
\begin{equation}
    q(x) = \sum_{i, j = 0}^3 q_{i j} x_i x_j,
\end{equation}
with $q_{i j} = q_{j i}$.  Imposing that the first five points belong to this quadric implies that $q_{0 0} = q_{1 1} = q_{2 2} = q_{3 3} = 0$ and $q_{0 1} + q_{0 2} + q_{0 3} + q_{1 2} + q_{1 3} + q_{2 3} = 0$.  The conditions that the last four points belong to the quadric are linear constraints on the $(q_{0 1}, q_{0 2}, q_{0 3}, q_{1 2}, q_{1 3}, q_{2 3})$ which can be put in the form of a $4 \times 6$ matrix
\begin{equation}
    \begin{pmatrix}
    1 & 1 & 1 & 1 & 1 & 1 \\
    \alpha_6 \beta_6 & \alpha_6 \gamma_6 & \alpha_6 \delta_6 & \beta_6 \gamma_6 & \beta_6 \delta_6 & \gamma_6 \delta_6 \\
    \alpha_7 \beta_7 & \alpha_7 \gamma_7 & \alpha_7 \delta_7 & \beta_7 \gamma_7 & \beta_7 \delta_7 & \gamma_6 \delta_7 \\
    \alpha_8 \beta_8 & \alpha_8 \gamma_8 & \alpha_8 \delta_8 & \beta_8 \gamma_8 & \beta_8 \delta_8 & \gamma_8 \delta_8
    \end{pmatrix}
\end{equation}
Then we can check that with the values in eq.~\eqref{eq:eightpt-rational} all the $4 \times 4$ minors of this matrix vanish.  This means that the eighth point is automatically contained in any quadric that contains the other seven.

Hence, it is possible to parametrize the on-shell kinematics rationally, in a sense from the ``inside out''.  That is, we first pick seven points as above and an eighth point whose coordinates are given by eq.~\eqref{eq:eightpt-rational}.  Then we find three quadrics such that these eight points are their intersection.  Then, in each quadric we pick three lines, member of the same ruling and not containing any of the eight intersection points.  Finally, we take these nine lines as the external lines $A_i B_i$ for $i = 1, \dotsc, n$.  The ability to rationally parametrize the on-shell spaces is connected with the possibility of computing the integral in terms of polylogarithms (see ref.~\cite{Duhr:2020gdd}).  It would be interesting to understand better the relation between this parametrization and those in terms of cluster variables (see App.~\ref{app:explicitparam}).

Finally, let us discuss possible degenerations of these configurations of eight points (which are the on-shell space of the diagram in fig.~\ref{fig:three-pentagons}).  This has been studied in Dolgachev \& Ortland (see ref.~\cite[p.~176]{MR1007155}).  They found three classes of possible degenerations:
\begin{enumerate}
    \item two of the eight points become coincident.
    \item four points become coplanar (when this happens the other four points also become coplanar as follows from Gale duality).
    \item the eight points lie on a twisted cubic.  This is a codimension two condition.
\end{enumerate}
The physical interpretation of the second and third possibilities is mysterious, with third in particular able to correspond to a codimension two Landau singularity. The first, on the other hand, is exactly the Landau locus of eq.~\eqref{eq:landau-cond-hexagon}, as we confirm below.

When two points $P_7$ and $P_8$ of a Cayley octad become coincident, we have $\langle m, n, 7, 8\rangle = 0$. Therefore by Gale duality, the four-bracket of the complementary points $\langle i j k l\rangle = 0$ where the points $\{P_m, P_n, P_7, P_8\} \cup \{P_i, P_j, P_k, P_l\} = \{P_1, \dotsc, P_8\}$.  Since this is true for all groups of four points in $\{P_1, \dotsc, P_6\}$, this means that the points $P_1, \dotsc, P_6$ belong to the same plane.  In fact, they also belong to the same conic, since if they didn't then they would not belong to a quadric (since the intersection of a quadric with a plane is a conic).  Six points do not generically belong to a conic, a conic being determined by five points (in general position in the same plane).

Since the points $P_1, \dotsc, P_6$ belong to the same plane we can pick them without loss of generality to have zero fourth component.  We take
\begin{gather}
  P_1 = (1 : 0 : 0 : 0), \qquad
  P_2 = (0 : 1 : 0 : 0), \\
  P_3 = (0 : 0 : 1 : 0), \qquad
  P_4 = (1 : 1 : 1 : 0), \\
  P_5 = (a_1 : b_1 : c_1 : 0), \qquad
  P_6 = (a_2 : b_2 : c_2 : 0).
\end{gather}

If the equation of the conic is
\begin{equation}
  c_{0 0} x_0^2 + 2 c_{0 1} x_0 x_1 + 2 c_{0 2} x_0 x_2 +
  c_{1 1} x_1^2 + 2 c_{1 2} x_1 x_2 + c_{2 2} x_2^2 = 0\,,
\end{equation}
then imposing that the points $P_1, \dotsc, P_6$ belong to it implies that $c_{0 0} = c_{1 1} = c_{2 2} = 0$, $a_{1 2} = -a_{0 1} - a_{0 2}$.  The conditions for the last two points to belong to this conic are
\begin{gather}
  \frac {c_{0 1}}{c_{0 2}} =
  -\frac {c_1 (a_1 - b_1)}{b_1 (a_1 - c_1)} =
  -\frac {c_2 (a_2 - b_2)}{b_2 (a_2 - c_2)}.
\end{gather}
This first equality can be used to determine the coefficients $c_{i j}$ (up to a multiplicative factor) while the last equality can be used to determine $c_2$.

Without loss of generality we take the remaining (coinciding) two points to be
\begin{equation}
  P_7 = P_8 = (0 : 0 : 0 : 1).
\end{equation}
Note that this parametrization can not be obtained from the one in eq.~\eqref{eq:eightpt-rational} for finite values of $\alpha, \beta, \gamma, \delta$. This not unexpected when one considers that a configuration where two points coincide lies on the boundary of the space of those parameters.

A quadric $Q \subset \mathbb{P}^3$ is defined by a general equation
\begin{equation}
  \sum_{0 \leq i \leq j \leq 3} q_{i j} x_i x_j.
\end{equation}
Imposing the conditions that the points $P_1, \dotsc, P_6 \in Q$ we obtain that $q_{i j} = c_{i j}$ if $i, j \in \{0, 1, 2\}$.  If we impose the condition that $P_7, P_8 \in Q$ we obtain $q_{3 3} = 0$.  The coefficients $q_{0 3}$, $q_{1 3}$ and $q_{2 3}$ remain undetermined.

We therefore obtain three natural quadrics which contain all eight points
\begin{gather}
  Q_1(x) = \sum_{0 \leq i \leq j \leq 2} c_{i j} x_i x_j + q_{0 3} x_0 x_3, \\
  Q_2(x) = \sum_{0 \leq i \leq j \leq 2} c_{i j} x_i x_j + q_{1 3} x_1 x_3, \\
  Q_3(x) = \sum_{0 \leq i \leq j \leq 2} c_{i j} x_i x_j + q_{2 3} x_2 x_3.
\end{gather}
It can be checked that these quadrics are generically smooth, by computing the determinants of the associated symmetric matrices.  We can set the coefficients $q_{0 3}$, $q_{1 3}$ and $q_{2 3}$ to one without loss of generality.

Next, we need to choose three skew lines in each of these quadrics.  To find a line in a smooth quadric $Q = \sum_{i, j = 0}^3 q_{i j} x_i x_j$ we first pick a point on $Q$ with coordinates $(\xi_0 : \xi_1 : \xi_2 : \xi_3)$.  We can find this point by, for example, picking $\xi_0, \xi_1, \xi_2$ and solving for $\xi_3$, which is a linear equation.

The (projective) tangent plane to $Q$ at $\xi$ is given by the equation $\sum_{i j} q_{i j} x_i \xi_j = 0$.  The intersection of this tangent plane with the quadric consists of two lines.  Doing this three times we obtain six lines and we can choose three which are skew.  Then we repeat this construction for $Q_1$, $Q_2$ and $Q_3$ to obtain the lines $A_i B_i$ for $i = 1, \dotsc, 9$.

In practice, we can build these lines using the Segre map.  For example, we have
\begin{equation}
  Q_1(x) = x_0 (c_{0 1} x_1 + c_{0 2} x_2 + x_3) + c_{1 2} x_1 x_2.
\end{equation}
Then, the condition imposed by this quadric can be solved by
\begin{gather}
  x_0 = \alpha_0 \beta_0, \qquad
  c_{0 1} x_1 + c_{0 2} x_2 + x_3 = \alpha_1 \beta_1, \\
  -c_{1 2} x_1 = \alpha_0 \beta_1, \qquad
  x_2 = \alpha_1 \beta_0.
\end{gather}
Then for every point $C$ on the surface defined by $Q_1$ we can find coordinates $(\bar{\alpha}_0 : \bar{\alpha}_1)$ and $(\bar{\beta}_0 : \bar{\beta}_1)$.  The lines through $C$ are given by $\alpha$ being constant or $\beta$ being constant. The points where $Q_1$ vanishes can then be parametrized by
\begin{gather}
  x_0 = \alpha_0 \beta_0, \qquad
  x_1 = -\frac {\alpha_0 \beta_1}{c_{1 2}}, \\
  x_2 = \alpha_1 \beta_0, \qquad
  x_3 = \alpha_1 \beta_1 + \frac {c_{0 1}}{c_{1 2}} \alpha_0 \beta_1 - c_{0 2} \alpha_1 \beta_0.
\end{gather}

For $Q_2$ we have
\begin{equation}
  Q_2(x) = x_1 (c_{0 1} x_0 + c_{1 2} x_2 + x_3) + c_{0 2} x_0 x_2.
\end{equation}
We can then solve this constraint by
\begin{gather}
  x_0 = \alpha_0 \beta_0, \qquad
  x_1 = \alpha_0 \beta_1, \\
  x_2 = -\frac {\alpha_1 \beta_1}{c_{0 2}}, \qquad
  x_3 = \alpha_1 \beta_0 - c_{0 1} \alpha_0 \beta_0 + c_{1 2} \frac {\alpha_1 \beta_1}{c_{0 2}}.
\end{gather}

Finally, for $Q_3$ we have
\begin{equation}
  Q_3(x) = x_2 (c_{0 2} x_0 + c_{1 2} x_1 + x_3) + c_{0 1} x_0 x_1,
\end{equation}
which can be solved by
\begin{gather}
  x_0 = \alpha_0 \beta_0, \qquad
  x_1 = -\frac {\alpha_1 \beta_1}{c_{0 1}}, \\
  x_2 = \alpha_0 \beta_1, \qquad
  x_3 = \alpha_1 \beta_0 - c_{0 2} \alpha_0 \beta_0 + c_{1 2} \frac {\alpha_1 \beta_1}{c_{0 1}}.
\end{gather}

Next, we pick $C$ to be one of the points $P_1, \dotsc, P_8$ and find the points $D_1, D_2, D_3$.  The point $C$ can be either one of the two coinciding points $P_7 = P_8$ or one of the six conconic points $P_1, \dotsc, P_6$. We consider one example of each kind.

If we pick $C = P_7 = (0 : 0 : 0 : 1)$, then we can take
\begin{gather}
  D_1^\pm = \begin{cases}
    (0 : 0 : \beta_0 : \beta_1 - c_{0 2} \beta_0), \\
    (0 : -\alpha_0 : 0 : c_{1 2} \alpha_1 + c_{0 1} \alpha_0),
  \end{cases} \\
  D_2^\pm = \begin{cases}
    (0 : 0 : -\beta_1 : c_{0 2} \beta_0 + c_{1 2} \beta_1), \\
    (\alpha_0 : 0 : 0 : \alpha_1 - c_{0 1} \alpha_0),
  \end{cases} \\
  D_3^\pm = \begin{cases}
    (0 : -\beta_1 : 0 : c_{0 1} \beta_0 + c_{1 2} \beta_1), \\
    (\alpha_0 : 0 : 0 : \alpha_1 - c_{0 2} \alpha_0).
  \end{cases}
\end{gather}

We can in principle analyze all of these possibilities, but to illustrate let us just pick $D_1^+$, $D_2^+$ and $D_3^+$.  Then, we can show that eq.~\eqref{eq:landau-cond-hexagon} holds as follows.  We first notice that $\langle C D_2 L_{21}\rangle = \langle C D_1 L_{21}\rangle$ (up to a multiplicative factor) since $C, D_1, D_2$ are collinear.  Then, $\langle C D_1 L_{21}\rangle = 0$ since $L_{21}$ is a line transversal to $C D_1$ in $Q_2$.  Similarly, we have $\langle C D_1 L_{12}\rangle = \langle C D_2 L_{12}\rangle = 0$.

We can instead pick $C = P_1 = (1 : 0 : 0 : 0)$.  Then we find
\begin{gather}
    D_1^\pm = \begin{cases}
        (\alpha_0 : 0 : \alpha_1 : -c_{0 2} \alpha_1), \\
        (\beta_0 : -\frac{\beta_1}{c_{1 2}} : 0 : \frac{c_{0 1} \beta_1}{c_{1 2}}),
    \end{cases} \\
    D_2^\pm = \begin{cases}
        (\alpha_0 : 0 : 0 : \alpha_1 - c_{0 1} \alpha_0), \\
        (\beta_0 : \beta_1 : 0 : -c_{0 2} \beta_0),
    \end{cases} \\
    D_3^\pm = \begin{cases}
        (\alpha_0 : 0 : 0 : \alpha_1 - c_{0 2} \alpha_0), \\
        (\beta_0 : 0 : \beta_1 : -c_{0 2} \beta_0).
    \end{cases}
\end{gather}

As before, let us pick the case $D_1^+$, $D_2^+$ and $D_3^+$ to analyze in detail.  Here we have that $D_2^+ = D_3^+$.  Then, we have $\langle C D_2 L_{1 3}\rangle = \langle C D_3 L_{1 3}\rangle = 0$ and $\langle C D_3 L_{ 1 2}\rangle = \langle C D_2 L_{1 2}\rangle = 0$.  Plugging this in eq.~\eqref{eq:landau-cond-hexagon} we find that it is satisfied.

Checking for the other cases, we find that in general, making two of the points $P_1\ldots P_8$ coincident lands us on the Landau locus, as expected.

\newpage
\section{\texorpdfstring{\emph{Exempli Gratia}}{Exempli Gratia}: A Sextic in Two Dimensions}
\label{sec:sextic}

Let us now consider the following integral for two-dimensional kinematics: 
\vspace{6pt}\eq{\fwbox{0pt}{\hspace{-20pt}\begin{tikzpicture}[scale=1.4*\figScale,baseline=-3.05]\useasboundingbox ($(-2,-1.75)$) rectangle ($(2,1.75)$);\draw[int,line width=0.1,red,draw=\boundingDraw] ($(-2.,-1.5)$) rectangle ($(2.,1.5)$);
\coordinate (v0) at (0,0);
\coordinate (ell1) at ($(v0)+(150:0.55)$);
\coordinate (ell2) at ($(v0)+(30:0.55)$);
\coordinate (ell3) at ($(v0)+(-90:0.55)$);
\coordinate (x1) at ($(v0)+(-90:1.45)$);
\coordinate (x2) at ($(v0)+(150:1.45)$);
\coordinate (x3) at ($(v0)+(30:1.45)$);
\coordinate (a1) at ($(v0)+(-150:1.45)$);
\coordinate (a2) at ($(v0)+(90:1.45)$);
\coordinate (a3) at ($(v0)+(-30:1.45)$);
\coordinate (p1) at ($(v0)+(-150:1.85)$);
\coordinate (p2) at ($(v0)+(90:1.85)$);
\coordinate (p3) at ($(v0)+(-30:1.85)$);
\draw[int] (v0)--(a1);\draw[int] (v0)--(a2);\draw[int] (v0)--(a3);
\draw[int](a1)--(a2);\draw[int](a2)--(a3);\draw[int](a3)--(a1);
\leg{(a1)}{-150}{$p_1$}\leg{(a2)}{90}{$p_2$}\leg{(a3)}{-30}{$p_3$}
\draw[directedEdge] (a1)--(v0);\draw[directedEdge] (a2)--(v0);\draw[directedEdge] (a3)--(v0);\draw[directedEdge] (a2)--(a1);\draw[directedEdge] (a3)--(a2);\draw[directedEdge] (a1)--(a3);\draw[directedEdge] (p1)--(a1);\draw[directedEdge] (p2)--(a2);\draw[directedEdge] (p3)--(a3);
\node[anchor=north] at ($(a1)!.5!(a3)$) {$q_2$};\node[anchor=south east] at ($(a1)!.5!(a2)$) {$q_3$};\node[anchor=south west] at ($(a3)!.5!(a2)$) {$q_1$};
\node[anchor=east,inner sep=1pt] at ($(a2)!.7!(v0)$) {$\ell_2$};\node[anchor=north west,inner sep=1pt] at ($(a1)!.7!(v0)$) {$\ell_1$};\node[anchor=south west,inner sep=1pt] at ($(a3)!.7!(v0)$) {$\ell_3$};
\end{tikzpicture}%
\bigger{\Leftrightarrow}\begin{tikzpicture}[scale=1.4*\figScale,baseline=-3.05]\useasboundingbox ($(-2,-1.75)$) rectangle ($(2,1.75)$);\draw[int,line width=0.1,red,draw=\boundingDraw] ($(-2.,-1.5)$) rectangle ($(2.,1.5)$);
\coordinate (v0) at (0,0);
\coordinate (ell1) at ($(v0)+(150:0.55)$);
\coordinate (ell2) at ($(v0)+(30:0.55)$);
\coordinate (ell3) at ($(v0)+(-90:0.55)$);
\coordinate (x2) at ($(v0)+(-90:1.45)$);
\coordinate (x3) at ($(v0)+(150:1.45)$);
\coordinate (x1) at ($(v0)+(30:1.45)$);
\coordinate (a1) at ($(v0)+(-150:1.45)$);
\coordinate (a2) at ($(v0)+(90:1.45)$);
\coordinate (a3) at ($(v0)+(-30:1.45)$);
\coordinate (p1) at ($(v0)+(-150:1.85)$);
\coordinate (p2) at ($(v0)+(90:1.85)$);
\coordinate (p3) at ($(v0)+(-30:1.85)$);
\dimLines\draw[int] (v0)--(a1);\draw[int] (v0)--(a2);\draw[int] (v0)--(a3);
\draw[int](a1)--(a2);\draw[int](a2)--(a3);\draw[int](a3)--(a1);
\leg{(a1)}{-150}{}\leg{(a2)}{90}{}\leg{(a3)}{-30}{}
\draw[directedEdge] (a1)--(v0);\draw[directedEdge] (a2)--(v0);\draw[directedEdge] (a3)--(v0);\draw[directedEdge] (a2)--(a1);\draw[directedEdge] (a3)--(a2);\draw[directedEdge] (a1)--(a3);\draw[directedEdge] (p1)--(a1);\draw[directedEdge] (p2)--(a2);\draw[directedEdge] (p3)--(a3);
\restoreDark
\draw[int](ell2)--(ell1);\draw[int](ell3)--(ell2);\draw[int](ell1)--(ell3);\draw[int] (x3)--(ell1);\draw[int] (x1)--(ell2);\draw[int] (x2)--(ell3);\draw[int](x2)--(x3);\draw[int](x3)--(x1);\draw[int](x1)--(x2);
\draw[directedEdge](ell1)--(ell2);\draw[directedEdge](ell2)--(ell3);\draw[directedEdge](ell3)--(ell1);\draw[directedEdge] (x3)--(ell1);\draw[directedEdge] (x1)--(ell2);\draw[directedEdge] (x2)--(ell3);\draw[directedEdge](x2)--(x3);\draw[directedEdge](x3)--(x1);\draw[directedEdge](x1)--(x2);
\node[ddot]at(ell1){};\node[ddot]at(ell2){};\node[ddot]at(ell3){};\node[ddot]at(x2){};\node[ddot]at(x3){};\node[ddot]at(x1){};\node[anchor=north]at(x2){$x_2$};\node[anchor=south east,inner sep=1pt]at(x3){$x_3$};\node[anchor=south west,inner sep=1pt]at(x1){$x_1$};\node[anchor=south,inner sep=3pt]at(ell2){$y_1$};\node[anchor=west,inner sep=3pt]at(ell3){$y_2$};\node[anchor=north east,inner sep=1pt]at(ell1){$y_3$};
\end{tikzpicture}}\vspace{-4pt}}
This integral involves six propagators and three, two-dimensional loop momenta; as such, its leading singularities would correspond to residues around which all propagators become on-shell. How many solutions to the cut equations do we get?

Let us consider the case where all external momenta are massive, with $p_i^2\equivL M_i^2$, and for the `all-mass' case of internal propagators: where the $q_i$ propagators have poles at $q_i^2\equivL m_i^2$ and the $\ell_i$ propagators have poles at $\ell_i^2\equivL \mu_i^2$. 

To determine the leading singularities of this Feynman integral, it is useful to introduce dual coordinates $p_1\equivL x_3{-}x_2$, etc.\ and $\ell_1\equivL y_3{-}y_2$, etc.;\footnote{We apologize that this labeling, motivated by the case of three particles, does not follow the usual conventions for dual coordinates.} in terms of these, we have $q_i= y_i{-}x_i$. Without loss of generality, we may translate the $x_i$'s and express them in light-cone coordinates so that 
\eq{x_1\equivR\big(0,0\big)\,,\qquad x_2=\big(M_3/M_2\,,M_3\,M_2\big)\,,\qquad x_3\equivL\big(x_3^+\,, M_2^2/x_3^+\big)\,,}
where 
\eq{x_3^+ + \frac 1 {x_3^+}\equivR \frac {M_2^2 + M_3^2 - M_1^2}{M_2 M_3}.}
Next, we may parameterize the solutions to the on-shell conditions for $q_1$, $\ell_2$ and $\ell_3$ by expressing
\eq{y_1\equivR\big(y_1^+,m_1^2/y_1^+\big)\,,\qquad y_2\equivL\big(\psi_2^+\,,\mu_3^2/\psi_2^+\big){+}\,y_1\,,\qquad y_3\equivL\big(\psi_3^+\,, \mu_2^2/\psi_3^+\big){+}\,y_1\,.}
In terms of these variables, the on-shell condition for $\ell_1$ ($\ell_1^2=\mu_1^2$) now reads
\eq{\ell_1^2=\big(y_3{-}y_2\big)^2=(\psi_3^+{-}\psi_2^+\big)\left(\frac{\mu_3^2}{\psi_2^+}{-}\frac{\mu_2^2}{\psi_2^+}\right)\,,}
which can be re-expressed as the condition that 
\eq{\frac {\mu_3 \psi_3^+}{\mu_2 \psi_2^+} + \Bigl(\frac {\mu_3 \psi_3^+}{\mu_2 \psi_2^+}\Big)^{-1} = \frac {\mu_2^2 + \mu_3^2 - \mu_1^2}{\mu_2 \mu_3}\,.}
This leaves only the final two on-shell conditions,
\eq{\mbox{$q_2^2=m_2^2=(x_2{-}y_2)^2$}\qquad\text{and} \qquad\mbox{$q_3^2=m_3^2=(x_3{-}y_3)^2$}\label{final_cut_conditions}}
to solve.

Let us define
\eq{a\equivR x_3^+\,,\quad b\equivR\frac{\mu_3\psi_3^+}{\mu_2\psi_2^+}\,,\qquad\text{and let}\qquad x\equivR\frac{\mu_2}{M_2\psi_2^+}\,,\quad y\equivR\frac{M_2y_1^+}{m_1}\,.}
We may think of $a,b$ as being fixed by external kinematics since
\eq{a+a^{-1}=\frac{M_2^2{+}M_3^2{-}M_1^2}{M_2\,M_3}\quad\text{and}\quad b+b^{-1}=\frac{\mu_2^2{+}\mu_3^2{-}\mu_1^2}{\mu_2\,\mu_1}\,,}
leaving us only with $x$ and $y$ to determine using the final equations (\ref{final_cut_conditions}). 

In terms of $x,y$, the final cut-conditions (\ref{final_cut_conditions}) are given by
\begin{align}
m_2^2=\,&M_3^2{+}\mu_3^2{+}m_1^2{+}m_1 \mu_3 \left(x y{+}\frac 1 {x y}\right){-}M_3 \mu_3 \left(x{+}\frac 1 x\right){-}m_1 M_3 \left(y{+}\frac 1 y\right)\,,\label{final_cut_eqns}\\
m_3^2=\,& M_2^2{+}\mu_2^2{+}m_1^2{+}m_1 \mu_2 \left(\frac {x y} b{+}\frac b {x y}\right){-}M_2 \mu_2 \left(\frac b {a x} + \frac {a x} b\right){-}m_1 M_2 \left(\frac y a + \frac a y\right)\,.\nonumber
\end{align}

To count the number of solutions to these cut equations, we may proceed as follows. Notice that the equations (\ref{final_cut_eqns}) take the following form in $y$:
\eq{\begin{split}
0=\,& A_1(x) + B_1(x) y + C_1(x) y^{-1}\\
0=\,&A_2(x) + B_2(x) y + C_2(x) y^{-1}\,
\end{split}}
where $A_i,B_i,C_i$ are some rational functions of $x$. Such a system which is linear in $y$ and $y^{-1}$ may be solved with the compatibility condition
\eq{   \begin{vmatrix}
    A_1 & C_1 \\ A_2 & C_2
    \end{vmatrix}
    \begin{vmatrix}
    B_1 & A_1 \\ B_2 & A_2
    \end{vmatrix} =
    \begin{vmatrix}
    B_1 & C_1 \\ B_2 & C_2
    \end{vmatrix}^2
}
which takes the general form (in terms of $x$) of
\eq{\label{eq:degree-six-equation}
    c_3 x^3 + c_2 x^2 + c_1 x + c_0 + c_{-1} x^{-1} + c_{-2} x^{-2} + c_{-3} x^{-3} = 0\,}
where the coefficients $c_{-3},\ldots,c_3$ are rational functions of the external kinematic variables $a,b,\mu_i,m_i,M_i$. It is clear that there are generally six solutions in $x$ to (\ref{eq:degree-six-equation}), and for each of them we may uniquely specify the corresponding point $y$. Therefore, there are six roots in $(x,y)$ to the final cut conditions. 

Explicit expressions for the coefficients $c_i$ in (\ref{eq:degree-six-equation}) can be given, but turn out to be rather cumbersome; for example, 
\eq{c_3= -\frac{M_2^2 M_3 a^2 m_1^2 \mu_2 \mu_3^2}{b} + \frac{M_2^2 M_3 a^2 m_1^2 \mu_2^2 \mu_3}{b^2} + \frac{M_2 M_3^2 a m_1^2 \mu_2 \mu_3^2}{b} - \frac{M_2 M_3^2 a m_1^2 \mu_2^2 \mu_3}{b^2}\,.\nonumber}
In all, the coefficients $c_{-3},\ldots,c_3$ involve sums of $4,29,86,124,86,29,4$ monomials, respectively. We have checked (using generic numerical values for the masses) that the roots of the sextic (\ref{eq:degree-six-equation}) are not expressible in terms of radicals. 

The Landau equations imply that singularities of the diagram occur when some of these six roots coincide, and this happens when the discriminant of the degree-six equation in eq.~\eqref{eq:degree-six-equation} vanishes.  The discriminant of a polynomial of degree $n$ is homogeneous in the coefficients with degree $2 n - 2$, so for a degree-six polynomial we obtain a discriminant of degree ten.  For our case the discriminant is a degree-ten polynomial in $c_{-3}, \dotsc, c_{3}$ with $246$ terms.\footnote{While it would be interesting to compare our derivation to one using the methods described in Ref.~\cite{Mizera:2021icv}, the sheer size of this discriminant suggests this would not be feasible. For this example (and perhaps more generally for leading Landau loci involving a set of isolated points), our approach is thus substantially more efficient.}

Let us now discuss the Landau loop equations. The Landau equations read
\eq{\begin{split}
  0=\,&\alpha_2 q_2 + \beta_3 \ell_3 - \beta_1 \ell_1\, \\
  0=\,&\alpha_1 q_1 + \beta_2 \ell_2 - \beta_3 \ell_3\,\\
  0=\,&\alpha_3 q_3 + \beta_1 \ell_1 - \beta_2 \ell_2\,.
\end{split}}
Okun \& Rudik (see Ref.~\cite{OKUN1960261}) take the cross-products\footnote{In light-cone coordinates, the cross-product reads $[p q]\equivR \frac 1 2 (p^+ q^-{-}\,p^- q^+)$. And because \mbox{$p \cdot q = \frac 1 2 (p^+ q^-{+}\,p^- q^+)$}, we have
$(p \cdot q)^2{-}[p q]^2 = p^2 q^2$.} with $q_i$, which removes the terms dependent on $\alpha$.  Doing this we obtain
\eq{\begin{split}
  \beta_3 [\ell_3 q_2] = \beta_1 [\ell_1 q_2]\,, \\
  \beta_2 [\ell_2 q_1] = \beta_3 [\ell_3 q_1]\,, \\
  \beta_1 [\ell_1 q_3] = \beta_2 [\ell_2 q_3]\,.
\end{split}}
Taking the product and simplifying we obtain
\eq{
  [\ell_1 q_3] [\ell_2 q_1] [\ell_3 q_2] = [\ell_1 q_2] [\ell_2 q_3] [\ell_3 q_1]\,.
}
This has the geometrical interpretation that the lines $x_i\, y_i$ intersect in a single point.  Let us show that this is indeed the case.

The lines through the points $(x_i, y_i)$ intersect in a single point if there exist $t_1$, $t_2$ and $t_3$ such that
\begin{equation}
  y_1 + t_1 (x_1 - y_1) = y_2 + t_2 (x_2 - y_2) = y_3 + t_3 (x_3 - y_3).
\end{equation}
Using the fact that $q_i = y_i{-}x_i$, $\ell_1 = y_3{-}y_2$, $\ell_2 = y_1{-}y_3$ and $\ell_3 = y_2{-}y_1$ we find
\eq{\begin{split}
  y_1 - y_2 + t_1 (x_1 - y_1) - t_2 (x_2 - y_2) = 0\,, \\
  y_2 - y_3 + t_2 (x_2 - y_2) - t_3 (x_3 - y_3) = 0\,, \\
  y_3 - y_1 + t_3 (x_3 - y_3) - t_1 (x_1 - y_1) = 0\,,
\end{split}}
which implies
\eq{\begin{split}
  -\ell_3 - t_1 q_1 + t_2 q_2 = 0\,, \\
  -\ell_1 - t_2 q_2 + t_3 q_3 = 0\,, \\
  -\ell_2 - t_3 q_3 + t_1 q_1 = 0\,.
\end{split}}

Taking the cross-product with the $l_i$, separating the remaining terms in the left-hand side and right-hand side and taking their product we obtain
\begin{equation}
  [q_1 \ell_3] [q_2 \ell_1] [q_3 \ell_2] = [q_2 \ell_3] [q_3 \ell_1] [q_1 \ell_2].
\end{equation}
This equation and its derivation look similar to eq.~\eqref{eq:landau-cond-hexagon} arising in the $\mathcal{N}\!=\!4$ super Yang-Mills theory.

We have presented this example as two-dimensional (massive) cousin of our four-dimensional (massless) examples; indeed, the pentagon loops in four dimensions are similar to triangle loops in two dimensions.  Due to the fact that it arises already at three points, it is possible to analyze this example in full detail, with fewer algebraic complexities.  The geometry of the problem is also much easier to understand thanks to the two-dimensional nature of the problem.

This example also serves to show that higher-order polynomials arise quite generically once the equations enforcing the on-shell condition are sufficiently coupled. As such, we expect such examples to become more common in the literature as the community investigates more complicated diagrams.

\section{Conclusions and Discussion}
As we have shown, leading singularities in Feynman diagrams can indeed involve cubic or higher roots, contrary to the naive expectation one might have from Landau's analysis. However, Landau's analysis is still correct: as we have argued, these cubic or higher roots do not lead to forbidden behavior on co-dimension one singularities, because any such singularity will only make two roots coincide. To make more roots coincide requires a singularity of higher co-dimension. We have illustrated this behavior in three concrete examples, a cubic root in a diagram in planar $\mathcal{N}\!=\!4$ super Yang-Mills, roots of a pair of octic polynomials for a more general diagram in the same theory, and roots of a sextic polynomial for massive scalars in two dimensions.

The existence of cubic root and more unusual singularities in higher co-dimension limits may have several implications. It would be interesting to see how their existence interacts with approaches that attempt to derive, not just leading, but iterative singularities of general Feynman diagrams~\cite{Bourjaily:2020wvq,Hannesdottir:2021kpd,Gong:2022erh}, in which one would expect to need to take into account these singularities of higher co-dimension in some way. They should also be relevant for series expansions of Feynman diagrams. In particular, there should be special kinematic limits in which these co-dimension two limits are uncovered as leading behavior. It would be interesting to see if there is a kinematic limit of physical interest in which these singularities are especially relevant.
\\

\newpage\noindent\textbf{Acknowledgments}\\[-0pt]
\noindent We thank Andrew McLeod, Chi Zhang, and Zhenjie Li for helpful discussions. For our access to the computer package \texttt{Magma}, we appreciate the generosity of the Simons Foundation. 
This work was supported by the Danish National Research Foundation (DNRF91), the research grant 00015369 from Villum Fonden, a Starting Grant (No. 757978) from the European Research Council, and by a grant from the US Department of Energy \mbox{(No.\ DE-SC00019066)}.

\newpage
\appendix

\section{Extended Discussion of the Leading Singularity}
\label{app:cussextdis}

Due to our choice of a blue vertex in the center of the on-shell diagram in eq.~(\ref{eleven_cut_diagram}), we have $\ell_1 = A \wedge B$, $\ell_2 = B \wedge C$ and $\ell_3 = A \wedge C$.  This solves the on-shell conditions for the internal lines.  We are left with the following on-shell conditions
\begin{gather}
    \langle A B 1 2\rangle = \langle A B 3 4\rangle = 0, \\
    \langle B C 4 5\rangle = \langle B C 6 7\rangle = \langle B C 8 9\rangle = 0, \\
    \langle A C 1 2\rangle = \langle A C 8 9\rangle = \langle A C, 10, 11\rangle = 0.
\end{gather}
Consider the following two equalities $\langle B C 8 9\rangle = \langle A C 8 9\rangle = 0$.  If $C$ and $z_8$ and $z_9$ are not collinear, then we have that $A$ and $B$ belong to the plane spanned by $C$ and $z_8$ and $z_9$.  Equivalently, we can say that $z_8$ and $z_9$ belong to the plane spanned by $A$, $B$ and $C$.

Similarly, from $\langle A C 1 2\rangle = \langle A B 1 2\rangle = 0$ if $A$ and $z_1$ and $z_2$ are not collinear, then we have that $B$ and $C$ belong to the plane spanned by $A$ and $z_1$ and $z_2$.  Equivalently, we have that $z_1, z_2$ belong to the plane spanned by $A$, $B$ and $C$.  This means that the lines $z_8 \wedge z_9$ and $z_1 \wedge z_2$ belong to the same plane $A \wedge B \wedge C$, which means that they must intersect.  But this is only possible if $\langle 1 2 8 9\rangle = 0$.

So instead, let us take $A$ on the line $z_1 \wedge z_2$ and $C$ on the line $z_8 \wedge z_9$.  This means that the two three-point vertices neighboring the middle three-point vertex are white, which implies that the lines $A \wedge B$ and $A \wedge C$ and $z_1 \wedge z_2$ intersect in a point and similarly, the lines $A \wedge C$, $B \wedge C$ and $z_8 \wedge z_9$ intersect in a point.

We have the following constraints left to satisfy
\begin{gather}
    \langle B C 4 5\rangle = \langle B C 6 7\rangle = \langle A C, 10, 11\rangle = \langle A B 3 4\rangle = 0.
\end{gather}
Since $A$ belongs to the line $z_1 \wedge z_2$ we have $A = z_2 + \r{\alpha} z_1 = \hat{2}$ for some complex $\r{\alpha}$.  Next, the constraint $\langle A C, 10, 11\rangle = 0$ can be solved by taking $C$ to belong to the plane $A \wedge z_{10} \wedge z_{11}$.  Since $C$ also belongs to the line $z_8 \wedge z_9$, we have $C = (8 9) \cap (\hat{2}, 10, 11) = -\hat{9}$.

The three remaining constraints involve the point $B$: $\langle B C 4 5\rangle = \langle B C 6 7\rangle = \langle A B 3 4\rangle = 0$.  Geometrically this means that $B$ belongs to the planes $C \wedge z_4 \wedge z_5$, $C \wedge z_6 \wedge z_7$ and $A \wedge z_3 \wedge z_4$, so it must belong to their intersection
\[
    B = (C 4 5) \cap (C 6 7) \cap (A 3 4) = (\hat{9} 4 5) \cap (\hat{9} 6 7) \cap (\hat{2} 3 4).
\]
If desired, the intersection $(\hat{9} 4 5) \cap (\hat{9} 6 7)$ can be expanded as
\[
    (\hat{9} 4 5) \cap (\hat{9} 6 7) = z_{\hat{9}} \wedge z_4 \langle 5 \hat{9} 6 7\rangle - z_{\hat{9}} \wedge z_5 \langle 4 \hat{9} 6 7\rangle = - z_{\hat{9}} \wedge z_{\hat{5}},
\]
where we have introduced $z_{\hat{5}} = (5 4) \cap (6 7 \hat{9}) = z_5 \langle 4 6 7 \hat{9}\rangle - z_4 \langle 5 6 7 \hat{9}\rangle$. 

Each hatted variable is linear in $\r{\alpha}$ so $B$ is cubic in $\r{\alpha}$.  The final on-shell condition we impose is $\langle A B 2 3\rangle = 0$ which is $\r{\alpha} \langle 1 2 3 B\rangle = 0$.  If we take $\r{\alpha} \neq 0$ we obtain a cubic polynomial in $\r{\alpha}$ from $\langle 1 2 3 B\rangle = 0$.

\newpage
\section{Explicit Parameterization of the Cubic Root}
\label{app:explicitparam}

In this appendix we present a few expressions that were too long for the main text, regarding the cluster parameterization in section~\ref{subsec:cluster}.

First, we give our parameterization for the momentum twistors in terms of our cluster chart, as these were too long for the main text:
{\allowdisplaybreaks
\myTwistors{2 e_6+1,e_7 \left(e_2 \left(e_8+1\right)+5\right),0,0
,1,0,0,0,
0,0,0,1,
0,0,\frac{3}{2} e_2 e_4 e_{10},\left(e_2 \left(2 e_7 e_6+e_6+1\right) e_4+e_4+1\right) e_{10}+1,
0,-e_1 e_2^2 e_3 e_4 e_8 \left(e_2 \left(e_8+1\right)+5\right) e_{10},\nonumber\\ &\quad\quad\frac{3}{2} e_2 e_4 \left(\left(e_3 \left(e_5+1\right) e_1+e_1+1\right) e_8 e_2+e_2+5\right) e_{10},\nonumber\\
&\quad\quad e_2+\big(\left(e_2+5\right) \left(e_2 \left(2 e_7 e_6+e_6+1\right) e_4+e_4+1\right)\nonumber\\
&\quad\quad+e_2 \big(e_4 \left(e_3+e_2 \left(\left(e_3+1\right) \left(e_6+1\right)+2 \left(e_3 \left(e_5+1\right)+1\right) e_6 e_7\right)+1\right) e_1\nonumber\\
&\quad\quad+e_1+e_4+e_2 e_4 \left(2 e_7 e_6+e_6+1\right)+1\big) e_8\big) e_{10}+5,
-e_2  e_3 e_4^2 e_8 e_9^2 e_{10},\nonumber\\
&\quad\quad-e_2 e_3 e_4 \left(e_2 \left(e_8+1\right)+5\right) \left(\left(e_8 \left(\left(e_7+1\right) e_9 e_4+e_4+1\right)+1\right) e_{10} e_9+e_9+5\right),\nonumber\\
&\quad\quad\frac{3}{2} e_2 e_4 \big(e_9+e_3 \big(\left(e_5+1\right) \left(e_9+5\right)\nonumber\\
&\quad\quad+e_9   \left(e_8 \left(e_4 \left(e_9+1\right)+1\right) e_5+e_5+e_8+1\right) e_{10}\big)+5\big),\nonumber\\
&\quad\quad e_4 \big(\left(e_3+e_2 \left(\left(e_3+1\right) \left(e_6+1\right)+2 \left(e_3 \left(e_5+1\right)+1\right) e_6 e_7\right)+1\right) \left(e_9+5\right)\nonumber\\
&\quad\quad+e_3 e_9   \big(e_2 \big(e_8+e_6 \big(e_8+e_7 \big(2 \left(e_8+1\right)\nonumber\\
&\quad\quad+e_5 \left(e_8 \left(e_4\left(e_9+2\right)+2\right)+2\right)\big)+1\big)+1\big)+1\big) e_{10}\big)+5,
-e_2 e_3 e_4^2 e_8 e_9^2 e_{10},\nonumber\\
&\quad\quad -e_2 e_3 e_4   \left(e_2 \left(e_8+1\right)+5\right) \left(\left(e_8 \left(\left(e_7+1\right) e_9 e_4+e_4+1\right)+1\right) e_{10} e_9+e_9+4\right),\nonumber\\
&\quad\quad\frac{3}{2} e_2 e_4 \big(e_9+e_3 \big(\left(e_5+1\right) \left(e_9+4\right)\nonumber\\
&\quad\quad+e_9 \left(e_8 \left(e_4   \left(e_9+1\right)+1\right) e_5+e_5+e_8+1\right) e_{10}\big)+4\big),\nonumber\\
&\quad\quad e_4 \big(\left(e_3+e_2\left(\left(e_3+1\right)   \left(e_6+1\right)+2\left(e_3 \left(e_5+1\right)+1\right) e_6 e_7\right)+1\right) \left(e_9+4\right)\nonumber\\
&\quad\quad+e_3 e_9 \big(e_2 \big(e_8+e_6   \big(e_8+e_7 \big(2 \left(e_8+1\right)\nonumber\\
&\quad\quad+e_5 \left(e_8 \left(e_4
   \left(e_9+2\right)+2\right)+2\right)\big)+1\big)+1\big)+1\big)e_{10}\big)+4,-\left(e_2+4\right) e_4 e_9,
   -\left(e_2   \left(e_8+1\right)+5\right) \left(\left(e_2+4\right)e_4\left(\left(e_7+1\right) e_9+1\right)+4\right),\nonumber\\
&\quad\quad\frac{3}{2} e_5   \left(\left(e_2+4\right) e_4\left(e_9+1\right)+4\right)+6,\nonumber\\
&\quad\quad e_6 \left(e_7 \left(e_5 \left(\left(e_2+4\right) e_4   \left(e_9+2\right)+8\right)+8\right)+4\right)+3,
-\left(e_2+3\right) e_4 e_9,-\left(e_2 \left(e_8+1\right)+5\right)   \left(\left(e_2+3\right) e_4 \left(\left(e_7+1\right) e_9+1\right)+3\right),\nonumber\\
&\quad\quad\frac{3}{2} \left(e_5 \left(\left(e_2+3\right) e_4   \left(e_9+1\right)+3\right)+3\right),\nonumber\\
&\quad\quad e_6 \left(e_7 \left(e_5 \left(\left(e_2+3\right) e_4
   \left(e_9+2\right)+6\right)+6\right)+3\right)+2,-1,-\frac{1}{3} \left(3 e_7+2\right) \left(e_2   \left(e_8+1\right)+5\right),e_5,\frac{1}{3} e_5 e_6 e_7,-4,-2 \left(2 e_7+1\right) \left(e_2 \left(e_8+1\right)+5\right),3   e_5,0}}

Second, we presented our cubic polynomial in the cluster coordinates in eq.~\ref{eq:clustercubic} in terms of coefficients $c_i$. These were also too long for the main text and are presented here:
{\allowdisplaybreaks
\begin{align}
   \hspace{-400pt}\fwboxR{0pt}{c_0=\,}&\fwboxL{0pt}{e_4 e_2^2+e_1 e_4 e_8 e_2^2+e_1 e_3 e_4 e_8 e_2^2+e_4 e_8 e_2^2+e_4 e_9 e_2^2+e_4 e_7 e_9 e_2^2+e_1 e_4 e_8 e_9 e_2^2}\\
   &\fwboxL{0pt}{+e_1 e_3 e_4
   e_8 e_9 e_2^2+e_4 e_8 e_9 e_2^2+e_1 e_4 e_7 e_8 e_9 e_2^2+e_1 e_3 e_4 e_7 e_8 e_9 e_2^2+e_4 e_7 e_8 e_9 e_2^2}\nonumber\\
   &\fwboxL{0pt}{+e_1 e_3 e_4 e_5 e_7 e_8
   e_9 e_2^2+2 e_4 e_2+e_1 e_8 e_2+e_1 e_4 e_8 e_2+e_1 e_3 e_4 e_8 e_2+e_4 e_8 e_2+e_8 e_2}\nonumber\\
   &\fwboxL{0pt}{+2 e_4 e_9 e_2+2 e_4 e_7 e_9 e_2+e_1 e_4 e_8
   e_9 e_2+e_1 e_3 e_4 e_8 e_9 e_2+e_4 e_8 e_9 e_2+e_1 e_4 e_7 e_8 e_9 e_2}\nonumber\\
   &\fwboxL{0pt}{+e_1 e_3 e_4 e_7 e_8 e_9 e_2+e_4 e_7 e_8 e_9 e_2+e_1 e_3 e_4
   e_5 e_7 e_8 e_9 e_2+e_2+e_4+e_4 e_9+e_4 e_7 e_9+1\,}\nonumber\\
   c_1=&e_4 e_2^2+4 e_4 e_6 e_2^2+e_1 e_4 e_8 e_2^2+e_1 e_3 e_4 e_8 e_2^2+e_4 e_8 e_2^2+4
   e_1 e_4 e_6 e_8 e_2^2+4 e_1 e_3 e_4 e_6 e_8 e_2^2\nonumber\\
   &+4 e_4 e_6 e_8 e_2^2+e_4 e_9 e_2^2+4 e_4 e_6 e_9 e_2^2+4 e_4 e_6 e_7 e_9 e_2^2+e_1
   e_4 e_8 e_9 e_2^2+e_1 e_3 e_4 e_8 e_9 e_2^2\nonumber\\
   &+e_4 e_8 e_9 e_2^2+4 e_1 e_4 e_6 e_8 e_9 e_2^2+4 e_1 e_3 e_4 e_6 e_8 e_9 e_2^2+4 e_4 e_6
   e_8 e_9 e_2^2+4 e_1 e_4 e_6 e_7 e_8 e_9 e_2^2\nonumber\\
   &+4 e_1 e_3 e_4 e_6 e_7 e_8 e_9 e_2^2+4 e_4 e_6 e_7 e_8 e_9 e_2^2+4 e_1 e_3 e_4 e_5 e_6
   e_7 e_8 e_9 e_2^2+2 e_4 e_2+10 e_4 e_6 e_2\nonumber\\
   &+5 e_6 e_2+e_1 e_8 e_2+e_1 e_4 e_8 e_2+e_1 e_3 e_4 e_8 e_2+e_4 e_8 e_2+5 e_1 e_6 e_8 e_2+5
   e_1 e_4 e_6 e_8 e_2\nonumber\\
   &+5 e_1 e_3 e_4 e_6 e_8 e_2+5 e_4 e_6 e_8 e_2+5 e_6 e_8 e_2+e_8 e_2+2 e_4 e_9 e_2+10 e_4 e_6 e_9 e_2\nonumber\\
   &+10 e_4 e_6 e_7
   e_9 e_2+e_1 e_4 e_8 e_9 e_2+e_1 e_3 e_4 e_8 e_9 e_2+e_4 e_8 e_9 e_2+5 e_1 e_4 e_6 e_8 e_9 e_2\nonumber\\
   &+5 e_1 e_3 e_4 e_6 e_8 e_9 e_2+5 e_4 e_6
   e_8 e_9 e_2+5 e_1 e_4 e_6 e_7 e_8 e_9 e_2+5 e_1 e_3 e_4 e_6 e_7 e_8 e_9 e_2\nonumber\\
   &+5 e_4 e_6 e_7 e_8 e_9 e_2+5 e_1 e_3 e_4 e_5 e_6 e_7 e_8
   e_9 e_2+e_2+e_4+6 e_4 e_6+6 e_6+e_4 e_9\nonumber\\
   &+6 e_4 e_6 e_9+6 e_4 e_6 e_7 e_9+1\\
   c_2=&e_6 \big(2 e_4 e_6^2 e_2^2+e_4 e_2^2+4 e_4 e_6 e_2^2+2 e_1
   e_4 e_6^2 e_8 e_2^2+2 e_1 e_3 e_4 e_6^2 e_8 e_2^2+2 e_4 e_6^2 e_8 e_2^2\nonumber\\
   &+e_1 e_4 e_8 e_2^2+e_1 e_3 e_4 e_8 e_2^2+e_4 e_8 e_2^2+4 e_1
   e_4 e_6 e_8 e_2^2+4 e_1 e_3 e_4 e_6 e_8 e_2^2+4 e_4 e_6 e_8 e_2^2\nonumber\\
   &+2 e_4 e_6^2 e_9 e_2^2+e_4 e_9 e_2^2+4 e_4 e_6 e_9 e_2^2+2 e_4 e_6^2
   e_7 e_9 e_2^2+3 e_4 e_6 e_7 e_9 e_2^2+2 e_1 e_4 e_6^2 e_8 e_9 e_2^2\nonumber\\
   &+2 e_1 e_3 e_4 e_6^2 e_8 e_9 e_2^2+2 e_4 e_6^2 e_8 e_9 e_2^2+e_1
   e_4 e_8 e_9 e_2^2+e_1 e_3 e_4 e_8 e_9 e_2^2+e_4 e_8 e_9 e_2^2\nonumber\\
   &+4 e_1 e_4 e_6 e_8 e_9 e_2^2+4 e_1 e_3 e_4 e_6 e_8 e_9 e_2^2+4 e_4 e_6
   e_8 e_9 e_2^2+2 e_1 e_4 e_6^2 e_7 e_8 e_9 e_2^2\nonumber\\
   &+2 e_1 e_3 e_4 e_6^2 e_7 e_8 e_9 e_2^2+2 e_4 e_6^2 e_7 e_8 e_9 e_2^2+2 e_1 e_3 e_4 e_5
   e_6^2 e_7 e_8 e_9 e_2^2+3 e_1 e_4 e_6 e_7 e_8 e_9 e_2^2\nonumber\\
   &+3 e_1 e_3 e_4 e_6 e_7 e_8 e_9 e_2^2+3 e_4 e_6 e_7 e_8 e_9 e_2^2+3 e_1 e_3 e_4
   e_5 e_6 e_7 e_8 e_9 e_2^2+8 e_4 e_6^2 e_2+4 e_6^2 e_2\nonumber\\
   &+3 e_4 e_2+14 e_4 e_6 e_2+6 e_6 e_2+4 e_1 e_6^2 e_8 e_2+4 e_1 e_4 e_6^2 e_8 e_2+4
   e_1 e_3 e_4 e_6^2 e_8 e_2\nonumber\\
   &+4 e_4 e_6^2 e_8 e_2+4 e_6^2 e_8 e_2+e_1 e_8 e_2+e_1 e_4 e_8 e_2+e_1 e_3 e_4 e_8 e_2+e_4 e_8 e_2+6 e_1 e_6
   e_8 e_2\nonumber\\
   &+6 e_1 e_4 e_6 e_8 e_2+6 e_1 e_3 e_4 e_6 e_8 e_2+6 e_4 e_6 e_8 e_2+6 e_6 e_8 e_2+e_8 e_2+8 e_4 e_6^2 e_9 e_2\nonumber\\
   &+3 e_4 e_9 e_2+14
   e_4 e_6 e_9 e_2+8 e_4 e_6^2 e_7 e_9 e_2+10 e_4 e_6 e_7 e_9 e_2+4 e_1 e_4 e_6^2 e_8 e_9 e_2\nonumber\\
   &+4 e_1 e_3 e_4 e_6^2 e_8 e_9 e_2+4 e_4 e_6^2
   e_8 e_9 e_2+e_1 e_4 e_8 e_9 e_2+e_1 e_3 e_4 e_8 e_9 e_2+e_4 e_8 e_9 e_2\nonumber\\
   &+6 e_1 e_4 e_6 e_8 e_9 e_2+6 e_1 e_3 e_4 e_6 e_8 e_9 e_2+6 e_4
   e_6 e_8 e_9 e_2+4 e_1 e_4 e_6^2 e_7 e_8 e_9 e_2\nonumber\\
   &+4 e_1 e_3 e_4 e_6^2 e_7 e_8 e_9 e_2+4 e_4 e_6^2 e_7 e_8 e_9 e_2+4 e_1 e_3 e_4 e_5
   e_6^2 e_7 e_8 e_9 e_2+4 e_1 e_4 e_6 e_7 e_8 e_9 e_2\nonumber\\
   &+4 e_1 e_3 e_4 e_6 e_7 e_8 e_9 e_2+4 e_4 e_6 e_7 e_8 e_9 e_2+4 e_1 e_3 e_4 e_5 e_6
   e_7 e_8 e_9 e_2+e_2+8 e_4 e_6^2+8 e_6^2\nonumber\\
   &+2 e_4+12 e_4 e_6+12 e_6+8 e_4 e_6^2 e_9+2 e_4 e_9\nonumber\\
   &+12 e_4 e_6 e_9+8 e_4 e_6^2 e_7 e_9+8 e_4 e_6
   e_7 e_9+2\big)\\
   c_3=&e_6^2 \big(e_4 e_2^2+2 e_4 e_6 e_2^2+e_1 e_4 e_8 e_2^2+e_1 e_3 e_4 e_8 e_2^2+e_4 e_8 e_2^2+2 e_1 e_4 e_6 e_8 e_2^2\nonumber\\
   &+2
   e_1 e_3 e_4 e_6 e_8 e_2^2+2 e_4 e_6 e_8 e_2^2+e_4 e_9 e_2^2+2 e_4 e_6 e_9 e_2^2+2 e_4 e_6 e_7 e_9 e_2^2+e_1 e_4 e_8 e_9 e_2^2\nonumber\\
   &+e_1 e_3
   e_4 e_8 e_9 e_2^2+e_4 e_8 e_9 e_2^2+2 e_1 e_4 e_6 e_8 e_9 e_2^2+2 e_1 e_3 e_4 e_6 e_8 e_9 e_2^2+2 e_4 e_6 e_8 e_9 e_2^2\nonumber\\
   &+2 e_1 e_4 e_6
   e_7 e_8 e_9 e_2^2+2 e_1 e_3 e_4 e_6 e_7 e_8 e_9 e_2^2+2 e_4 e_6 e_7 e_8 e_9 e_2^2+2 e_1 e_3 e_4 e_5 e_6 e_7 e_8 e_9 e_2^2\nonumber\\
   &+4 e_4 e_2+8
   e_4 e_6 e_2+4 e_6 e_2+2 e_1 e_8 e_2+2 e_1 e_4 e_8 e_2+2 e_1 e_3 e_4 e_8 e_2+2 e_4 e_8 e_2\nonumber\\
   &+4 e_1 e_6 e_8 e_2+4 e_1 e_4 e_6 e_8 e_2+4
   e_1 e_3 e_4 e_6 e_8 e_2+4 e_4 e_6 e_8 e_2+4 e_6 e_8 e_2+2 e_8 e_2\nonumber\\
   &+4 e_4 e_9 e_2+8 e_4 e_6 e_9 e_2+8 e_4 e_6 e_7 e_9 e_2+2 e_1 e_4 e_8
   e_9 e_2+2 e_1 e_3 e_4 e_8 e_9 e_2+2 e_4 e_8 e_9 e_2\nonumber\\
   &+4 e_1 e_4 e_6 e_8 e_9 e_2+4 e_1 e_3 e_4 e_6 e_8 e_9 e_2+4 e_4 e_6 e_8 e_9 e_2+4
   e_1 e_4 e_6 e_7 e_8 e_9 e_2\nonumber\\
   &+4 e_1 e_3 e_4 e_6 e_7 e_8 e_9 e_2+4 e_4 e_6 e_7 e_8 e_9 e_2+4 e_1 e_3 e_4 e_5 e_6 e_7 e_8 e_9 e_2+2 e_2+4
   e_4+8 e_4 e_6\nonumber\\
   &+8 e_6+4 e_4 e_9+8 e_4 e_6 e_9+8 e_4 e_6 e_7 e_9+4\big)\,.
\end{align}}

\end{document}